\documentclass[reprint, aps, prx, longbibliography, twocolumn]{revtex4-1}
\usepackage{epsf}
\usepackage{graphicx}
\usepackage{subfigure}
\usepackage{bbm}
\usepackage{amsmath}
\usepackage[colorlinks,linkcolor=red]{hyperref}
\usepackage{bm}
\usepackage{graphicx}
\usepackage{picinpar}
\usepackage{times}
\usepackage{indentfirst}
\usepackage{textcomp}
\usepackage{endnotes}
\usepackage{centernot}
\usepackage{array}
\usepackage{fancyhdr}
\usepackage{float}
\usepackage{multirow}
\newcommand{\cut}[1]{{}}
\newcommand{\new}{\textcolor{black}}
\newcommand{\hanlin}{\textcolor{black}}
\newcommand{\DS}{\textcolor{black}}
\newtheorem{theorem}{Theorem}

\newtheorem{fact}[theorem]{Fact}
\graphicspath{ {./Images/} }
\begin{document}
\title{Competition, Collaboration, and Optimization in Multiple Interacting Spreading Processes}
\author{Hanlin Sun}
\email{hanlin.sun@qmul.ac.uk} 
\affiliation{School of Mathematical Sciences,
	Queen Mary University of London, Mile End Rd, London E1 4NS, United Kingdom}
	
		\author{David Saad}
	\email{d.saad@aston.ac.uk}
	\affiliation{Non-linearity and Complexity Research Group, Aston University, Birmingham, B4 7ET, United Kingdom}
	
		\author{Andrey Y. Lokhov}
	\email{lokhov@lanl.gov}
	\affiliation{Theoretical Division, Los Alamos National Laboratory, Los Alamos, NM 87545, United States}
	
\begin{abstract}
	Competition and collaboration are at the heart of multi-agent probabilistic spreading processes. The battle on public opinion and competitive marketing campaigns are typical examples of the former, while the joint spread of multiple diseases such as HIV and tuberculosis demonstrates the latter. These spreads are influenced by the underlying network topology, the infection rates between network constituents, recovery rates and, equally importantly, the interactions between the spreading processes themselves. Here, \new{for the first time we derive dynamic message-passing equations that provide an exact description of the dynamics of two interacting \DS{unidirectional} spreading processes on tree graphs, and develop systematic low-complexity models that predict the spread on general graphs.}
   \new{We also develop a theoretical framework for an optimal control of interacting spreading processes through \cut{an} optimized resource allocation under budget constraints and within a finite time window. Derived algorithms can be used to maximize the desired spread in the presence of a rival competitive process, and to limit the spread through vaccination in the case of coupled infectious diseases.} We demonstrate the efficacy of the framework and optimization method on both synthetic and real-world networks. 
\end{abstract}
\maketitle
\section{Introduction}

Spreading processes have become increasingly more important in the fast moving modern world, where physical mobility is cheaper and easier than ever before, and where information is passed instantaneously on multilayered interwoven webs of contacts. Consequently, pandemics, whether physical or virtual in the form of computer viruses, Internet rumors or marketing campaigns spread very rapidly. \new{For instance, the\cut{an} ongoing COVID-19 pandemic \cite{wu2020new, zhou2020pneumonia} caused an unprecedented disruption in world's functioning.} Previously, \hanlin{an outbreak} of African swine fever caused by \emph{Asfivirus} in China~\cite{Zhou2018} in 2018 posed the risk of spreading globally. In 2017, a worldwide cyberattack by the WannaCry ransomware cryptoworm was estimated to have affected more then 200,000 computers across 150 countries, with total damages ranging from hundreds of millions to billions of dollars~\cite{Mohurle2017}.

\new{Very often, spreading processes do not diffuse on their own, but instead show a complex dynamics, characterized by a collaboration or competition between them.} An example for a collaborative spreading process is the co-infection of HIV and tuberculosis, the latter being \new{a major factor that influences death rates} of AIDS patients~\cite{TubAIDS}. The risk of developing tuberculosis is estimated to be 16-27 times higher in people living with HIV than among uninfected individuals~\cite{Bruchfeld2015}. AIDS patients are more susceptible to tuberculosis, due to their weakened immune system, and tuberculosis can also activate the replication of HIV virus. Epidemiological studies have also shown that the co-infection also exists between HIV and many other diseases including malaria parasites~\cite{Jegede2017}, \new{herpes~\cite{Munawwar2016}}, fungal~\cite{LIMPER2017e334} and bacteria~\cite{PMID:8089471}, and between Zika and Dengue viruses~\cite{Dupont-Rouzeyrol2015}. \new{\hanlin{An example} of asymmetric collaborative spreading is given by Hepatitis D that can only transmit to people who already infected with Hepatitis B.}

An example for a competitive process can be the spreading of rumors. Recently, the ``Anti-Vaxx movement'' in the US, has attracted the attention of the public, for instance through Twitter messages, leading to the belief of a growing number of parents that vaccination is a violation of human-rights, that vaccines cause autism, brain damage and do not benefit the health and safety of society. The idea has been spreading rapidly through social media. As a result, the measles virus, which was declared to be eliminated in 2000, is making a comeback~\cite{Hotez2016}; consequently, the World Health Organization declared Vaccine hesitancy as one of the top 10 global threats~\cite{WHOprogrammeofWork} and social network platforms have been requested to block the spread of related information~\cite{Guardian2019}. \new{A realistic competitive scenario is given by a situation where} ``valid information'' and ``unsubstantiated rumors'' spread on the network simultaneously, and where individuals exposed to one tend to believe in its content and are less susceptible to the other.

\new{Two important problems that recurrently arise in the analysis of spreading processes are \emph{forecasting} \cut{of} the dynamics, and the optimal use of resources to \emph{control} the dynamics, for instance \hanlin{to} maximize or minimize the spread. Forecasting is based on a probabilistic modeling and inference of the system state, such as prediction of infection probability over time for given initial conditions and interaction type. Optimization is often refereed to resource allocation tasks such as the initial choice of best spreaders or to the best vaccination strategy to contain an outbreak of a disease. In this paper, we develop novel methods that address current gaps pertaining to both inference and optimization of interaction spreading processes.}

There exists a large body of work on probabilistic modeling of spreading processes.
Most commonly studied models include single spreading processes that follow the SIR (Susceptible, Infected, Recovered/Removed), SI (Susceptible, Infected), and SIS (Susceptible, Infected, \new{Susceptible) dynamics~\cite{RevModPhys.87.925,Kiss2017},} where variables can take up a small set of \hanlin{statuses such as $S$, $I$, $R$} and transition from one \hanlin{status} to another depending on their original \hanlin{status} and that of their neighbors.
\new{Exact prediction of the spread within these models is NP-hard~\cite{shapiro2012finding}, and therefore the dynamics has been approximately analyzed using a variety of mean-field methods, see~\cite{anderson1992infectious,Boccaletti2006,Rogers2010,pastor2015epidemic,Gleeson2011,Gleeson2013,Kiss2017} for a review. A mean-field method of the message-passing type that is particularly suited for approximating dynamics of continuous and discrete SIR-type models on sparse network has been introduced in \cite{KarrerNewman2010, shrestha2014message, lokhov2014inferring}, which in particular gives an \emph{exact} prediction of the spread on tree graphs. When averaged over an ensemble of graphs, this method is equivalent to the Edge-based Compartmental Modeling (EBCM) method~\cite{Miller2011,Miller2012,Sherborne2018} derived using the cavity method type arguments and the correct choice of dynamic variables that allows one to close the system of equations. Yet another equivalent representation is given in terms of \emph{Dynamic Belief Propagation} (DBP) equations on time trajectories that \cut{has been discussed}\new{was presented}  in~\cite{altarelli2013optimizing, Lokhov2015, guggiola2015minimal}. A framework introduced in \cite{Lokhov2015} showed how starting from a DBP representation allows one to systematically derive closed-form \emph{Dynamic Message-Passing} (DMP) equations for any models with unidirectional dynamics \DS{(so that variable statuses cannot be revisited)}. In particular, the method of \cite{Lokhov2015} not only recovers previously known DMP equations for simple SIR-type models~\cite{KarrerNewman2010, Miller2012, lokhov2014inferring, lokhov2019scalable}, but also allows one to derive DMP equations for more complex models with multiple neighbor-dependent transitions, where guessing correct dynamic variables becomes incomparably harder.}

The analysis of multi-agent spreading is much more involved due the interaction between processes and its impact on the spreading dynamics. Numerical studies of multi-agent processes~\cite{Cai2015} have revealed the existence of new transitions, as a result of the cross-process interaction, and an aggressive spreading mode, which points to a percolation transition. These results highlight the risk of an unpredictable and violent outbreak in cooperative spreading scenarios. The most relevant studies to the current work focus on the analysis of multi-agent spreading in a competitive scenario on a specific network, using continuous equations similar to those of dynamic message-passing~\cite{PhysRevE.87.060801}; and on a two-stage infection process, which is a specific case for multi-agent spreading processes~\cite{10.1371/journal.pone.0071321}. \new{Also relevant to our work are studies of complex contagions~\cite{Gleeson2007, Miller2015}, characterized by the requirement for multiple transmissions before a network node changes \hanlin{status}. While this is not exactly the scenario of interacting processes that we analyze in this paper, its dynamics also depend on the infection-history similarly to the scenarios we examine. In this case, the interplay between topology and initial conditions may give rise to hybrid phase transitions when cascades are only possible for sufficiently prevalent initial infections. We can envisage similar phenomena in the scenarios studied here for some infection probabilities but have not observed it in the experiments carried out here as this has not been the focus of our study. Interestingly, recent work~\cite{HbertDufresne2019InteractingCA} shows a mapping between interacting multi-agent spreading processes and social reinforcement infections through multiple transmissions. }

\new{Competitive~\cite{PhysRevX.4.041005,PhysRevE.87.060801,PhysRevE.84.036106,PhysRevE.89.062817,Vasconcelos_2019,ventura2020role,7997937} and collaborative~\cite{Chen_2013,Hebert-Dufresne10551,PhysRevE.93.042316,PhysRevE.93.042303,Janssen_2016,PhysRevE.96.022301,10.2307/4096293,doi:10.1063/1.4996807,Chen_2017,chang_2018,doi:10.1063/1.5010002} spreading processes have been studied in different contexts and in a variety of scenarios. The foci of many of these studies have been the fixed-point properties of the system, such as a phase diagram~\cite{PhysRevE.84.036106,Chen_2013,PhysRevE.89.062817,PhysRevE.84.036106,ventura2020role}, describing regimes where one spreading process dominates the infection map or where both processes co-infect the system nodes and the type of transition between phases, epidemic thresholds and the infection cluster size~\cite{doi:10.1063/1.5010002,Chen_2017,7997937,PhysRevX.4.041005,PhysRevE.87.060801,Hebert-Dufresne10551,PhysRevE.93.042316,PhysRevE.93.042303,Janssen_2016,PhysRevE.96.022301}. These analyses mostly do not require a full solution of the dynamics. Other studies focus on dynamical properties of the fraction of infected network nodes by one of the processes or both~\cite{ventura2020role,PhysRevX.4.041005} by investigating the corresponding differential equations to identify phenomena such as hysteresis~\cite{doi:10.1063/1.4996807} particularly in the SIS model scenario, the relation between topology and dynamics~\cite{Hebert-Dufresne10551,PhysRevE.96.022301} and the emergence of infected clusters~\cite{sajjadi2020impact} linked to temporal correlations. Additionally, most studies focus on analyzing networks of different degree distributions~\cite{PhysRevE.96.022301, PhysRevX.4.041005,PhysRevE.87.060801,PhysRevE.84.036106,Chen_2013,PhysRevE.93.042316,Janssen_2016,doi:10.1063/1.5010002}, rather than specific network instances.}

\new{A recent attempt to extend \cut{DMP}\new{message passing} equations to the case of cooperative epidemic spreading~\cite{min2020message}, which is most relevant to the current study, only focused on a particular case where transmission is independent \hanlin{of} the \hanlin{status} of the target node. \new{Additionally, the work studies different degree distributions, rather than specific instances, and concentrates on the fixed point properties such as the phase diagram and infected cluster size, falling short of a complete description of the dynamics.} Hence, unlike in the case of single dynamics, exact equations for \new{describing the complete dynamics of} general interacting processes on tree graphs remain unknown. In this work, \emph{our first major contribution consists in a derivation of DMP equations for interacting \DS{unidirectional} spreading processes that are exact on tree graphs, including low-complexity message-passing equations for the case of collaborative interactions. Moreover, we study approximate schemes to these equations that result in simplified expressions that can be applied on general sparse networks.}}

Optimal resource deployment in various spreading settings has been mostly investigated in the case of a single spreading process. One of the most commonly studied \hanlin{problems} is identifying the most influential spreaders, on which the deployment of resource at time zero would maximize the spread at a given end time. Most of these studies rely on the network's topological properties and selection strategies are based on high-degree nodes~\cite{pastor2002immunization}, neighbors of randomly selected vertices~\cite{cohen2003efficient}, betweenness centrality~\cite{holme2002attack}, random-walk~\cite{holme2004efficient}, graph-partitioning~\cite{chen2008finding}, and k-shell decomposition~\cite{kitsak2010identification}. These approaches mostly ignore important dynamical aspects that impact on performance~\cite{borge2012absence, hebert2013global}. A related approach termed network dismantling~\cite{morone2015influence,mugisha2016identifying,braunstein2016network} aims at identifying the nodes which, if removed, lead to the fragmentation of the giant component and prevent the global percolation. The optimal deployment of immunization has been addressed using a belief propagation algorithm~\cite{altarelli2014containing}, based on cavity method techniques developed previously for deterministic threshold models~\cite{altarelli2013optimizing, guggiola2015minimal}. Several scenarios that incorporate the dynamical properties of the spreading process, such as the optimal seeding problem, where one allocates the set of initially infected nodes that maximizes the spread asymptotically, have been studied and analyzed~\cite{domingos2001mining,kempe2003maximizing,chen2013information}. The optimal seeding problem has been analyzed for the simple diffusion models of Independent Cascade (IC) and Linear Threshold, and the optimal seeding problem has been shown to be NP-hard~\cite{Kempe2003} for both, i.e., there are no deterministic algorithmic solutions that grow polynomially with the system size. A different perspective is given by the study of scenarios with a finite time horizon as studied for the 
IC~\cite{du2013scalable} and other spreading models~\cite{nowzari2015analysis}. Most relevant to the current study is the application of a recurrent optimization framework~\cite{Lokhov2017} to the DMP-based probabilistic modeling of spreading processes~\cite{Lokhov2015}. The framework also facilitates both open-loop resource allocation (a one-off preplanned  assignment) and a closed-loop (dynamical resource deployment with feedback) under a limited remedial budget. We utilize a similar framework to investigate and optimize the dynamics of \DS{multiple} spreading processes.

To the best of our knowledge, no analysis and optimization algorithms have been offered to address the \emph{general} case of multi-process modeling and optimization, namely incorporating both detailed topologies and dynamical properties within a fixed time window for both inference and optimization. \new{Special cases, such as optimal seeding, have been addressed mostly via linearized fixed-point analysis~\cite{min2020message} and for a simple dynamic that lends itself to single time-step optimization.} Moreover, most optimization algorithms for single-agent processes \DS{follow the spread on a static network topology and} cannot fully capture the intricate dynamics of \DS{multiple spreading} processes\DS{; they are} therefore less effective for \DS{the} optimization tasks \DS{we aim to solve}. \new{\emph{As a second major contribution, we build an optimization framework for both competitive and collaborative scenarios based on the derived DMP equations that enable an accurate probabilistic forecasting}}.

We demonstrate that the inference method that we construct in this paper provides an accurate dynamical description of both competitive and collaborative scenarios on both toy and large-scale problems; it is asymptotically exact on tree-like networks and provides a good approximation on networks with loops. We develop a related optimization algorithm for maximizing the spread within a given time window against a competing spread, as well as the containment of spreads in a collaborative spreading scenario through an optimized vaccination strategy that curbs one of the spreading processes. We demonstrate the efficacy of the suggested algorithm, offering excellent results with a scalable computational complexity.

The paper is organized as follows: In Section~\ref{sec:model}, we derive exact and approximate DMP equations for general models of multiple interacting spreading processes. In Section~\ref{sec:validation}, we validate the efficacy of the probabilistic modeling by comparing the results with Monte Carlo simulations on synthetic and real networks. The optimization algorithm is introduced in Sec.~\ref{sec:optimization} for both competitive and cooperative scenarios and is tested on synthetic networks in Sec.~\ref{sec:toyopt}. In Section~\ref{sec:realworld} we apply the optimization algorithm to real world networks for demonstrating its usefulness to more realistic scenarios. These include both competitive and collaborative cases, and the optimal deployment of vaccines to contain an epidemic. A summary and outlook are provided in Sec.~\ref{sec:summary}.

\section{Model and Dynamic Message-Passing Equations}
\label{sec:model}
\new{Spreading models studied in this work} are based on the \new{discrete-time} SI process, where a couple of spreading agents are active in parallel and interact with one another. The implication is that the \hanlin{status} of a network vertex determines its susceptibility to be infected by either (or both) of the spreading processes. For instance, in \emph{mutually exclusive competitive processes} that describes, for instance, the battle for public opinion, once a vertex has been infected by one process it cannot be infected by any other and retains its \hanlin{status}, which is also termed a ``cross-immunity''~\cite{10.1371/journal.pone.0071321}. 
In a \emph{collaborative spreading} scenario that describes, for example, the spread of multiple diseases, being infected by one process increases the susceptibility of being infected by another according to some predefined conditional probability. Another variant of the model considered here is that of vaccination in the presence of collaboratively spreading diseases, where vaccination against one agent affects the spread of both processes. Although the framework for the various scenarios is similar, it does include some important differences and will therefore be developed separately. \new{It is important to point out that the introduction of cooperative/competitive spreading processes cannot simply be reduced to a stochastic process with more \hanlin{statuses}\DS{, which would include co-infection statuses}; the new interactions between \hanlin{statuses} complicate the dynamics due to the dependence of the interaction probability on the \hanlin{status} of neighboring variables.}

\new{
In both competitive and collaborative scenarios, we assume that two SI-type processes are spreading in discrete time on a graph $G \!=\! (V, E)$ comprising the set of vertices $V$ and edges $E$ such that each node $i$ can generally be found in one of four \hanlin{statuses} at any time \new{step} $t$: susceptible ($\sigma_i(t) \!=\! S$), infected by disease $A$, ($\sigma_i(t) \!=\! A$), infected by disease $B$, ($\sigma_i(t) \!=\! B$), or activated by both processes $A$ and $B$, ($\sigma_i(t) \!=\! AB$). In what follows, we define the spreading model in two scenarios, and derive the corresponding DMP equations.
}

\subsection{From Dynamic Belief Propagation to Dynamic Message Passing}

\new{
DMP equations can be hard to guess beyond simplest models, but can be systematically derived starting from the general Dynamic Belief Propagation algorithm that approximates probabilities of time trajectories of individual nodes \cite{Lokhov2015}. This is the approach that we will be undertaking here to obtain exact equations on tree graphs. For the two-processes dynamics considered in this work, the dynamics of a single node $i$ is fully described by a pair of activation times, $(\tau_i^A,\tau_i^B)$, where $\tau_i^A$ denotes the first time when node $i$ is found in the \hanlin{status} $A$, and similarly for $\tau_i^B$. For instance, $\tau_i^A = 0$ means that node $i$ was initially in the active \hanlin{status} $A$, and we will denote by $\tau_i^A = *$ the situation where node $i$ did not get $A$-activated before some final observation time, in other words $*$ absorbs all the history that happens after the end of the observation window. For the convenience of presentation, in what follows we will consider two separate ``observation windows'', for $A$ and $B$ processes.}

\new{
The starting point for our derivations are the general \cut{Dynamic Belief Propagation}DBP equations~\cite{kanoria2011majority, altarelli2013large, Lokhov2015} on the interaction graph, where the goal is to approximate the probability $m^i_{T_A+1,T_B+1}(\tau_i^A,\tau_i^B)$ that node $i$ \cut{has}exhibits a trajectory $(\tau_i^A,\tau_i^B)$ during the observation time window of length $T_A$ for process $A$ and $T_B$ for process $B$ \DS{(we keep the flexibility of having two separate time windows although in most situations we use $T_A = T_B = t$)}. Exact equations that compute \hanlin{the} probability $m^i_{T_A+1,T_B+1}(\tau_i^A,\tau_i^B)$ of a given time trajectory $(\tau_i^A,\tau_i^B)$ of node $i$ are explained in Appendix~A. Due to the properties of Belief Propagation algorithm \cite{MezardMontanari2009, wainwright2008graphical}, the fixed point solution of the DBP equations is guaranteed to be exact on tree graphs, and provides good estimates of marginal probabilities on loopy but sparse graphs \cite{Lokhov2015}.
}

\new{
Given the computed value of the marginals $m_{T_A,T_B}^i(\tau_i^A,\tau_i^B)$, one can straightforwardly define quantities of interest, such as probabilities for a given node $i$ to be found in a given \hanlin{status}:
\begin{align}
    &P^i_S(t) = \hanlin{\sum_{\tau_i^A > t} \sum_{\tau_i^B > t}} m^i_{T_A,T_B}(\tau_i^A,\tau_i^B),
    \label{eq:Marginal_Probability_S}
    \\
    &P^i_A(t) = \hanlin{\sum_{\tau_i^A \leq t} \sum_{\tau_i^B}} m_{T_A,T_B}^i(\tau_i^A,\tau_i^B),
    \label{eq:Marginal_Probability_A}
    \\
    &P^i_B(t) = \hanlin{\sum_{\tau_i^B \leq t} \sum_{\tau_i^A}} m_{T_A,T_B}^i(\tau_i^A,\tau_i^B),
    \label{eq:Marginal_Probability_B}
    \\
    &P^i_{AB}(t) = \hanlin{\sum_{\tau_i^A \leq t} \sum_{\tau_i^B \leq t}} m_{T_A,T_B}^i(\tau_i^A,\tau_i^B).
    \label{eq:Marginal_Probability_AB}
\end{align}
\DS{The definition of $P_A^i(t)$ ($P_B^i(t)$) indicates that at time $t$ the vertex $i$ is $A$-activated ($B$-activated) irrespective of the other process. Therefore $P_A^i(t)$ ($P_B^i(t)$) represents the probability that at time $t$ vertex $i$ is in status $A$ ($B$) alone, or in status $AB$. Initial condition probabilities where the node is exclusively found is status $A$ ($B$) will be denoted as $P_{A^*}^i(0)$ ($P_{B^*}^i(0)$).} In principle, one can solve the DBP equations to obtain $m^i_{T_A+1,T_B+1}(\tau_i^A,\tau_i^B)$, and then use the expressions \eqref{eq:Marginal_Probability_S}-\eqref{eq:Marginal_Probability_AB} to obtain final aggregated expressions for the dynamic messages $P^i_S(t)$, $P^i_A(t)$, $P^i_B(t)$, and $P^i_{AB}(t)$. However, due to the generality of the DBP equations that are valid for any dynamics, this approach may not be the most efficient one: computing a single marginal $m^i_{t+1,t+1}(\tau_i^A,\tau_i^B)$ may require as many as $O(t^{2d})$ operations for a single marginal, where $d$ is the degree of the node $i$. On the other hand, for the concrete dynamics such as the processes considered here that often have a special structure, \cut{and} it is beneficial from \cut{the}a computational point of view to explore this structure in order to drastically reduce the complexity of \cut{the computation of}computing the dynamic marginals $P^i_S(t)$, $P^i_A(t)$, $P^i_B(t)$, and $P^i_{AB}(t)$. \cut{If}Where this special structure \cut{can be}is exploited to produce closed-form algebraic equations for computing $P^i_S(t)$, $P^i_A(t)$, $P^i_B(t)$, and $P^i_{\hanlin{AB}}(t)$ with low algorithmic complexity, the resulting computational scheme will be referred to as Dynamic Message-Passing equations for the given process.
}

\new{
Therefore, the procedure \cut{that} we adopt below for deriving DMP equations will be as follows: (i) specify DBP equations based off a given two-processes dynamics; (ii) \cut{if}where possible, exploit the structure in the dynamics to derive low-complexity closed-form DMP equations that iteratively compute the quantities of interest $P^i_S(t)$, $P^i_A(t)$, $P^i_B(t)$, and $P^i_{AB}(t)$ starting with the algebraic definitions \eqref{eq:Marginal_Probability_S}-\eqref{eq:Marginal_Probability_AB}, which will inherit exactness of prediction of tree graphs; and (iii) to further reduce the computational complexity or the algebraic form of the exact equations, derive \emph{approximate} DMP equations that could be used as an algorithm for inference or optimization problems on general graphs.
}

\subsection{Mutually Exclusive Competitive Processes}
\DS{The dynamics in mutually exclusive competitive
processes can be made explicit by listing the allowed transitions
and their respective probabilities at every discrete time step:}
\begin{eqnarray}
 \label{eq:infecdynamics}
	S(i) + A(j) &\stackrel{\alpha^A_{ji}}{\longrightarrow}& A(i) + A(j)
\\
	S(i) + B(j) &\stackrel{\alpha^B_{ji}}{\longrightarrow}& B(i) + B(j)~~. \nonumber
\end{eqnarray}
In other words, the two infection processes $A$ and $B$ are mutually exclusive: any node can be infected by a neighboring nodes, assuming \emph{one of the two \hanlin{statuses}, but once infected by one of the two processes}, it cannot change its \hanlin{status}. Since the infection is \new{based on a two-vertex interaction through an edge}, the processes in~(\ref{eq:infecdynamics}) seem deceptively as two completely independent parallel processes; however, they clearly interact through the graph topology and the exclusivity of the adopted \hanlin{statuses}.
Indeed, we assume that at any given time step, the infection probabilities of process $A$ (denoted $\alpha^A_{ji}$) or process $B$ (denoted $\alpha^B_{ji}$) are \new{treated as} independent, but the probability of being infected by both $A$ and $B$ simultaneously is forced to be 0 \new{(thus creating probabilistic dependence between the two processes)}. For closing the \new{update rule} we need to define what happens in the case where both processes jointly infect the vertex \hanlin{in the same time step}, resulting in an invalid \hanlin{status}. There are \new{many possible ways to deal with this case that could be accommodated in both analysis and simulations, depending on the needs of a particular application. For the sake of simplicity, we consider the rule where the probabilities of transitioning to either of the \hanlin{statuses} $A$ or $B$, or staying in the \hanlin{status} $S$ is proportionally renormalized in such a way that they sum to one. Alternatively, in simulation one could think of this procedure as \emph{resampling} in the case where the joint infection occurs: If a joint infection \hanlin{status} by both processes $A$ and $B$ is sampled, it is rejected and resampling is carried out.}
\new{This is done repeatedly until a valid \hanlin{status} without progressing the dynamics, such that no spurious probabilistic dependencies emerge. According to the vanilla dynamic rules defined above, the probability of transition to the \hanlin{status} $A$ from \hanlin{status} $\hanlin{S}$ for a node $i$ is given by
\begin{eqnarray}
    v^i_A(t) = 1 - \prod_{j \in \partial i} (1 - \alpha^{A}_{ji} \mathbbm{1}[\sigma_j(t) = A]),
\end{eqnarray}
where $\partial i$ denotes the set of neighbors of node $i$, and $\mathbbm{1}$ is an indicator function. Similarly, define
\begin{align}
    &v^i_B(t) = 1 - \prod_{j \in \partial i} (1 - \alpha^{B}_{ji} \mathbbm{1}[\sigma_j(t) = B]),
    \\
    &Z_i = 1 - v^i_A(t) v^i_B(t).
\end{align}
Then under the resampling procedure explained above, the final renormalized transition probabilities at time step $t$ read:
\begin{align}
    & q^i_{S \to A} (t) = \frac{v^i_A(t)(1-v^i_B(t))}{Z_i(t)},
    \label{eq:Transition_S_A}
    \\
    & q^i_{S \to B} (t) = \frac{v^i_B(t)(1-v^i_A(t))}{Z_i(t)},
    \label{eq:Transition_S_B}
    \\
    & q^i_{S \to S} (t) = \frac{(1 - v^i_A(t) - v^i_B(t) + v^i_A(t)v^i_B(t))}{Z_i(t)},
    \label{eq:Transition_S_S}
\end{align}
where notation \hanlin{$\sigma_i(t)=A/B/S$} refers to a node \hanlin{$i$} at time $t$, being in \emph{one} of the \hanlin{statuses} $A$, $B$ or $S$.}

\new{
Given expressions \eqref{eq:Transition_S_A}-\eqref{eq:Transition_S_S}, we can straightforwardly specify the respective DBP equations for the mutually exclusive competing dynamics of two processes that are given in Appendix~B. In Table~\ref{tab:competitive_DBP_MC}, we numerically verify that the DBP equations are exact on tree graphs. With a naive implementation, DBP marginals can be computed explicitly in time $\max(O(\vert E \vert T^{2(c-1)}),O(N T^{2c}))$, where $c$ is the maximum degree of the graph, and $T$ is the final observation time. It is important to note that transition probabilities defined as in \eqref{eq:Transition_S_A}-\eqref{eq:Transition_S_S} reflect the complexity of interaction between processes that results from the renormalization of probabilities. Indeed, these expressions depend on the particular realization of \hanlin{statuses} for all neighbors, and hence lack any iterative structure at each time step. Due to this lack of structure, exact low-complexity DMP equations can not be straightforwardly derived for the chosen dynamics.
}

\new{
Although DBP equations for the mutually exclusive competing processes are exact on trees (see Table \ref{tab:competitive_DBP_MC}), their polynomial but potentially high computational complexity makes them a less attractive choice in applications. To address this issue, we notice that real application problems are typically defined on sparse, but non-tree graphs\DS{. Even} if low-complexity DMP equations were available, they would still yield only approximate solutions on general loopy network instances. This observation motivates us to search for a tractable approximation of the message-passing equations on tree graphs: if the approximation is good, the resulting error on tree-like loopy networks may be similar compared to the application of exact DBP or DMP equations. In Appendix~C, we implement this strategy and derive \emph{approximate} DMP equations for the mutually exclusive competing scenario with a computational complexity $O(\vert E \vert T)$, where $\vert E \vert$ is the number of edges in the graph and $T$ is the observation window, which makes them scalable for very large sparse networks with millions of nodes. The approximation that we use is inspired by the fact that in the absence of renormalization the dynamics of each processes follows the dynamics of the usual SI-type process, and hence it is natural to try to perform the renormalization procedure at the level of dynamic marginals. In what follows, we present numerical tests that illustrate the accuracy of the employed approximation. 
}

\begin{widetext}
\centering
    \begin{table*}
    \new{
    \begin{tabular}{ |c|c|c|c|c|c|c| }
    \hline
     & \multicolumn{2}{|c|}{Case 1} & \multicolumn{2}{|c|}{Case 2} & \multicolumn{2}{|c|}{Case 3}\\
     \cline{2-7}
     & DMP & MC &  DMP & MC & DMP & MC\\ 
     \hline
     ~~~$P^i_S(t)$~~~ & ~~~0.07986111~~~ & ~~~0.0799122~~~ & ~~~0.0077903~~~ & ~~~0.0078026~~~ & ~~~0.02302213~~~ & ~~~0.0230145~~~    \\ 
     ~~~$P^i_A(t)$~~~ & ~~~0.13194444~~~ & ~~~0.1318928~~~ & 0.00257234  & 0.0025947 & 0.00967836 & 0.0096883  \\ 
     ~~~$P^i_B(t)$~~~ & ~~~0.78819444~~~ & ~~~0.788195~~~ & 0.98963737 & 0.9896027 & 0.967299512 & 0.9672972  \\
     \hline
    \end{tabular}
    \caption{Demonstration of exactness of DBP equations for mutually exclusive competitive processes via numerical comparison with the results from $10^8$ Monte Carlo simulations on random trees. {\bf Case 1:} 4-node random tree with uniform parameters $\alpha_A=0.5$, $\alpha_B=0.5$ and the initial conditions $P_A^2(0)=1$ and $P_B^3(0)=1$. Marginal probabilities for node $i=0$ at time $t=3$ are presented. {\bf Case 2:} 4-node random tree with uniform parameters $\alpha_A=0.2$, $\alpha_B=0.8$ and the initial conditions $P_A^2(0)=1$ and $P_B^3(0)=1$. Marginal probabilities for node $i=0$ at time $t=3$ are presented. {\bf Case 3:} 5-node random tree with uniform parameters $\alpha_A=0.4$, $\alpha_B=0.6$ and the initial conditions $P_A^2(0)=1$ and $P_B^3(0)=1$. Marginal probabilities for node $i=0$ at time $t=4$ are presented. \DS{Here and in the following tables and figures, the index of the chosen node has no significance but comes to emphasize that a {\em specific} node is investigated.}}
    \label{tab:competitive_DBP_MC}
    }
    \end{table*}
 \end{widetext}

\subsection{Collaborative Process}
\new{In the two-processes collaborative scenario considered here, when a node is infected by one process, its susceptibility to the activation by another process increases (or decreases). Unlike in the mutually exclusive case, a node can be infected by both processes and the influence of the different processes is not necessarily symmetric. The dynamics in collaborative processes can be made explicit by listing the possible transitions and their respective probabilities at every discrete time step:
\begin{align}
    \label{eq:main_Dynamical_Rules0}
    S(i) + A(j) &\xrightarrow{\alpha^{A}_{ji}} A(i) + A(j),\\
    S(i) + B(j) &\xrightarrow{\alpha^{B}_{ji}} B(i) + B(j),\\
    A(i) + B(j) &\xrightarrow{\alpha^{BA}_{ji}} AB(i) + B(j),\\
    A(i) + B(j) &\xrightarrow{\alpha^{AB}_{ij}} A(i) + AB(j).
    \label{eq:main_Dynamical_Rules}
\end{align}
}
\new{
In this scenario, \hanlin{status} $AB$ can be regarded as a combination of \hanlin{status} $A$ and \hanlin{status} $B$.
The non-triviality of the interaction between both processes comes from the fact that $\alpha^{AB}_{ij}$ and $\alpha^{BA}_{ji}$ are different from $\alpha^{A}_{ji}$ and $\alpha^{B}_{ji}$, respectively: when a node is infected by one process and becomes more (or less) vulnerable to another, and vice versa. Notice that unlike the mutually exclusive scenario where the \hanlin{status} $AB$ is forbidden, under the general collaborative scenario the process $S(i) \longrightarrow AB(i)$ is allowed, and in discrete time the rule
\begin{equation}
    S(i) + AB(j) \xrightarrow{\alpha^{A}_{ji} \times \alpha^{B}_{ji}} AB(i) + AB(j)
\end{equation}
follows from the transition rules above, simply as co-activation that happens at the same time.
}

\new{
Following the scheme outlined above, we can start by forming a dynamic transition kernel that encapsulates the various transition rules and allows us to write the DBP equations for this spreading model. The resulting DBP equations are given in Appendix~D.
The special structure of the dynamic kernel is written as a sum of possible transition sequences factorized over the neighbors of a given node, making it possible to derive low-complexity and \emph{exact} DMP equations for the collaborative model. The algebraic form of equations for the dynamic marginals is given a simple and intuitive meaning. The probability of finding node $i$ in the \hanlin{status} $S$ can be written as 
\begin{equation}
    P_S^i(t) = P_S^i(0)\prod_{k \in \partial i \backslash j}\theta_{A,B}^{k \rightarrow i}(t, t),
    \label{eq:main_P_S}
\end{equation}
where $\theta_{A,B}^{k \rightarrow i}(t, t)$ is an aggregated dynamic message defined through the fundamental messages on time trajectories:
\begin{align}
        \DS{\theta_{A,B}^{k \rightarrow i}(t,t) = \sum_{\tau_k^A} \sum_{\tau_k^B} \prod_{t'=0}^{t-1} (1-\alpha_{ki}^A\mathbbm{1}[\tau_k^A \leq t'])}&\\
        \DS{\times \prod_{t''=0}^{t-1} (1-\alpha_{ki}^B\mathbbm{1}[\tau_k^B \leq t'']) m^{k \rightarrow i}_{T_A,T_B}(\tau_k^A,\tau_k^B).}&
        \label{eq:main_theta_A_B_definition}
\end{align}
From the definition of $\theta_{A,B}^{k \rightarrow i}(t, t)$, it is easy to ``read off'' its physical meaning: it corresponds to the probability that node $k$ did not send activation signals \hanlin{either} $A$ \hanlin{or} $B$ before time $t$, while $i$ follows a fixed $(\tau_i^A, \tau_i^B) = (*,*)$ dynamics, i.e.\hanlin{,} it does not activate until time $t$ (\DS{see Appendices \ref{app:DBP} and \ref{app:collaborative_exactDMP} for more details}). This conveys the following meaning to the expression \eqref{eq:main_P_S}: $P_S^i(t)$ is given by the probability $i$ is in the \hanlin{status} $S$ at initial time, times the probability that \hanlin{none} of its neighbors has activated it with any of the processes until time $t$ (which exactly factorizes over neighbors on a tree graph).
}

\new{
In a similar way, the marginals corresponding to \hanlin{statuses} $A$ and $B$ can be expressed as activations by the time $t$:
\begin{align}
    P_A^i(t) = \sum_{t' \leq t}\mu_A^i(t'),
    \label{eq:main_P_A}\\
    P_B^i(t) = \sum_{t' \leq t}\mu_B^i(t'),
    \label{eq:main_P_B}
\end{align}
where reduced marginals $\mu_A^i(t')$ and $\mu_B^i(t')$ are defined as follows:
\begin{align}
    \mu^{i}_A(t) = \sum_{\tau_i^B} m_{t,t}^i(t,\tau_i^B) = P^i_A(t) - P^i_A(t-1),\\
    \mu^{i}_B(t) = \sum_{\tau_i^A} m_{t,t}^i(\tau_i^A,t) = P^i_B(t) - P^i_B(t-1).
    \label{eq:main_reduced_marginals}
\end{align}
Finally, using the normalization of probabilities \DS{(notice that by definition $P_{AB}^i(t)$ is contained in both $P_A^i(t)$ and $P_B^i(t)$)}, we finally get
\begin{equation}
    P_{AB}^i(t) = P_A^i(t) + P_B^i(t) + P_S^i(t) - 1.
    \label{eq:main_P_AB}
\end{equation}
\hanlin{The} exact form of the DMP equations for collaborative processes along with detailed derivations are provided in the Appendix~D. Due to exploitation of the structure properties, the DMP equations have a much lower computational complexity $O(\vert E \vert T^2)$ for a final observation time $T$ compared to the DBP equations, while still providing exact predictions on tree graphs. We numerically verify this fact on a number of problem instances as shown in Table~\ref{tab:collaborative}.
}

\new{
In the search for simplified equations that are easier to use for inference and optimization purposes, as well as reduced computational complexity in time, we implement a similar strategy to the one used for the mutually exclusive scenario and derive \emph{approximate} equations that could be used in lieu of exact DMP equations on general graph. To do so, we notice that the special case of transmission probabilities $\alpha_{ij}^{AB} = \alpha_{ij}^{A}$ and $\alpha_{ij}^{BA} = \alpha_{ij}^{B}$ for collaborative dynamics \eqref{eq:main_Dynamical_Rules0}-\eqref{eq:main_Dynamical_Rules} corresponds to non-interacting spreading processes: activation by one process does not change the activation dynamics for the other. In this case, the DMP equations should simplify into the product of two independent Susceptible-Infected (SI) like processes:
\begin{align}
    & m_{t,t}^i(\tau_i^A, \tau_i^B) = \mu_A^i(\tau_i^A) \mu_B^i(\tau_i^B)
    \\
    & \text{if  } \alpha_{ij}^{AB} = \alpha_{ij}^{A} \text{  and  } \alpha_{ij}^{BA} = \alpha_{ij}^{B}.
\end{align}
We use this observation to produce a simplified version of DMP equations, expanding the exact equations for interacting spreading processes around the non-interacting point. We keep certain first-order corrections in the update equations only, so that the resulting equations are similar to the SI-type equations. This approximate version is expected to be good as long as $\alpha_{ij}^{AB}$ and $\alpha_{ij}^{A}$, $\alpha_{ij}^{BA}$ and $\alpha_{ij}^{B}$ are similar. \DS{It would clearly break down if $\alpha_{ij}^{A}=0$ or $\alpha_{ij}^{B}=0$ while the other infection probabilities remain finite.} Full derivation and expressions for the approximate DMP equations are provided in Appendix~E.
}

\begin{widetext}
\centering
    \begin{table*}
    \new{
\begin{tabular}{ |c|c|c|c|c|c|c| } 
    \hline
     & \multicolumn{2}{|c|}{Case 1} & \multicolumn{2}{|c|}{Case 2} & \multicolumn{2}{|c|}{Case 3}\\
     \cline{2-7}
     & DMP & MC &  DMP & MC & DMP & MC\\ 
     \hline
     ~~~$P^i_S(t)$~~~ & ~~~0.22111846~~~ & ~~~0.22111299~~~ & ~~~0.0152~~~ & ~~~0.0152037~~~ & ~~~0.06912~~~ & ~~~0.06908835~~~    \\ 
     ~~~$P^i_A(t)$~~~ & ~~~0.02838733~~~ & ~~~0.02837872~~~ & 0.02859 & 0.02859528 & 0.11808 & 0.1180704  \\ 
     ~~~$P^i_B(t)$~~~ & ~~~0.07848602~~~ & ~~~0.07845333~~~ & 0.67248 & 0.67248319 & 0.12416 & 0.12412187  \\
     ~~~$P^i_{AB}(t)$~~~ & ~~~0.67200819~~~ & ~~~0.67205496~~~ & 0.28373 & 0.28371783 & 0.68864 & 0.68871938 \\
     \hline
    \end{tabular}
    \caption{Demonstrating the exactness of the DMP equations for collaborative processes via numerical comparison against results from $10^8$ Monte Carlo simulations on random trees.
    {\bf Case 1:} 6-node random tree with uniform parameters $\alpha_A=0.1$, $\alpha_B=0.2$, $\alpha_{AB}=0.8$, $\alpha_{BA}=0.9$, and the initial conditions $P_A^2(0)=1$ and $P_B^0(0)=1$. Marginal probabilities for node $i=3$ at time $t=5$ are presented.
    {\bf Case 2:} 6-node random tree with uniform parameters $\alpha_A=0.6$, $\alpha_B=0.5$, $\alpha_{AB}=0.2$, $\alpha_{BA}=0.3$, and the initial conditions $P_A^2(0)=1$ and $P_B^0(0)=1$. Marginal probabilities for node $i=3$ at time $t=5$ are presented.
    {\bf Case 3:} 5-node random tree with uniform parameters $\alpha_A=0.5$, $\alpha_B=0.4$, $\alpha_{AB}=0.8$, $\alpha_{BA}=0.6$, and the initial conditions $P_A^2(0)=1$ and $P_B^0(0)=1$. Marginal probabilities for node $i=3$ at time $t=3$ are presented. }
    \label{tab:collaborative}
    }
    \end{table*}
 \end{widetext}

\section{Inference using Approximate Dynamic Message-Passing Equations}
\label{sec:validation}

\new{
In Section~\ref{sec:model}, we considered DBP and DMP equations that are exact on tree graphs, which follows from their derivation and supporting numerical checks. In this Section, our goal is to numerically establish the validity of \emph{approximate} DMP equations introduced in the previous Section. These equations enjoy an improved computational complexity and a simpler algebraic form, and are expected to provide a good approximation in the regimes discussed above.
}

To validate the \new{approximate} DMP equations obtained for competitive/collaborative processes we test the accuracy of the inferred node values against numerical results obtained via Monte Carlo simulations. Testing is carried out on both synthetically generated networks and real instances.

\subsection{Inference in Competitive Spreading}
Validation is carried out on two synthetic networks generated by the package NetworkX, a tree network and a network with loops; and on a real world undirected benchmark Polbooks~\cite{Unpublish} network. The latter is provided as an example of a sparse network. With the given initial condition, we apply both Monte Carlo simulation and the DMP method to all models.

Exhaustive numerical experiments reveal that DMP-based inference provides an accurate marginal posterior probabilities for the variable \hanlin{statuses}, with the exception of very high infection parameters values (very close to 1). Therefore, we do not examine cases with extreme infection parameter values in the examples provided. The location of seeds initializing the processes and target nodes to be observed are chosen randomly and have no significance.

In Fig.~\ref{fig:graphmodels} we show two synthetically generated networks (a) a toy tree network of 10 nodes, (b) a network of 10 nodes with loops and (c) the Polbooks~\cite{Unpublish} network with 105 nodes. The choice of network has no significance, it is a standard benchmark network used in the literature, of books about US politics sold by Amazon~\cite{Unpublish}. The network comprises two sparsely connected hubs, where edges represent books (vertices) bought jointly by the same individuals.
To validate the efficacy of DMP in modeling competitive scenarios we compare results obtained from running equations~(A4)-(A9) against results obtained using Monte Carlo simulations. Simulations are carried out 10 times for gathering statistics, each round includes $10^3$ samples per node (about $10^5$ samplings in total, depending on the network size). The parameters used in the toy model tests are $\alpha_A\!=\!0.3$ and $\alpha_B\!=\!0.7$, and we observed the marginal posterior probabilities $P^{i=3}_{A}(t)$ in both Figs.~\ref{fig:toymodelsvalidation}(a) and (c), and $P^{i=3}_{S}(t)$ in Figs.~\ref{fig:toymodelsvalidation}(b) and (d) for the tree and loopy graphs, respectively. The seeds initializing the processes are placed on nodes 7 for process $A$ and on nodes 2 for process $B$. The choice of these particular nodes is arbitrary and has no significance. The results obtained show excellent agreement between theory and simulations. We also tested the accuracy of the method on the benchmark Polbooks network as shown in Fig.~\ref{fig:toymodelsvalidation}(e) and (f) for the parameters $\alpha_A\!=\!0.2$, $\alpha_B\!=\!0.2$. In this case, process $A$ starts from nodes 1 and 2, process $B$ from nodes 4 and 37 (again, both are arbitrary choices). The observed probabilities $P^{i=0}_{A}(t)$ in Fig.~\ref{fig:toymodelsvalidation}(e) and $P^{i=0}_{S}(t)$ in Fig.~\ref{fig:toymodelsvalidation}(e) show good agreement between theory and simulations. 

\subsection{Inference in Collaborative Spreading}
Similar experiments were run for a collaborative process on the toy tree network, graphs with loops and the benchmark Football network. The latter is an undirected network of American football games between colleges during season Fall 2000~\cite{Girvan2002} which is quite uniformly connected.

\begin{widetext}

	\begin{figure}[!ht]
	    \centering
		\includegraphics[width=6.3in]{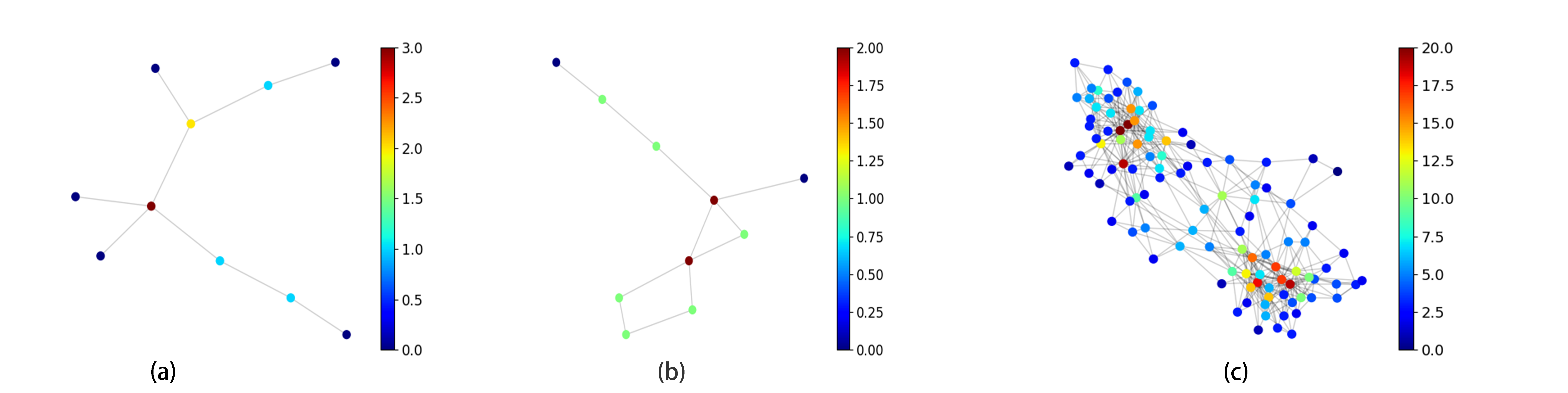}
		\caption{Networks used for validation: (a) a toy tree network of 10 nodes; (b) a toy network with loops of 10 nodes;  (c) the Polbooks~\cite{Unpublish} with 105 nodes. The color scale represents the degree of nodes.\label{fig:graphmodels}} 
	\end{figure}
	
\begin{figure}[!h]
    \centering
    \includegraphics[width=6.4in]{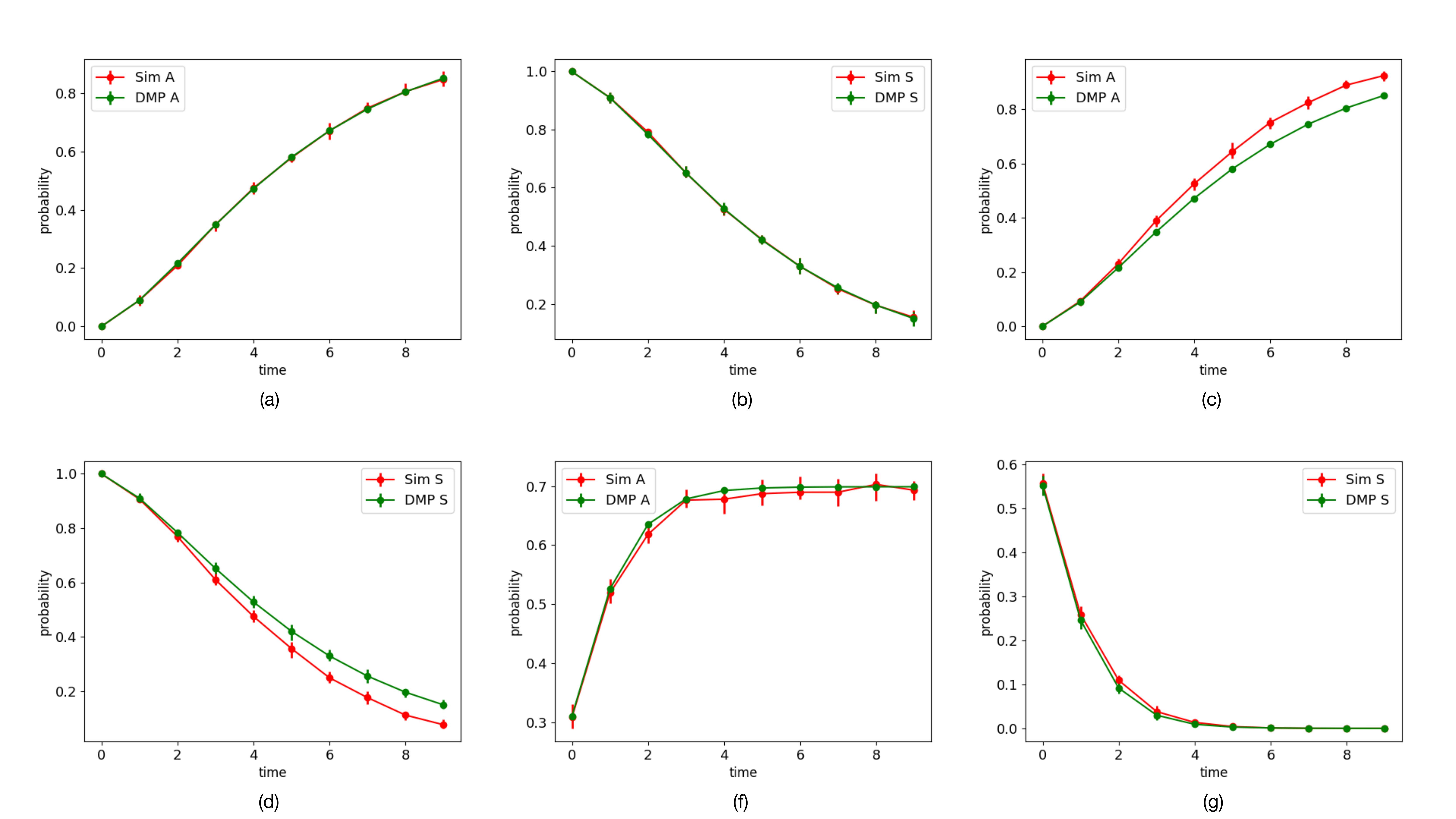}
	\caption{Comparison of DMP-based results and Monte Carlo simulations for competitive processes. For each of the graphs $10^5$ samples have been used and both mean values and error-bars are shown, unless they are smaller than the symbol size. (a) For a competitive process on a tree-like network using the parameters  $\alpha_A\!=\!0.3$, $\alpha_B\!=\!0.7$ and observing $P^{i=3}_{A}(t)$. The seeds initializing the processes are placed in nodes 7 for process $A$ and in nodes 2 for process $B$. (b) The same as in (a) but observing $P^{i=3}_{S}(t)$. (c) Comparing results from DMP method and Monte Carlo simulation for the network with loops using the same as (a) and observing $P^{i=3}_{A}(t)$. (d) The same as (c) but observing $P^{i=3}_{S}(t)$. For the Polbooks experiment we observe (e) $P^{i=0}_{A}(t)$ and (f) $P^{i=0}_{S}(t)$; the parameters used are $\alpha_A\!=\!0.2$, $\alpha_B\!=\!0.2$ and the seeds initializing the processes are placed in nodes 1 and 2 for process $A$ and on nodes 4 and 37 for process $B$.}
	\label{fig:toymodelsvalidation}
\end{figure}

The results are shown in Fig.~\ref{fig:collaborationvalidation} for the various cases.

\begin{figure}[!h]
\centering
 \includegraphics[width=6.4in]{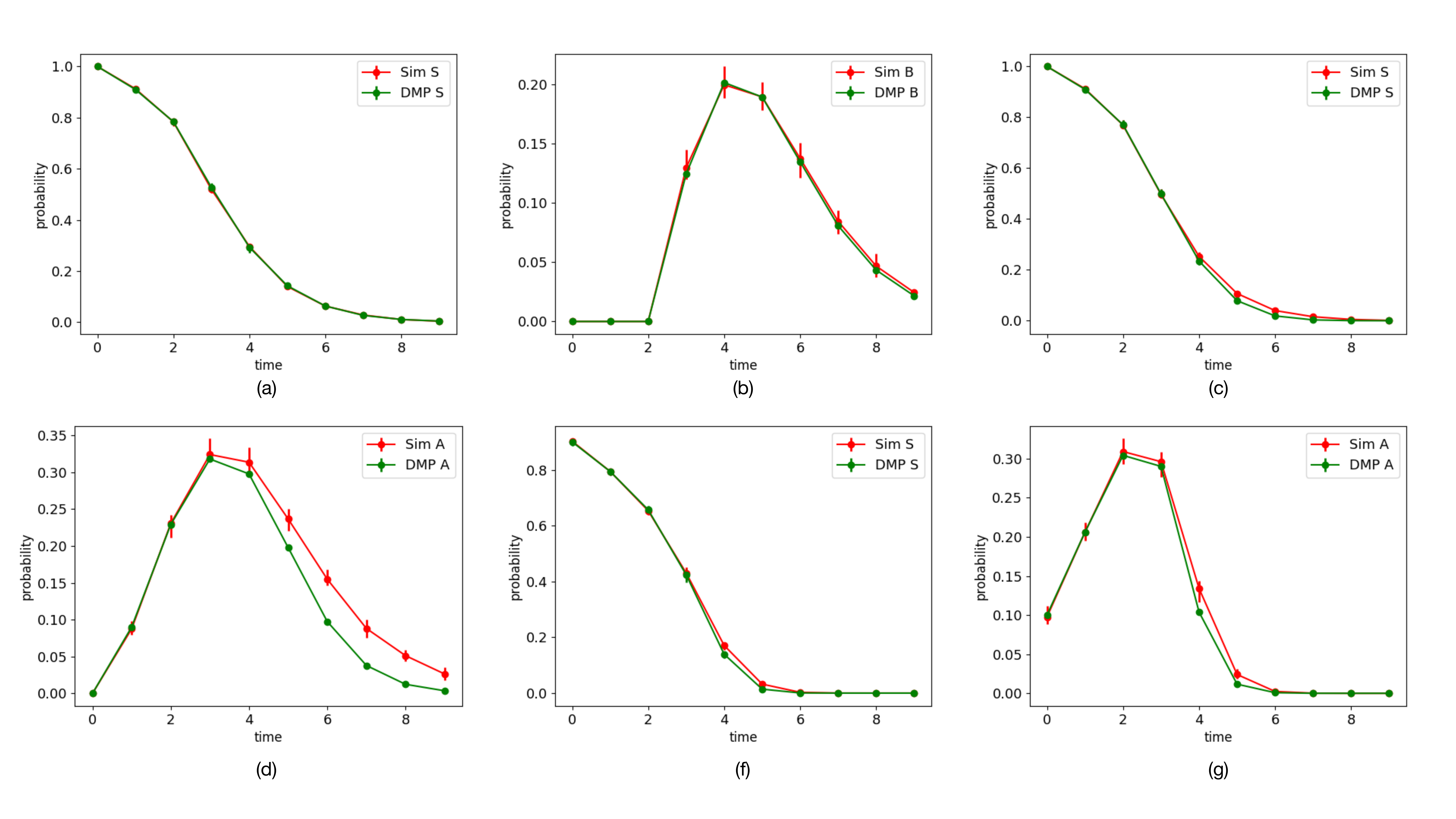}
\caption{Comparison of DMP-based results and Monte Carlo simulations for collaborative processes. For each of the graphs $10^3$ samples have been used per node (about $10^5$ samples in total) and both mean values and error-bars are shown (unless they are smaller than the symbol size). (a) For a collaborative process on a tree-like network using the parameters $\alpha_A\!=\!0.3$, $\alpha_B\!=\!0.7$, $\alpha_{AB}\!=\!0.6$ and $\alpha_{BA}\!=\!0.6$, observing $P^{i=3}_{S}(t)$ (the node to monitor has been selected arbitrarily). Process $A$ was seeded at node 7 and process $B$ at node 2. 
(b) The same as in (a) but observing $P^{i=3}_{B}(t)$. (c) Comparing results for the network with loops using the same parameters as in (a), observing $P^{i=3}_{S}(t)$ and (d) $P^{i=3}_{A}(t)$. (e) Comparing results obtained from the DMP method and Monte Carlo simulations of a collaborative process on the Football network. The parameters used are $\alpha_A\!=\!0.1$, $\alpha_B\!=\!0.2$, $\alpha_{AB}\!=\!0.3$ and $\alpha_{BA}\!=\!0.4$. The probabilities represent observations of node 2, while process $A$ was seeded at nodes 3 and 4, and process $B$ at nodes 0 and 1 (there is no significance to any of these choices).}
\label{fig:collaborationvalidation}	
\end{figure}
\end{widetext}	
	
The experimental results indicate that modeling based on \new{approximate} DMP equations is very accurate on tree-like networks, as expected for message-passing algorithms. It is less accurate on small loopy graphs at longer times, as expected, due to the small loops that violate the cavity method's assumption (the specific 10-nodes network used includes 2 small loops). This effect is suppressed to some extent on larger networks where loops are typically longer as demonstrated on the real benchmark network examples.

\new{As discussed in Appendix~E, the approximation quality is expected to degrade when $\alpha_A$ and $\alpha_B$ are very different from $\alpha_{AB}$ and $\alpha_{BA}$, respectively. Indeed, in this case the prediction of the dynamics by the approximate DMP equations can become inaccurate. To illustrate this point, let us consider an extreme example on a chain of three nodes and two links, with node 2 connected to nodes 1 and 3. Let us assume that $\alpha_A=1$, $\alpha_B=0$, $\alpha_{AB}=\alpha_{BA}=1$, i.e.\hanlin{,} that the infection $B$ can be transmitted only to nodes that are already infected with $A$. Interestingly, this scenario is relevant for several disease pairs such as Hepatitis D that can only be transmitted to individuals already infected with Hepatitis B. Given the initial condition $P_A^1(0)=1$, $P_S^2(0)=1$ and $P_B^3(0)=1$, node $1$ will become infected with disease B at time $t=3$: it first infects node $2$ with process $A$ at time $t=1$; this leads to infection of node $2$ by infection $B$ coming from node $3$ at time $t=2$; and finally node $2$ transmits infection $B$ to node $1$ at time $t=3$. This example represents an extreme case where the approximate DMP equations are not valid, and indeed preclude node $1$ from being infected by process $B$, which illustrates that they may not be exact even on tree graphs when the approximation criterion is not satisfied. However, it is easy to check that exact DMP equations provide an accurate answer in this case as well, as it should be. This counter example re-iterates the trade-off between the exactness and the computational complexity between exact and approximate DMP equations discussed in the previous section. However, as illustrated on numerical examples, in applications with \DS{plausible} parameters one could expect that the approximate DMP method provides a good description for cooperative spreading processes on sparse networks.}

\section{DMP-based Optimization Method}
\label{sec:optimization}
Competition and collaboration of spreading processes on graphs can be optimized through a judicious use of resource. We will demonstrate how managing a spreading process against an adversarial competing agent can be optimized within a given time frame; and how joint collaborative processes can be affected through best use of resource. The latter can take the form of spreading maximization while making use of the process interdependencies, or of containment through vaccination to impact on the spread of both processes.

We outline a general procedure for optimization in this section. Details for specific optimization problems we address here are given in \new{Appendix~F} for competitive processes and in \new{Appendix~H} for collaborative processes.

The core approach for optimization is based on a discretized variational method, whereby a functional over a time window (Lagrangian) is optimized through changes in control parameters throughout the time interval. The dynamics, resource constraints, initial conditions and other restrictions on the parameters used are enforced through the use of Lagrange multipliers. A similar method is used in optimal control.

We denote the components of the Lagrangian function used in a similar way to~\cite{Lokhov2017}:
\begin{equation}
\mathcal{L} \!=\! \underbrace{\mathcal{O}}_{\text{objective}}\hspace{-0.23cm} + \underbrace{\mathcal{B} +  \mathcal{P} + \mathcal{I} + \mathcal{D}}_{\text{constraints}}.
\label{eq:Lagrangian}
\end{equation}
where in the ${\cal L}$ is the Lagrangian function, ${\cal O}$ is the objective to be optimized, ${\cal B}$ is the budget or constraints on the resource used,  ${\cal I}$ represents the component that forces initial conditions, ${\cal P}$ are restrictions on the probabilities used and ${\cal D}$ represent the \new{dynamical constraints that here take form of approximate DMP equations}. All of the terms ${\cal B}$, ${\cal I}$,${\cal P}$ and ${\cal D}$ are forced through the use of Lagrange multipliers.

For different problems, the constraints and objectives vary. We take the competitive process as an example. In this problem, we would like to find an optimal allocation of a limited number of spreaders (representing a budget, potentially time dependent) for process $A$, which minimizes the spreading of process $B$. These could represent a competition in a political or commercial setting. The objective function in this case is 
\begin{equation}
\label{eq:Objective}
{\cal O} \!=\! \sum_i(1-P_i^B(T))~,
\end{equation}
where $T$ is the end of the set time window. Our goal is to maximize this objective function, thus minimizing the spread of process $B$. The resources at our disposal for seeding nodes with process $A$ are represented by the budget constraint at time zero (although more elaborate budget constraint could be accommodated as in~\cite{Lokhov2017} and in some of the examples that follow)
\begin{equation}
B_{\nu} \!=\! \sum_i{\nu^i(0)}~,
\end{equation}
where $\nu^i(0)$ is the deployment of a fraction of the budget for process $A$ on node $i$.  The budget constraint is forced through the Lagrange multiplier $\lambda^{Bu}$, such that
\begin{equation}
\mathcal{B}  \!=\! \lambda^{Bu}(B_{\nu}-\sum_i{\nu^i(0)})~.
\end{equation}
The fraction/probability $\nu^i$ is kept within a certain range, determined by the upper and lower bounds  $\overline{\nu}$ and $\underline{\nu}$, respectively. Also this is enforced using a Lagrange multiplier
\begin{equation}
\mathcal{P}  \!=\! \epsilon \sum_i(\log(\overline{\nu} - \nu^i(0)) + \log(\nu^i(0)-\underline{\nu}))\label{Restriction}
\end{equation}

The term ${\mathcal{D}}$ is given by enforcing the approximate DMP equations for the competitive case using a set of Lagrange multipliers as detailed in \new{Appendix~H} for the competitive case. The remaining term ${\mathcal{I}}$ forces the initial conditions for the dynamics.

The extremization of the Lagrangian~\eqref{eq:Lagrangian} is done as follows. Variation of $\mathcal{L}$ with respect to the dual variables (Lagrange multipliers) results in the DMP equations starting from the given initial conditions, while derivation with respect to the primal variables (control and dynamic parameters) results in a second set of equations, coupling the Lagrange multipliers and the primal variable values at different times. End conditions for the forward dynamics provide the initial conditions for the backward dynamics. 

We solve the coupled systems of equations by forward-backward propagation, a widely used method in control and detailed in~\cite{Lokhov2017}. This method has a number of advantages compared to other localized optimization procedures.
It is simple to implement, of modest computational complexity \new{$O(ET)$, where $E$ the number of edges in the graph and $T$ the time window)} and does not require any adjustable parameters.
The forward-backward optimization provides resource (budget) values to be placed at time zero (or of at any time within the time window if we so wish) in order to optimize the objective function, e.g., that of Eq.~(\ref{eq:Objective}). \new{One potential drawback of the method is the possible non-convergence of the dynamics to an optimal solution. This can be mitigated to some extent by solving the equations for the backwards dynamics using other available solvers and by storing the best solutions found over time, \new{or approaching a fixed point via gradient descent}. In general, as the functions used become more nonlinear it will become more difficult to obtain optimal solutions, although we have not experienced significant problems in the cases studied here.} 

\section{Numerical Study of the Optimization Algorithm}
\label{sec:experiments}
To validate and demonstrate the efficacy of the optimization method we carry out experiments on both synthetic and realistic networks. Before embarking on a large-scale application we study the performance of the derived method on a tree-like network of 30 nodes.

\subsection{Validation of the Optimization Algorithm}
\label{sec:toyopt}

To validate the DMP optimization algorithm on a problem that could be exhaustively studied and intuitively presented, we restrict the study to a small exemplar tree-like synthetic model. Moreover, we select a small number of nodes (3) on which resource could be deployed. The objective is to maximize the spread of both agents or to minimize the spreading of one of them (the disease control scenario) in both competitive and collaborative processes. 

\begin{figure}[!ht]
\centering  
\includegraphics[width=0.5\textwidth]{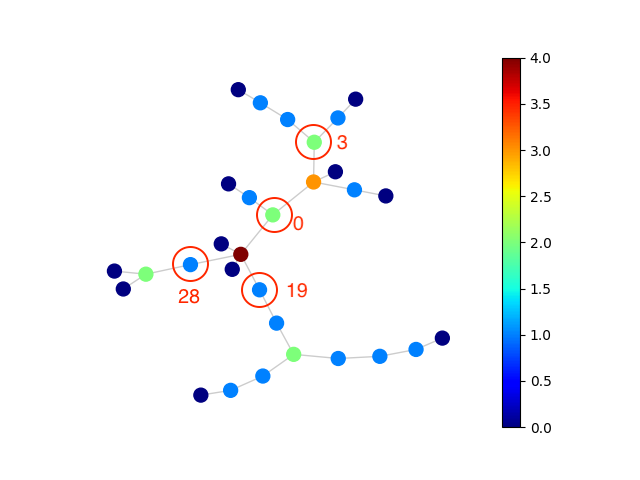}  
\caption{Tree network of 30 nodes used for carry out the experiments. The color scale represents the degree of nodes. Controllable nodes in the various experiments are marked.\label{fig:1}} 
\end{figure}
The 30-node network used for carrying out these experiments is presented in Fig.~\ref{fig:1}.
Comparison between the DMP-based optimization algorithm and the exhaustive search is implemented in the following way: We consider a scenario where the entire budget is available at time $t\!=\!0$. The optimization problem minimizes the spreading of process $B$ in a competitive processes through judicious budget allocation of the seeds for process $A$, i.e.\hanlin{,} we aim at minimizing $\sum_iP_i^B(T)$, where $T$ is the end time of the process. 

\begin{widetext}

\begin{figure}[h]	{\center
   \hspace*{-1.3cm} \includegraphics[width=6.5in]{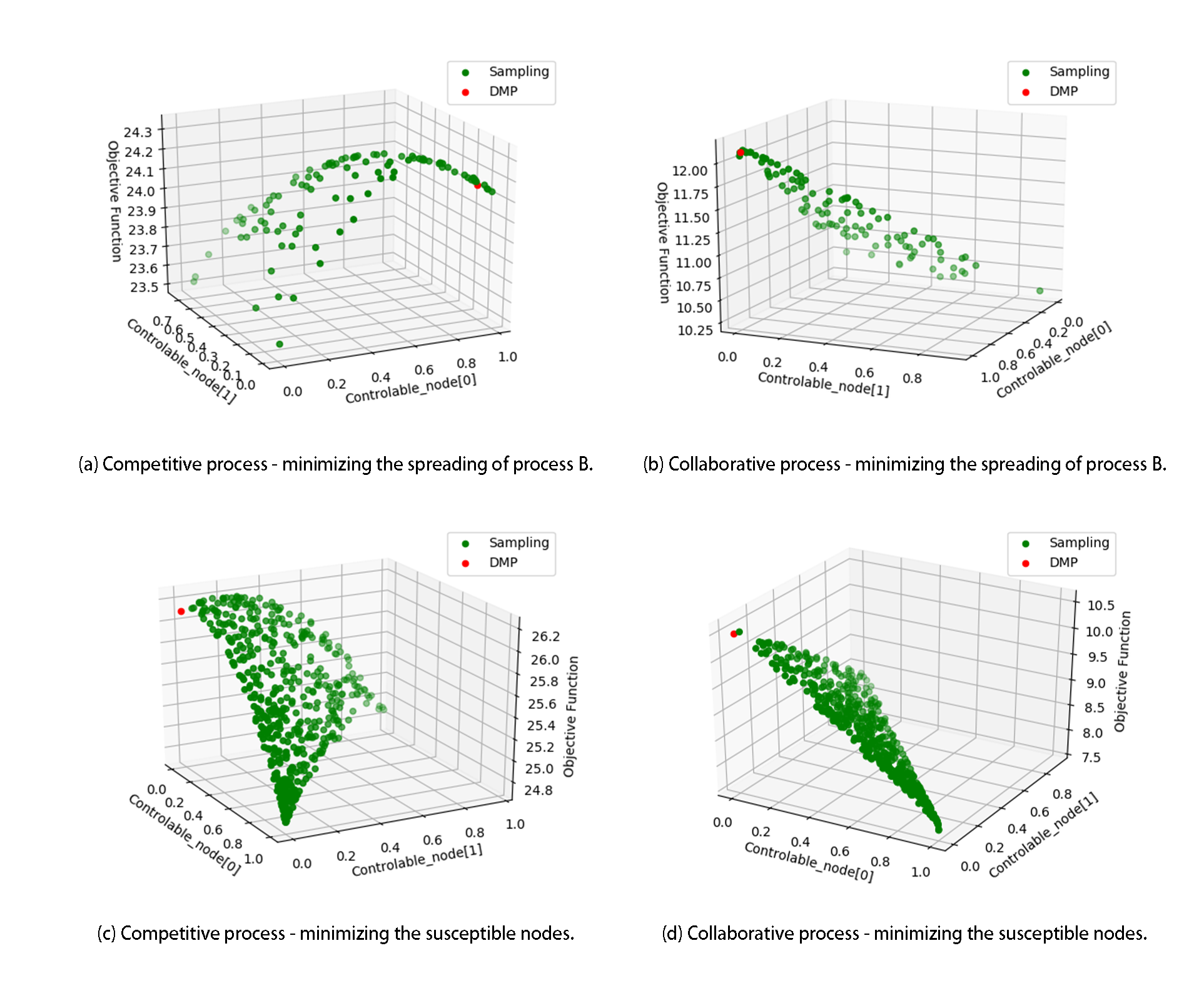} 
  \caption{Comparison of results obtained using the DMP-based optimization method and random sampling in parameter space (a) Competitive process with a time window $T\!=\!3$ and infection probabilities $\alpha_A\!=\!0.5$, $\alpha_B\!=\!0.5$. Sampled values for the various control parameters are marked by green points, while the DMP-optimal value is marked by a red point. Nodes $0$, $28$ and $19$ were infected by process $A$ as controllable nodes, with a total budget of one, while the fixed seed for process $B$ is node $3$. The objective in this case is containing the spread of process $B$, namely minimizing $\sum_i P_i^B(T)$. \new{The DMP-based optimization maximizes the objective function as $24.17$ while the optimal sampling result is $24.1$.}(b) A similar experiment to (a), with the same conditions and objective, for the collaborative process with double infection parameters $\alpha_{AB}\!=\!0.8$, and $\alpha_{BA}\!=\!0.8$. \new{The DMP optimal objective value is $12.11$ against $12.1$ from sampling}. (c) Competitive process using with time window and infection parameters as in (a), but a different objective function - maximizing the infection $\sum_i(1-P_i^S(T))$. Controllable nodes are $6$, $8$ and $15$ and node $3$ is infected by process $B$. \new{The optimal DMP objective function value is $26.11$ against $26.2$ from sampling.} (d) A similar experiment to (c) for the collaborative process with double infection parameters as is (b). \new{The DMP optimal result is $10.32$ and $10.3$ from sampling.}}
  \label{fig:subfig} }
\end{figure}
\end{widetext}
In the experiments, resources for process $A$ were deployed on three nodes: $0$, $19$ and $28$ (determined by the choices for nodes $0$ and $19$ due to the total budget constraint). The fixed seed for process $B$ is node $3$; these choices are arbitrary and insignificant. The objective function landscape has been explored by sampling for different parameter values as denoted by the green points in Fig.~\ref{fig:subfig}(a). The collaborative scenario with specific infection parameters is plotted in Fig.~\ref{fig:subfig}(b). A different set of controllable nodes is presented in Fig.~\ref{fig:subfig}(c) and Fig.~\ref{fig:subfig}(d) for the competitive and collaborative cases, respectively. The maximum value obtained for the objective function is contrasted with the results obtained using the DMP-based optimization procedure, marked by the red dot, in all cases. It is clear that there is good agreement between the DMP-optimal values obtained and the optima discovered through sampling\new{, although in some of the cases they are not identical. The corresponding objective function values for Fig.~\ref{fig:subfig} are denoted on the various subfigures.}

\subsection{Results on Real Networks}
\label{sec:realworld}
\subsubsection{Competitive processes}
We study the performance of our optimization algorithm on networks assuming that all resources are available at time zero. Other optimization strategies where resources' availability is time-dependent could also be considered~\cite{Lokhov2017} as shown later on.
One of the problems in comparing the performance of our method to other approaches is the limited number of \emph{dedicated} competing optimization methods for the multiple agents scenario, especially in cases where resource availability and its deployment are spread over the whole time window. 
We therefore compare the DMP-optimized spreading process with known heuristics for the single agent scenario such as a \emph{uniform allocation} of resources over all nodes (except seed nodes induced with process $B$), the high degree deployment strategy~\cite{PhysRevE.65.036104} \emph{HDA} and \emph{K-shell} decomposition~\cite{kitsak2010identification}. \emph{Free spreading} refers to an uncontrolled spread of process $B$. \emph{Blocking} refers to the allocation of a competing agent that is non-infectious ($p\!=\!0$). In all cases, resources are used to contain the spreading of process $B$.

Multi-agent spreading processes exist in different settings, from social network, energy grid and road network to the graph of random interaction between individuals. There is no specific network type that is more relevant to the scenarios we examine. We therefore chose a set of sparsely connected benchmark networks of different characteristics to test the efficacy of our algorithm compared to competing approaches. The description of the different networks and their specific properties appear in the corresponding references. We aim to show that the suggested methods work well on all of the sparse networks examined and we therefore expect it to work well on most other sparse networks.

In these experiments we allocated seeds of process $B$ on $0.05N$ nodes at time zero, where $N$ is the total number of nodes in the network. The same amount of resource was allocated to the controlled competing process $A$. Infection parameters are $\alpha_A \!=\!0.2$ and $\alpha_B\!=\!0.3$. This choice of parameters is arbitrary and the performance for other parameter choices provides qualitatively similar results. The free parameter that forces the upper/lower limits of the resource variables~\eqref{Restriction} is set to $\epsilon\!=\!0.1$ initially and decays exponentially with iteration steps. Experimental results show that decaying $\epsilon$ in this manner leads to a improved performance, arguably since it allows one to obtain solutions which are closer to the limit values.
The optimization procedure is iterated 10 times and the best result is selected. The objective is to minimize the spreading of process $B$ and the normalized total spreading ${\sum_iP_B^i(T)}/{N}$ at time $T\!=\!3$ is shown in Table~\ref{tab:competitive}. The short time window used is due to the small diameter of the networks. 

In competitive processes we observe that the containment of process $B$  is carried out effectively by optimal deployment of spreading agents of process $A$, compared to a static blocking, HDA deployment, K-shell and uniform seeding. This scenario corresponds to marketing, rumor spreading/fake news and opinion setting. We also evaluate the improvement obtained for a given budget in blocking the spread, which allows one to allocate the appropriate budget for containment (e.g., addressing the spread of fake news or antivaxxing rumors by releasing verifiable information). Clearly, key factors in determining the spread are the actual infection parameters associated with the various processes, which can be obtained through data analysis. The Football network is highly connected and we speculate that this is the reason for the success of uniformly allocating the spreading agents.

\begin{widetext}
	{\center
\begin{table}[H]
	\centering 
	\begin{tabular}{|c|c|c|c|c|c|c|c|}\hline
		Network  &  No.~of nodes & DMP & Uniform  & ~~K-shell~~ & ~~HDA~~  & Blocking & Free Spreading\\\hline	
		Football~\cite{Girvan2002} & 115 & 0.5829 & \bf{0.5674} & 0.8264 & 0.7731 & 0.7860 & 0.8264\\\hline
		Lesmis~\cite{Knuth1993} & 77 & \bf{0.0834} & 0.2211 &0.1962 & 0.0990 & 0.1016 & 0.3549\\\hline
		Karate~\cite{Zachary1977} & 35 & \bf{0.4206} & 0.4902 &0.5470 & 0.4212 & 0.4966& 0.5472\\\hline
		Power~\cite{Watts1998} & 4941 & \bf{0.0500} & 0.0621 &0.0641 & 0.0632 & 0.0501 & 0.0644  \\\hline
		Polbooks~\cite{Unpublish} & 105 & \bf{0.2243} & 0.3733 &0.3102 & 0.2779 & 0.4316 & 0.5744\\\hline		
	\end{tabular}
	\caption{\label{tab:competitive}Comparing different deployment methods for a competitive scenario on various networks.  On each network, we randomly choose $0.05N$ nodes as seeds of process $B$ and the same total budget for the competitive process $A$, where $N$ is number of nodes in the network.  Infection parameters are $\alpha_A\!=\!0.2$ and $\alpha_B\!=\!0.3$. The initial parameter $\epsilon$ which forces the  budget limits per node as in~\eqref{Restriction} is initially set to 0.1 but decays exponentially with the iteration steps. The optimization procedure is iterated 10 times and the best result is selected. The objective is to minimize the spreading of process $B$; the normalized total spreading ${\sum_iP_B^i(T)}/{N}$ at time $T\!=\!3$ is shown. In all methods budgets are available at time zero. \emph{Uniform Allocation} assumes all the budgets are uniformly allocated to all nodes at time zero (except those infected with process $B$). In \emph{Free Spreading}, no agents are allocated for process $A$. In \emph{Blocking,} competitive agents are not infectious ($p\!=\!0$), all the budget is used to contain the spreading of $B$. Best results are denoted by bold fonts.}
\end{table} }
\end{widetext}

When the budget allocation is done dynamically at different times, several optimization procedures can be used. \new{Conventional stochastic optimal control~\cite{Fleming1975} is based on \hanlin{planning} ahead for the entire time window, taking into account future uncertainties. An optimal solution in discrete time can be stated through a solution of the Bellman equation that in our setting would result in an algorithm with a high computational complexity due to the curse of dimensionality \cite{meyn2008control}. To mitigate this issue, we adopt a procedure that reminds updates in the closed loop control, which updates the resource allocation at every step based on the information on the realization of the stochastic dynamics. This approach does not strictly guarantee the solution optimality, but quantitatively takes into account realization of uncertainties while keeping computational complexity under control. We carry out the optimization for the end time $T$ at each step, on a shrinking time window, incorporating the newly available information. Considering} a case where one process spreads randomly while for the other a unit budget is available per time step to be deployed optimally. Given the information at hand for each time step, one may then consider the following dynamic allocation strategies: (a) \emph{DMP greedy} deploys the unit to optimize the objective function for the next time step; while (b) \emph{DMP optimal} optimizes the objective function for the end-time $T$, while incorporating information available at each time step. Note that the latter represents a \new{closed-loop like optimization in the sense that updated information on realization of dynamics at each time step is incorporated to produce a better resource allocation plan}.

\begin{widetext}

\begin{figure} [H]	{	\center
	\hspace*{-1.6cm}\includegraphics[width=7.5in]{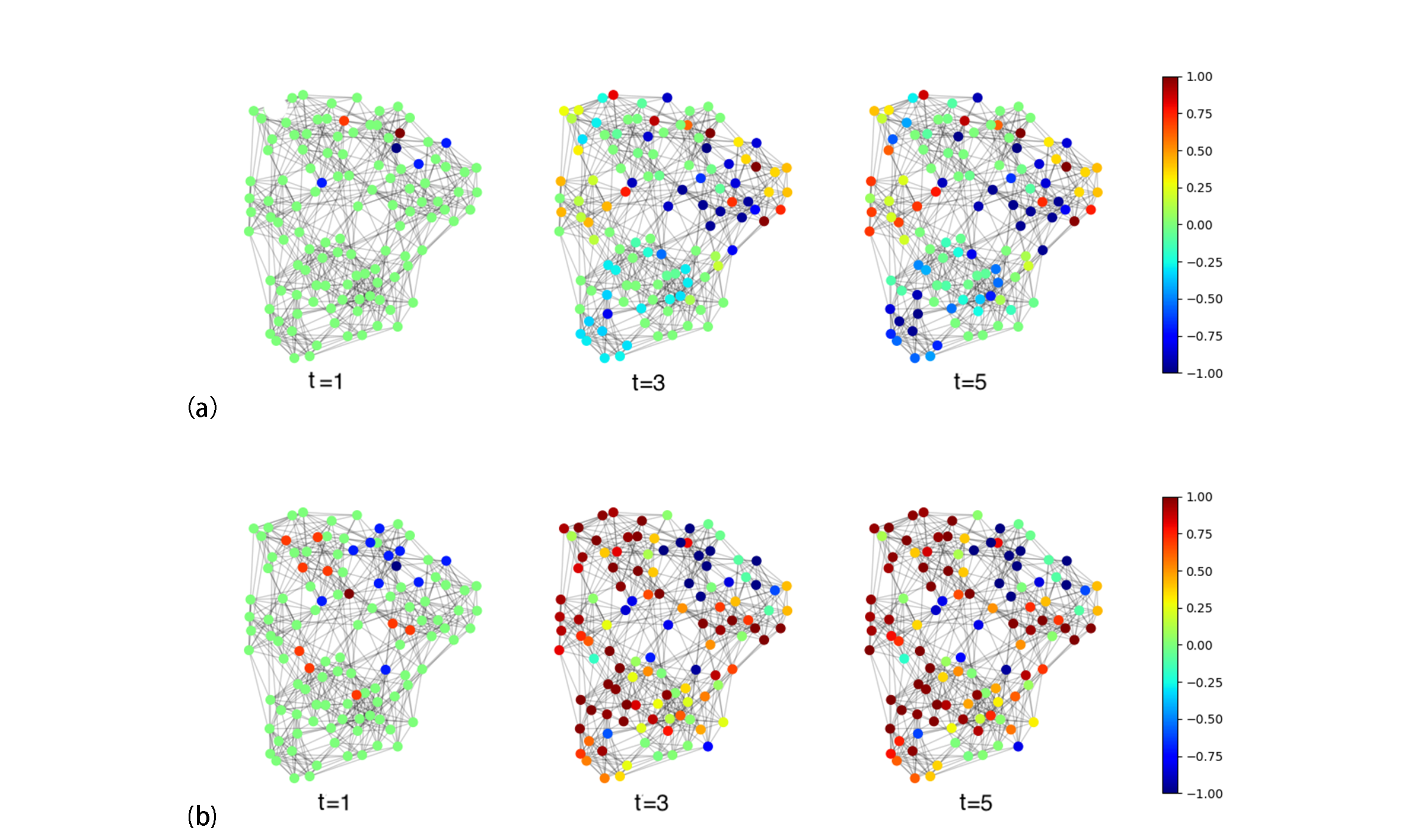} 
	\caption{Football network with infection parameters $\alpha_A \!=\! \alpha_B \!=\! 0.7$, the budget for $B$ is 1, allocated on node 1 at $t\!=\!0$, and a budget of 1 per time step is assigned optimally for process $A$. The figures represent the containment of process $B$ at different times $t\!=\!1,2,\ldots,5$ due to the judicious allocation of resources of process $A$.  (a) DMP-greedy; (b) DMP-optimal. The heat bar represents the dominating process per node through the value $P_A^i(t)\!-\!P_B^i(t)$. red/blue represent dominating processes $A/B$, respectively.}
	\label{fig:t125}}
\end{figure}
\end{widetext}

\begin{table}[!ht]
	\centering
	\begin{tabular}{|c|c|c|}\hline
		Time & DMP optimal & DMP greedy \\ \hline
		~~$t\!=\!1$~~ & ~~0.0767~~ & ~~\bf{0.0556}~~  \\ \hline
		~~$t\!=\!2$~~ & ~~0.2157~~ & ~~\bf{0.1871}~~  \\ \hline
		~~$t\!=\!3$~~ & ~~\bf{0.3109}~~ & ~~0.4336~~  \\ \hline 
		~~$t\!=\!4$~~ & ~~\bf{0.3211}~~ & ~~0.5615~~  \\ \hline
		~~$t\!=\!5$~~ & ~~\bf{0.3211}~~ & ~~0.5628~~  \\ \hline
	\end{tabular}
	\caption{\label{tab:opti_greedy} DMP based resource deployment methods for the Football network under the same condition as in Fig.~\ref{fig:t125}. The objective is to minimize the spreading of  $B$, i.e.\hanlin{,} minimize the fraction of $B$ nodes at time $T\!=\!5$, ${\sum_iP_B^i(T)}/{N}$.}
\end{table}
To demonstrate the efficacy of our method and the differences between the DMP-greedy and DMP-optimal resource deployment, we use the Football network~\cite{Girvan2002} and infect node 1 at time 0 (blue point left of the center, total budget for $B$ is 1); we then deploy optimally a budget of 1 for process $A$ at each time step $t\!=\!1,2,\ldots,5$. The infection parameters used are $\alpha_A \!=\! \alpha_B \!=\! 0.7$. The results are shown in Fig.~\ref{fig:t125} where the heat bar represents the dominating process per node through the value $P_A^i(t)\!-\! P_B^i(t)$. Red and blue represent dominating processes $A$ and $B$, respectively. It is clear that DMP-optimal (b) is much more effective than DMP-greedy (a) in restricting the spread of process $B$ by maximizing the spread of process $A$. Numerical comparison between the two methods for the same network and conditions are presented in Table~\ref{tab:opti_greedy}. It is clear that while DMP-greedy is successful at earlier time steps, DMP-optimal minimizes the spread of process $B$ at $T\!=\!5$. A second example on a more densely connected network is given in \new{Appendix~G}.

To demonstrate the improvement in reducing the spread of agent $B$ given a budget $b_B$ against an optimally deployed budget $b_A$ of process $A$, we plotted the ratio of infected $A$ and $B$ \hanlin{statuses} at time $T\!=\!5$ against the budget ratio $b_A / b_B$. This has been done for the Lesmis network~\cite{Knuth1993} (77 nodes, representing the co-appearance of characters in the novel Les Miserables) with infection probabilities $\alpha_A\!=\!\alpha_B\!=\!0.5$ and time window $T\!=\!5$. The results shown in Fig.~\ref{fig:phase_transition}(a) have been averaged over 5 instances for both uniform and optimal DMP-based deployment. It is clear from the figure that the optimal DMP-based deployment provides much better results than uniform deployment, which exhibits a linear increase. The saturation for high ratios is limited by the size of the graph.

We observed several instances in which the ratio of infected probabilities exhibits a fast transition at specific points. Figure~\ref{fig:phase_transition}(b) shows a fast transition towards a process-$A$ dominated network as the ratio between the budgets allocated exceeds a certain value $b_A / b_B\!\approx\!0.28$. The $y$-axis represents the ratio of process probabilities $ \frac{\sum_{i\!=\!1}^N P_A^i(T)}{\sum_{k\!=\!1}^N P_B^k(T)}$. These results clearly depend on the topology, budgets and infection rates used. The example given here is based on the Lesmis network with infection probabilities $\alpha_A\!=\!0.5$ and $\alpha_B\!=\!0.7$, fixed budget $b_B\!=\!1$, time window $T\!=\!5$ and optimal DMP-based deployment. 

\begin{widetext}
	{	\center
\begin{figure} [H] 
		\includegraphics[width=6.8in]{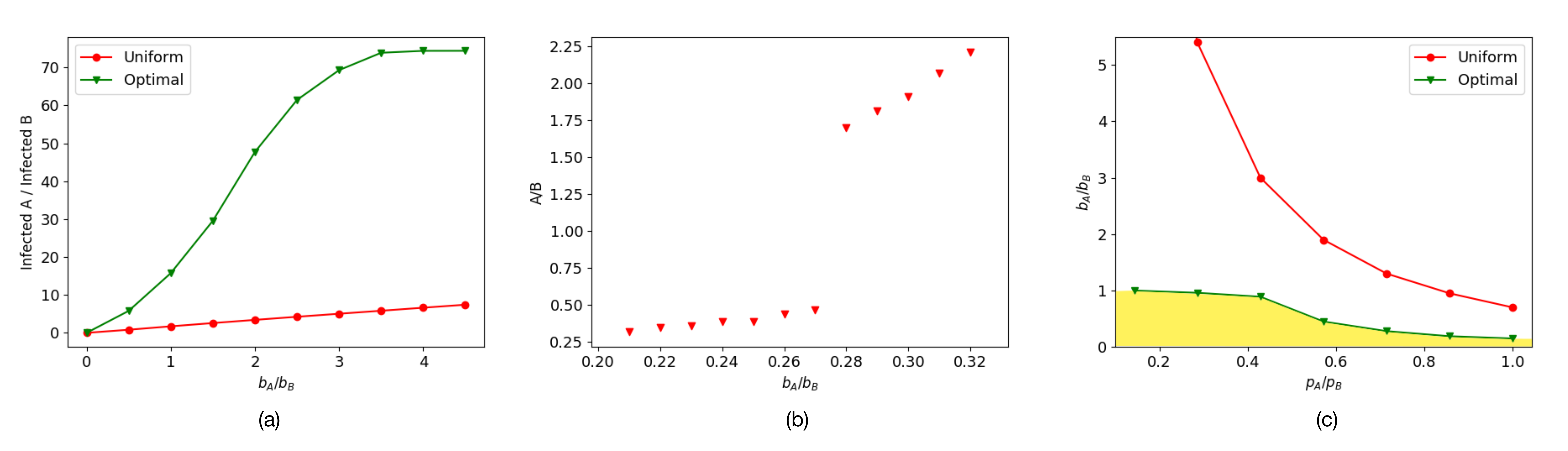}
	\caption{\label{fig:phase_transition} Optimized competitive scenario on the Lesmis network. (a) The ratio of nodes infected by processes $A$ and $B$ after $T\!=\!5$  time steps for a given initial budget ratio $b_A / b_B$. The infection probabilities used are $\alpha_A\!=\!\alpha_B\!=\!0.5$. The green curve represents the ratio in the case of DMP optimized deployment of the $A$ budget, while the red curve represents the uniform deployment case. (b) A fast transition in the ratio of expected infected \hanlin{statuses} $ \frac{\sum_{i\!=\!1}^N P_A^i(T)}{\sum_{k\!=\!1}^N P_B^k(T)}$ for infection probabilities $\alpha_A\!=\!0.5$ and $\alpha_B\!=\!0.7$ and time window $T\!=\!5$. The  budget for $B$ was fixed at $b_B\!=\!1$ and the budget deployment for $A$ was optimized using DMP. (c) The interplay between infection parameter values and budget allocated to each of the processes on the Lesmis network. The $x$-axis represents the ratio between infection parameters $\alpha_A/\alpha_B$ and the $y$-axis the ratio between budgets $b_A/b_B$.  The green line represent values for which the two processes have equal probabilities in the network at $T\!=\!5$ when DMP-optimized deployment of resource $A$ is used; the red line represents the same line for uniform deployment. Results are averaged over randomly chosen 5 initial positions for the seed of $B$.
	}
\end{figure}}
\end{widetext}

To evaluate the interplay between infection parameter values and budget allocated to each of the processes, we studied a competitive case on the Lesmis network, with a varying ratio between budgets ($b_B\!=\!1$, $b_A\!=\!0.5,\ldots,4.5$)  and infection probabilities ($\alpha_B\!=\!0.7$, $\alpha_A\!=\! 0.1,\ldots,0.7$). The points where the processes end up with equal probabilities at $T\!=\!5$ are plotted in Fig.~\ref{fig:phase_transition}(c) for the DMP-optimized deployment of resource $A$ (green line) and for uniform deployment (red line). The results are averaged over randomly chosen 5 initial position for the seed of $B$. From the figure it is clear that in this case DMP-optimized deployment can effectively mitigate significantly inferior infection rates or budget ratios (area above the green curve); only when both ratios are very low the $B$ process dominates the network after $T$ steps (yellow area). Uniform deployment results in much inferior performance (area above the red line).

\subsubsection{Collaborative processes}

Results obtained for collaborative scenarios, shown in Table~\ref{tab:collborative_DMP_MC}, exhibit a similar behavior, showing that collaborative processes can be optimized to spread quickly due to the mutually-supportive role played by the two processes. In this case, some nodes ($0.05N$) are infected by process $B$ and we allocate a given budget of process $A$ such that the joint spreading will be maximized and the number of non-infected nodes (in \hanlin{status} $S$) minimized. This could represent, for instance, the spread of opinions on the basis of political affiliation. Also in this case the DMP-based optimization algorithm works well, with the exception of the Football network where uniform spreading seems to be successful, arguably due to the same reasons as in the competitive case.

\begin{widetext}

\begin{table}[H]
	\centering
	\begin{tabular}{|c|c|c|c|c|c|c|c|}\hline
		Network & No.~of nodes & DMP Allocation & Uniform Allocation  & ~~K-shell~~ & ~~HDA~~ & Free Spreading\\\hline
		Football~\cite{Girvan2002} & 115 & \textbf{0.0536} & 0.0582 & 0.1736  &0.1543 & 0.1735\\\hline
		Lesmis~\cite{Knuth1993} & 77 & \textbf{0.2222} & 0.3832  & 0.3151 & 0.3051 & 0.6451\\\hline
		Karate~\cite{Zachary1977} & 35 & \textbf{0.2771} & 0.3743  & 0.4527 & 0.2481& 0.4529 \\\hline 
		Power~\cite{Watts1998} & 4941 & \textbf{0.7652} & 0.7930  & 0.8434 & 0.9024 & 0.9355\\\hline
		Polbooks~\cite{Unpublish} & 105 & \textbf{0.1524} & 0.2065  & 0.3347 & 0.2204 & 0.4255\\\hline
	\end{tabular}
	\caption{\label{tab:collborative_DMP_MC}Comparing different resource-allocation approaches for collaborative processes on various networks. On each network, we randomly select $0.05N$ nodes as seeds of process $B$ and the same total budget for process $A$ to be optimally allocated, where $N$ is total number of nodes in the network. The infection parameters chosen are $\alpha_A \!=\! 0.2$, $\alpha_B \!=\! 0.3$, $\alpha_{AB}\!=\!0.4$ and $\alpha_{BA}\!=\!0.5$. The initial parameter $\epsilon$ which forces the  budget limits per node as in~\eqref{Restriction} is initially set to 0.1 but decays exponentially with the iteration steps. The optimization procedure is iterated 10 times and the best result is selected. The objective function in this case is to maximize the spreading of processes $A$ and $B$, i.e., minimize the fraction of susceptible nodes ${\sum_iP_S^i(T)}/{N}$ at time $T\!=\!3$. In DMP allocation-based optimization, we assume the seeding budget for $B$ is available at time zero. Uniform allocation assumes that the budget for $A$ is allocated uniformly to all free nodes initially. HDA deployment~\cite{PhysRevE.65.036104}  and K-shell~\cite{kitsak2010identification} seeding are used as before;   
		in Free Spreading, no collaborative spreading budget $A$ is allocated. Lowest fraction of susceptible nodes is denoted by bold fonts. }
\end{table}
\end{widetext}

The final set of experiments we carried out relates to the vaccination policy in collaborative spreading processes. In this case, two collaborative processes $A$ and $B$ spread throughout the system and the task is to minimize the spread through a vaccination process to one of the two, say $B$. The role of vaccination in this case is in reducing the infection probability by the level of vaccination deployed. For instance, we adopt a simplistic model whereby if the vaccine deployed at a given node is $b$, its probability of being infected will reduce to $\alpha_B\!-\!b$. Similarly, the infection parameter $\alpha_{BA}$ will also decrease by the same amount to $\alpha_{BA}\!-\!b$. Other infection models can be easily accommodated. The optimization task is in the deployment of a given vaccination budget such that the fraction of non-infected sites ${\sum_iP_S^i(T)}/{N}$ at time $T$, is maximized. 

The experimental results demonstrate that the DMP-based optimization shows excellent performance.
\begin{widetext}	
	
\begin{table}[H]	
	\centering
\hspace*{5mm}	\begin{tabular}{|c|c|c|c|c|c|}\hline
		Network & Number of nodes & DMP Allocation & Uniformed Allocation & Free Spreading\\ \hline
		Football~\cite{Girvan2002} & 115 & \textbf{0.1887} & 0.0607 & 0.0294 \\ \hline
		Lesmis~\cite{Knuth1993} & 77 & \textbf{0.9455} & 0.5519 & 0.4941 \\ \hline
		Karate~\cite{Zachary1977} & 35 & \textbf{0.4676} & 0.3491 & 0.3123 \\ \hline 
		Power~\cite{Watts1998} & 4941 & \textbf{0.8891} & 0.8650 & 0.8356 \\ \hline
		Polbooks~\cite{Unpublish} & 105 & \textbf{0.5483} & 0.3071 & 0.2354 \\ \hline
	\end{tabular}
	\caption{\label{tab:vaccin} Different vaccine-allocation policies in a collaborative process on difference benchmark networks. On each network, we randomly choose $0.05N$ of the nodes as seeds for process $B$ and $0.01N$ nodes as seeds for process $A$; the total vaccination budget is $0.05N$, where $N$ is number of nodes in the network.  Infection parameters are arbitrarily set to $\alpha_A \!=\! 0.4$, $\alpha_B \!=\! 0.4$, $\alpha_{AB}\!=\!0.9$ and $\alpha_{BA}\!=\!0.9$. The parameter $\epsilon$ in  \eqref{Restriction} is set to $0.01$. The optimization procedure is iterated 10 times and the best result is selected. The objective is to minimize the spreading of processes $A$ and $B$, i.e., maximize the fraction of susceptible nodes ${\sum_iP_S^i(T)}/{N}$ at the end of the process (time $T\!=\!3$).  In DMP-based optimal allocation, we assume that all budgets are available at time zero. Uniform Allocation assumes all budgets to be uniformly allocated to all nodes at time zero. In Free Spreading, no vaccine is allocated. A detailed description of the vaccine allocation problem can be found in \new{Appendix~I}.}
\end{table}
\end{widetext}
	
Generally, for the different network topologies and size, DMP-based optimization appears to be more effective than other methods. The main cases where it does not offer the best result are in complex networks with bounded connectivity fluctuation, where it has been shown that uniformly applied immunization strategies are highly effective~\cite{PhysRevE.65.036104}.

\begin{widetext}
	
\begin{figure} [H]	\centering
	\includegraphics[width=6in]{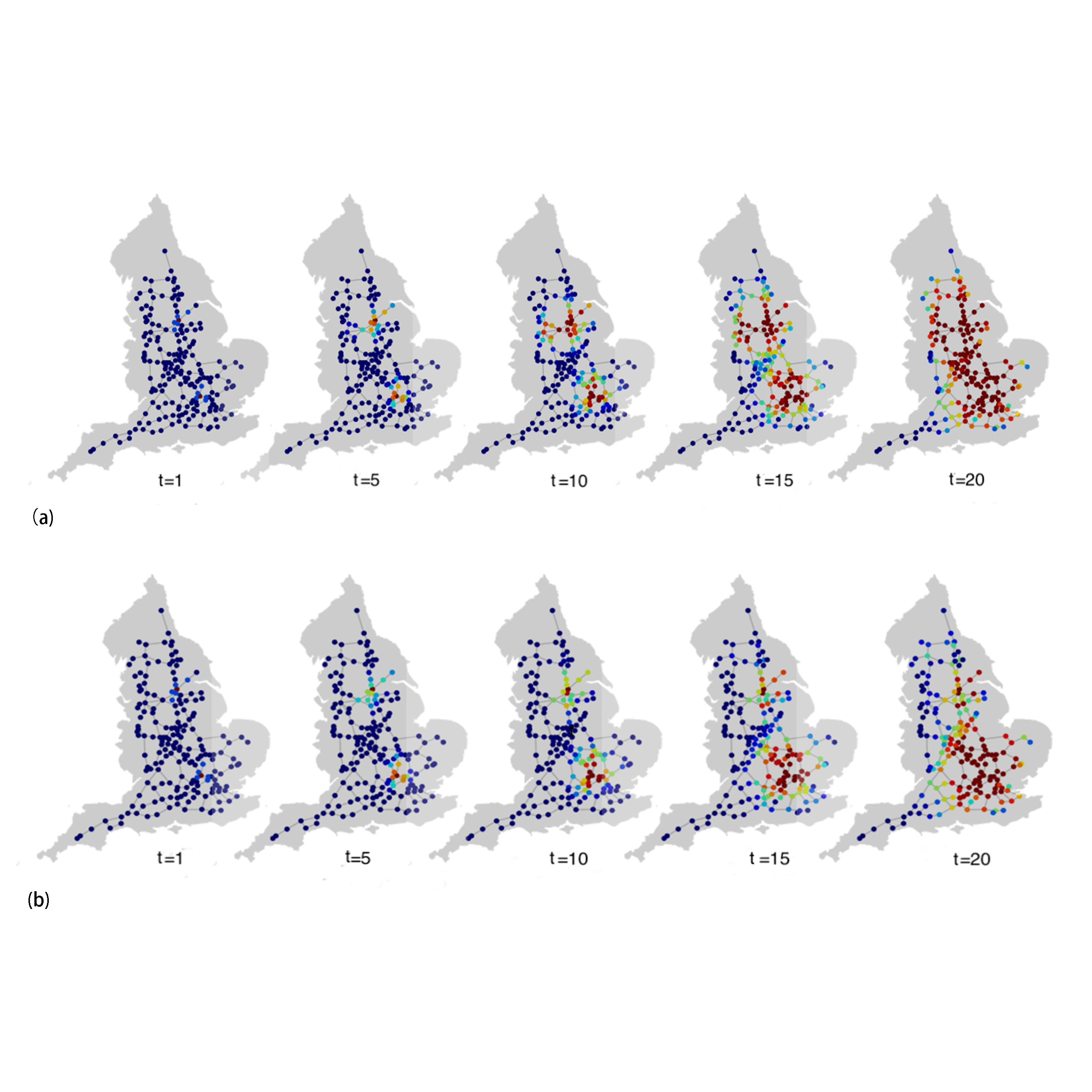} 
	\vspace*{-20mm}
	\caption{\label{fig:road_England} England road network, the collaborative spread of  highly infectious coupled processes.
		We consider the spread of highly infectious coupled processes through the road network starting from the areas of Greater London (node 10, process $A$)  and Leeds (node 11, process $B$); the budget for both $A$ and $B$  is one unit and infection parameters $\alpha_A\! =\! \alpha_B\! =\! 0.2$, $\alpha_{BA} \!=\! \alpha_{AB} \!= \!0.99$. The process is runs for 20 steps. The vaccination budget  is one unit per time step and is effective to suppress process $B$ only, while being ineffective for process $A$. The node color represents the value of  $1\!-\!P_S^i(t)$, where red and blue correspond to 1 and 0, respectively (the heat bar is similar to that of Fig.~\ref{fig:t125}). 
		(a) Free spread of the epidemics with no vaccination. (b)  A vaccination budget of one unit per time step against process $B$  is deployed at each time step using the DMP-optimal algorithm. We see that in (b) that the infection in the of London area spreads unhindered in both (a) and (b) as it is mainly infected by process $A$, while spread emanating from the Leeds area is effectively blocked when the vaccination is deployed.
	}
\end{figure}
\end{widetext}
To demonstrate the efficacy of DMP-optimal resource deployment for vaccination in the case of a collaborative spreading process we use the main road network of England~\cite{EnglandRoadData2019}. \new{Livestock} epidemics often spreads through the transport of infected livestock on the road network (as was the case in the 2001 Foot and Mouth epidemic in the UK). We consider the spread of highly infectious coupled processes through the road network starting from the areas of Greater London (node 10, process $A$)  and Leeds (node 11, process $B$); the budget for both $A$ and $B$  is one unit and infection parameters $\alpha_A\! =\! \alpha_B\! =\! 0.2$, $\alpha_{BA} \!=\! \alpha_{AB} \!= \!0.99$. The process is runs for 20 steps. The vaccination budget is one unit per time step and is effective against process $B$ only, while being ineffective for process $A$. The results shown in Fig.~\ref{fig:road_England} demonstrate the efficacy of the DMP-optimal vaccination strategy aimed at minimizing $\sum_i{(1\!-\!P_S^i(t))}$ (b) in contrast to the free spreading of both infections (a). Blue/red represent uninfected/infected \hanlin{statuses}, respectively; more specifically, the node color represents  $1\!-\! P_S^i(t)$, where red and blue correspond to 1 and 0, respectively. As we can see in Fig.~\ref{fig:road_England}, the infection spreading around London remains the same, with or without the deployment of vaccine as it is mainly infected by process $A$, and hence the vaccination has no effect, while the spread emanating from the Leeds area is effectively blocked.

\section{Summary and future work}
\label{sec:summary}
Competition and collaboration between spreading processes are prevalent on social and information networks, and on interaction-networks between humans or livestock, to name but a few. To better understand the expected spread and dynamics of diseases, marketing material, opinions and information it is essential to infer and forecast the spreading dynamics. Moreover, the judicious use of limited resource will help contain the spread of epidemics, win political and marketing campaigns, and better inform the public in the battle against unsubstantiated or misleading information. 

Most current studies of multi-spreading processes such as~\cite{Cai2015} heavily rely on computer simulations and heuristics with varying results from one instance to another. Obtaining reliable estimates therefore comes at a high computational cost in order to obtain meaningful statistics. We advocate the use of a principled probabilistic technique for the analysis and employ the same framework to develop and offer an optimization algorithm for best use of the resource available. 

Both competitive and collaborative spreading processes are investigated here using the probabilistic DMP framework and provide an accurate description of the spreading dynamics, as validated on both synthetic and real networks. \new{For the first time, we derive exact message-passing equations for two interacting \DS{unidirectional} dynamic processes, and study approximate DMP equations that benefit from a simpler form and a lower algorithmic complexity; this represents the first major novel contribution of this work.} This probabilistic description facilitates the study of multi-agent spreading processes on general sparse networks at a modest computational cost, opening up new possibilities for obtaining insights into the characteristics of such scenarios.

\new{As a second major contribution in this work, we employ a scalable optimization framework} that incorporates the DMP dynamics to deploy limited resource for an objective function to be optimized at the end of the time window. This is a difficult hard-computational problem since the resource is deployed at earlier times, the two processes dynamically interact throughout the time window and the aim is to optimize the objective function at its end. 

One of the scenarios we considered is that of resource deployment in a mutually exclusive competitive case where the deployment of an agent's resources aims to maximize its spread at the end of the time window with respect to the spread of an adversarial process. This can be seen as a competition between adversarial agents, each of which aims to contain the spread of the other, making it highly relevant in the battle for public opinion. The collaborative scenarios we optimized include constructive interaction between the spreading processes, such that infection by one process facilitates the infection by another. The aim in this case is to exploit the exposure optimally in order to maximize the spread of both processes.  Finally, we also examined optimal vaccination strategies in the case of collaborative spreading processes, where vaccination is effective with respect to one process only but helps to reduce the spread of both due to the interaction between them.

The optimization algorithm has been tested in several different scenarios and on a range of small-scale and real networks, showing excellent performance with respect to the existing alternatives of high-degree allocation, K-shell based approaches, free spreading and uniform budget allocation. We demonstrated that the optimized deployment makes a \emph{huge} difference in the use of limited resources and allows for a balanced competition even with significantly inferior resource availability or lower infection rates. We see a significant potential in the use of such principled algorithms to make the most of limited resource. 

Moreover, the suggested framework is highly adaptive and can accommodate: targeted spreading~\cite{Lokhov2017}, where only specific vertices are available and at specific times (as in the case of critical vote and time-sensitive campaigns); temporal deployment of resources, either of the process of interest or of the adversarial process, where resources are available at different times within the time window (e.g., due to limited production or shipment restrictions); determining  vaccination policies and for more complex collaborative scenarios. \new{We anticipate that the developed methods will also be useful for development of scalable~\cite{wilinski2020scalable} algorithms for learning~\cite{lokhov2015efficient, lokhov2016reconstructing} of spreading models in the context of multiple interacting spreading processes}.

The application of our method \new{for suggesting vaccination policy to contain the spread of collaborative health conditions, for process containment (e.g., the spread of fake news, anti-vaxxing messages), and for effective  information dissemination (e.g., marketing or opinion-setting material)}, are promising and timely applications that should be explored. 

\acknowledgments
\new{We would like to thank Ho Fai Po for compiling the UK road network data. H.S. and D.S acknowledge support by the Leverhulme trust (RPG-2018-092) and the EPSRC Programme Grant TRANSNET (EP/R035342/1). A.Y.L. acknowledges support from the Laboratory Directed Research and Development program of Los Alamos National Laboratory under project number 20200121ER.}

%%%%%%%%%%%%%%%%%% Appendices %%%%%%%%%%%%%%
\begin{widetext}
\appendix
\section{Dynamic Belief Propagation}
\label{app:DBP}
\new{
In the two-process dynamics, the dynamics of a single node $i$ is described by a pair of activation times, $(\tau_i^A,\tau_i^B)$, where $\tau_i^A$ denotes the first time when node $i$ is found in the \hanlin{status} $A$, and similarly for $\tau_i^B$. For instance, $\tau_i^A = 0$ means that node $i$ was initially in the active \hanlin{status} $A$, and we will denote by $\tau_i^A = *$ the situation where node $i$ did not get $A$-activated before some final observation time, i.e.\hanlin{,} in some sense $*$ absorbs all the history that happens after the end of the observation window. For the convenience of presentation, in what follows we will consider two separate ``observation windows'', for $A$ and $B$ processes.
}

\new{
The starting point for our derivations are the general Dynamic Belief Propagation (DBP) equations \cite{kanoria2011majority, altarelli2013large, Lokhov2015} on the interaction graph, where the goal is to approximate the probability $m^i_{T_A+1,T_B+1}(\tau_i^A,\tau_i^B)$ that node $i$ has a trajectory $(\tau_i^A,\tau_i^B)$ during the observation time window of length $T_A$ for process $A$ and $T_B$ for process $B$:
\begin{equation}
    m^i_{T_A+1,T_B+1}(\tau_i^A,\tau_i^B) = \sum_{\{\tau_k^A,\tau_k^B\}_{k \in \partial i}} W^i(\tau_i^A,\tau_i^B; \{\tau_k^A,\tau_k^B\}_{k \in \partial i}) \prod_{k \in \partial i} m^{k \to i}_{T_A,T_B}(\tau_k^A,\tau_k^B \Vert \tau_i^A,\tau_i^B),
    \label{eq:DBP_marginals_initial}
\end{equation}
where $\partial i$ denotes neighbors of $i$ on the graph, and the sum runs over all possible values of $\tau_k^A$ and $\tau_k^B$, i.e.\hanlin{,} $\{0, 1, \ldots, t_{A / B}, *\}$. The transition kernel $W^i$ is defined through the dynamical rules for a given dynamics; we will provide an explicit expression for the kernel below. The probabilities $m_{T_A,T_B}^{i \to j}(\tau_i^A,\tau_i^B \Vert \tau_j^A,\tau_j^B)$ have a sense of the probability that node $i$ has a trajectory of length $T_A$ for process $A$ and $T_B$ for process $B$ (in what follows referred to as $(T_A,T_B)$-trajectory) equal to $(\tau_i^A,\tau_i^B)$ subject to a fixed trajectory $(\tau_j^A,\tau_j^B)$ of node $j$, and satisfy the following consistency relations:
\begin{equation}
    m^{i \to j}_{T_A+1,T_B+1}(\tau_i^A,\tau_i^B \Vert \tau_j^A,\tau_j^B) = \sum_{\{\tau_k^A,\tau_k^B\}_{k \in \partial i \backslash j}} W^i(\tau_i^A,\tau_i^B; \{\tau_k^A,\tau_k^B\}_{k \in \partial i}) \prod_{k \in \partial i \backslash j} m^{k \to i}_{T_A,T_B}(\tau_k^A,\tau_k^B \Vert \tau_i^A,\tau_i^B).
    \label{eq:DBP_messages_initial}
\end{equation}
The fixed point solution of equations \eqref{eq:DBP_marginals_initial}-\eqref{eq:DBP_messages_initial} is guaranteed to be exact on tree graphs, and provides good estimates of the marginal probabilities on loopy but sparse graphs~\cite{MezardMontanari2009, Lokhov2015}.
}

\new{
Notice that $m^{i \to j}_{T_A,T_B}(\tau_i^A,\tau_i^B \Vert \tau_j^A,\tau_j^B)$ is not a conditional probability (e.g.\hanlin{,} Bayes rule does not apply). Let us state the following fact that will be crucial for the derivations below. (This fact is due to the properties of the DBP equations and the dynamics specified by $W^i$.)
\begin{fact}[Fundamental property of messages]
For $\tau_i^A \neq *$ and $\tau_i^B \neq *$,
\begin{equation*}
    m_{T_A,T_B}^i(\tau_i^A,\tau_i^B) = m_{\tau_i^A,\tau_i^B}^i(\tau_i^A,\tau_i^B) \quad \forall \, T_A > \tau_i^A, \text{and } T_B > \tau_i^B.
\end{equation*}
A similar statement holds for messages.
\label{fact:reduction}
\end{fact}
}

\new{\DS{This means that future events cannot affect current or past events.}
Due to Fact~\ref{fact:reduction}, the DBP equations can be considerably simplified. Indeed, Eqs.~\eqref{eq:DBP_marginals_initial} and \eqref{eq:DBP_messages_initial} basically state that the right hand sides do not depend on the final times in the observation windows, i.e.\hanlin{,} on times $(T_A+1,T_B+1)$. In particular, this means that for $\tau_i^A = 0$ or $\tau_i^B = 0$, the right hand side does not depend on $\tau_i^A$ or $\tau_i^B$, respectively. For other cases, choosing $T_A = \tau_i^A - 1$ and $T_B = \tau_i^B - 1$, we see that since the time window of messages are $(\tau_i^A - 1,\tau_i^B - 1)$, they do not depend on the precise value of $\tau_i^A$ and $\tau_i^B$. Because of that, the following fact holds.
}

\new{
\begin{fact}[Simplified DBP equations]
For the dynamics considered, the following DBP equations hold:
\begin{equation}
    m^i_{T_A+1,T_B+1}(\tau_i^A,\tau_i^B) = \sum_{\{\tau_k^A,\tau_k^B\}_{k \in \partial i}} W^i(\tau_i^A,\tau_i^B; \{\tau_k^A,\tau_k^B\}_{k \in \partial i}) \prod_{k \in \partial i} m^{k \to i}_{T_A,T_B}(\tau_k^A,\tau_k^B \Vert *,*),
    \label{eq:DBP_marginals}
\end{equation}
\begin{equation}
    m^{i \to j}_{T_A+1,T_B+1}(\tau_i^A,\tau_i^B \Vert \tau_j^A,\tau_j^B) = \sum_{\{\tau_k^A,\tau_k^B\}_{k \in \partial i \backslash j}} W^i(\tau_i^A,\tau_i^B; \{\tau_k^A,\tau_k^B\}_{k \in \partial i}) \prod_{k \in \partial i \backslash j} m^{k \to i}_{T_A,T_B}(\tau_k^A,\tau_k^B \Vert *,*).
    \label{eq:DBP_messages}
\end{equation}
\label{fact:simplified_DBP}
\end{fact}
To simplify the notations, in what follows we will use the notation $m^{k \to i}_{T_A,T_B}(\tau_k^A,\tau_k^B) \equiv m^{k \to i}_{T_A,T_B}(\tau_k^A,\tau_k^B \Vert *,*)$. Given the computed value of the marginals $m_{T}^i(\tau_i^A,\tau_i^B)$, one can straightforwardly define quantities of interest, such as probabilities for a given node $i$ to be found in a given \hanlin{status}:
\begin{align}
    &P^i_S(t) = \hanlin{\sum_{\tau_i^A > t}\sum_{\tau_i^B > t}} m^i_{T_A,T_B}(\tau_i^A,\tau_i^B) = m^i_{t,t}(*,*),
    \label{eq:Marginal_Probability_S}
    \\
    &P^i_A(t) = \hanlin{\sum_{\tau_i^A \leq t}\sum_{\tau_i^B}} m_{T_A,T_B}^i(\tau_i^A,\tau_i^B) = \sum_{\tau_i^A \leq t} \sum_{\tau_i^B \in \{0, 1, \ldots, t, *\}} m_{t,t}^i(\tau_i^A,\tau_i^B) = \sum_{\tau_i^A \leq t} \left(m_{t,0}^i(\tau_i^A,0) + m_{t,0}^i(\tau_i^A,*)\right),
    \label{eq:Marginal_Probability_A}
    \\
    &P^i_B(t) = \hanlin{\sum_{\tau_i^B \leq t}\sum_{\tau_i^A}} m_{T_A,T_B}^i(\tau_i^A,\tau_i^B) = \sum_{\tau_i^B \leq t} \sum_{\tau_i^A \in \{0, 1, \ldots, t, *\}} m_{t,t}^i(\tau_i^A,\tau_i^B) = \sum_{\tau_i^B \leq t} \left(m_{0,t}^i(0,\tau_i^B) + m_{0,t}^i(*,\tau_i^B)\right),
    \label{eq:Marginal_Probability_B}
    \\
    &P^i_{AB}(t) = \hanlin{\sum_{\tau_i^A \leq t}\sum_{\tau_i^B \leq t}} m_{T_A,T_B}^i(\tau_i^A,\tau_i^B) = \sum_{\tau_i^A \leq t, \tau_i^B \leq t} m_{t,t}^i(\tau_i^A,\tau_i^B).
    \label{eq:Marginal_Probability_AB}
\end{align}
In the last expression, we used Fact~\ref{fact:reduction}, and in the previous expressions, we used the definition of $*$ and hence the equalities of the type
\begin{equation}
    m_{T_A,T_B}^i(\tau_i^A,*) = m_{T_A,T_B+1}^i(\tau_i^A,*) + m_{T_A,T_B+1}^i(\tau_i^A,T_B+1).
\end{equation}
}
% $m_{t}^i(*,*) = m_{t+1}^i(*,*) + m_{t+1}^i(*,t+1) + m_{t+1}^i(t+1,*) + m_{t+1}^i(t+1,t+1)$. 

\new{
Hence, in principle, to each node $i$ one can associate a $(t+1) \times (t+1)$ matrix of probabilities $m_{t,t}^i(\tau_i^A,\tau_i^B)$, and to each edge $(ij)$ associate a \hanlin{fourth} order tensor of probabilities $m_{t,t}^{i \to j}(\tau_i^A,\tau_i^B \Vert \tau_j^A,\tau_j^B)$; solve the fixed point equations \eqref{eq:DBP_messages}, e.g., through iteration using some initialization for the conditional probabilities; use \eqref{eq:DBP_marginals} to compute the quantities $m_{t,t}^i(\tau_i^A,\tau_i^B)$; and finally, use \eqref{eq:Marginal_Probability_S}-\eqref{eq:Marginal_Probability_AB} to assemble the probabilities of finding node $i$ in one of the \hanlin{statuses} $S$, $A$, $B$, or $AB$. However, the naive implementation of this scheme would require as many as $O(t^{2(d-1)})$ operations for an estimation of a single message in \eqref{eq:DBP_messages} and $O(t^{2d})$ operations for a single marginal \eqref{eq:DBP_marginals}, which still produces a polynomial-time algorithm for bounded-degree graphs, but can quickly become prohibitive for large $d$.
}

\new{
On the other hand, the argument above applies to general dynamics parametrized by the flipping times $\tau_i^A,\tau_i^B$ for node $i$, while dynamics of interest (such as processes considered here) often has a special structure, and it is beneficial from the computational point of view to explore this structure in order to drastically reduce the complexity of the computation of the marginal probabilities \eqref{eq:Marginal_Probability_S}-\eqref{eq:Marginal_Probability_AB}. Therefore, while the problem is conceptually solvable in polynomial time at the level of the basic DBP equations \eqref{eq:DBP_marginals}-\eqref{eq:DBP_messages}, below we take care of deriving a computational scheme that would allow for an efficient computation of marginal probabilities \eqref{eq:Marginal_Probability_S}-\eqref{eq:Marginal_Probability_AB}. We refer to this scheme as to the Dynamic Message-Passing (DMP) equations.
}

\section{Exact DBP Equations for Mutually Exclusive Competitive Processes}
\label{app:competitive_exactDBP}

\new{
As explained in the Main Text, the transition rules for competitive dynamics at time step $t$ are defined as follows:
\begin{align}
    S(i) &\rightarrow A(i), \quad \text{ with probability } q^i_{S \to A} (t) = v^i_A(t)(1-v^i_B(t))/Z_i(t),
    \label{eq:Transition_S_A}
    \\
    S(i) &\rightarrow B(i), \quad \text{ with probability } q^i_{S \to B} (t) = v^i_B(t)(1-v^i_A(t))/Z_i(t),
    \label{eq:Transition_S_B}
    \\
    S(i) &\rightarrow S(i), \quad \text{ with probability } q^i_{S \to S} (t) = (1 - v^i_A(t) - v^i_B(t) + v^i_A(t)v^i_B(t))/Z_i(t),
    \label{eq:Transition_S_S}
\end{align}
where
\begin{align}
    &v^i_A(t) = 1 - \prod_{j \in \partial i} (1 - \alpha^{A}_{ji} \mathbbm{1}[\sigma_j(t) = A]),
    \\
    &v^i_B(t) = 1 - \prod_{j \in \partial i} (1 - \alpha^{B}_{ji} \mathbbm{1}[\sigma_j(t) = B]),
    \\
    &Z_i = 1 - v^i_A(t) v^i_B(t),
\end{align}
and $\sigma_i(t)$ denotes the \hanlin{status} of node $i$ at time $t$. For two subsets $\Theta_A \subset \partial i$ and $\Theta_B \subset \partial i$, \DS{the} transmission probabilities introduced \DS{in \eqref{eq:Transition_S_A}-\eqref{eq:Transition_S_S}} can be explicitly written as follows:
\begin{align}
    & q^i_{S \to A} (t; \Theta_A, \Theta_B; \{\tau_k^A, \tau_k^B\}_{k \in \partial i})\nonumber
    \\
    & = \frac{(1 - \prod_{j \in \Theta_A} (1 - \alpha^{A}_{ji}\mathbbm{1}[\tau_j^A \leq t])) \prod_{l \in \Theta_B} (1 - \alpha^{B}_{li}\mathbbm{1}[\tau_l^B \leq t])}{\prod_{j \in \Theta_A}(1 - \alpha_{ji}^A\mathbbm{1}[\tau_j^A \leq t]) + \prod_{l \in \Theta_B}(1 - \alpha_{li}^B\mathbbm{1}[\tau_l^B \leq t]) - \prod_{j \in \Theta_A, l \in \Theta_B}(1 - \alpha_{ji}^A\mathbbm{1}[\tau_j^A \leq t])(1 - \alpha_{li}^B\mathbbm{1}[\tau_l^B \leq t])},
    \\
    & q^i_{S \to B} (t; \Theta_A, \Theta_B; \{\tau_k^A, \tau_k^B\}_{k \in \partial i})\nonumber
    \\
    & = \frac{\prod_{j \in \Theta_A} (1 - \alpha^{A}_{ji}\mathbbm{1}[\tau_j^A \leq t])(1 - \prod_{l \in \Theta_B} (1 - \alpha^{B}_{li}\mathbbm{1}[\tau_l^B \leq t]))}{\prod_{j \in \Theta_A}(1 - \alpha_{ji}^A\mathbbm{1}[\tau_j^A \leq t]) + \prod_{l \in \Theta_B}(1 - \alpha_{li}^B\mathbbm{1}[\tau_l^B \leq t]) - \prod_{j \in \Theta_A, l \in \Theta_B}(1 - \alpha_{ji}^A\mathbbm{1}[\tau_j^A \leq t])(1 - \alpha_{li}^B\mathbbm{1}[\tau_l^B \leq t])},
    \\
    & q^i_{S \to S} (t; \Theta_A, \Theta_B; \{\tau_k^A, \tau_k^B\}_{k \in \partial i})\nonumber
    \\
    & = \frac{\prod_{j \in \Theta_A} (1 - \alpha^{A}_{ji}\mathbbm{1}[\tau_j^A \leq t]) \prod_{l \in \Theta_B} (1 - \alpha^{B}_{li}\mathbbm{1}[\tau_l^B \leq t])}{\prod_{j \in \Theta_A}(1 - \alpha_{ji}^A\mathbbm{1}[\tau_j^A \leq t]) + \prod_{l \in \Theta_B}(1 - \alpha_{li}^B\mathbbm{1}[\tau_l^B \leq t]) - \prod_{j \in \Theta_A, l \in \Theta_B}(1 - \alpha_{ji}^A\mathbbm{1}[\tau_j^A \leq t])(1 - \alpha_{li}^B\mathbbm{1}[\tau_l^B \leq t])}.
\end{align}
}

\new{
Given the usual parametrization of marginals and messages $m^i(\tau_i^A,\tau_i^B)$, only the following messages can be non-zero for this dynamics: $m^i(\tau_i^A,*)$, $m^i(*,\tau_i^B)$, and $m^i(*,*)$. The dynamic messages are then defined as follows:
\begin{align}
    P^i_A(t) = \sum_{\tau_i^A \leq t} m^i_{t,t}(\tau_i^A,*) = P^i_A(t-1) + m^i_{t,t}(t,*),\\
    P^i_B(t) = \sum_{\tau_i^B \leq t} m^i_{t,t}(*,\tau_i^B) = P^i_B(t-1) + m^i_{t,t}(*,t),\\
    P^i_S(t) = m^i_{t,t}(*,*)
\end{align}
with the initial conditions $m^i_{t,t}(0,*) = P^i_A(0)$ and $m^i_{t,t}(*,0) = P^i_B(0)$.
}

\new{
As discussed in Appendix~\ref{app:DBP}, the complexity of a general DBP message evaluation requires $O(t^{2d})$ steps, where $d$ is the node degree. Under the mutually exclusive scenario, some of the combinations of flipping times are forbidden since node can not be simultaneously activated under both $A$ and $B$ processes. To reflect this structure, we rewrite DBP equations for $t>0$ in a computationally sub-optimal, but compact way:
\begin{align}
\notag
    m^i_{t,t}&(t,*)
    \\
    \notag
    & = P^i_S(0) \hspace{-0.3cm} \sum_{\substack{\Theta_A \subset \partial i;\\ \Theta_B \subset \partial i;\\  \Theta_S = \partial i \backslash (\Theta_A \cup \Theta_B)}} \hspace{-0.3cm} \sum_{\{\tau_k^A\}_{k \in \Theta_A}} \sum_{\{\tau_l^B\}_{l \in \Theta_B}} \left( \prod_{t' = 0}^{t-2} q^i_{S \to S} (t'; \Theta_A, \Theta_B; \{\tau_k^A, \tau_k^B\}_{k \in \partial i}) \right) q^i_{S \to A} (t-1; \Theta_A, \Theta_B; \{\tau_k^A, \tau_k^B\}_{k \in \partial i}) 
    \\
    & \times \prod_{k \in \Theta_A} m^{k \to i}_{t-1,t-1}(\tau_k^A,*) \prod_{l \in \Theta_B} m^{l \to i}_{t-1,t-1}(*,\tau_l^B) \prod_{n \in \Theta_S} m^{n \to i}_{t-1,t-1}(*,*),
    \\
    \notag
    m^i_{t,t}&(*,t)
    \\
    \notag
    & = P^i_S(0) \hspace{-0.3cm} \sum_{\substack{\Theta_A \subset \partial i;\\ \Theta_B \subset \partial i;\\  \Theta_S = \partial i \backslash (\Theta_A \cup \Theta_B)}} \hspace{-0.3cm} \sum_{\{\tau_k^A\}_{k \in \Theta_A}} \sum_{\{\tau_l^B\}_{l \in \Theta_B}} \left( \prod_{t' = 0}^{t-2} q^i_{S \to S} (t'; \Theta_A, \Theta_B; \{\tau_k^A, \tau_k^B\}_{k \in \partial i})  \right) q^i_{S \to B} (t-1; \Theta_A, \Theta_B; \{\tau_k^A, \tau_k^B\}_{k \in \partial i})
    \\
    & \times \prod_{k \in \Theta_A} m^{k \to i}_{t-1,t-1}(\tau_k^A,*) \prod_{l \in \Theta_B} m^{l \to i}_{t-1,t-1}(*,\tau_l^B) \prod_{n \in \Theta_S} m^{n \to i}_{t-1,t-1}(*,*),
    \\
    \notag
    m^i_{t,t}&(*,*) = P^i_S(0) \hspace{-0.3cm} \sum_{\substack{\Theta_A \subset \partial i;\\ \Theta_B \subset \partial i;\\  \Theta_S = \partial i \backslash (\Theta_A \cup \Theta_B)}} \hspace{-0.3cm} \sum_{\{\tau_k^A\}_{k \in \Theta_A}} \sum_{\{\tau_l^B\}_{l \in \Theta_B}} \left( \prod_{t' = 0}^{t-1} q^i_{S \to S} (t'; \Theta_A, \Theta_B; \{\tau_k^A, \tau_k^B\}_{k \in \partial i})  \right)
    \\
    & \times \prod_{k \in \Theta_A} m^{k \to i}_{t-1,t-1}(\tau_k^A,*) \prod_{l \in \Theta_B} m^{l \to i}_{t-1,t-1}(*,\tau_l^B) \prod_{n \in \Theta_S} m^{n \to i}_{t-1,t-1}(*,*).
\end{align}
DBP messages in the expressions above are computed in a similar way, where subsets $\Theta_A$ and $\Theta_B$ are chosen from $\partial i \backslash j$:
\begin{align}
\notag
    m^{i \to j}_{t,t}&(t,*)
    \\
    \notag
    &\hspace{-0.45cm}= P^i_S(0) \hspace{-0.3cm} \sum_{\substack{\Theta_A \subset \partial i \backslash j;\\ \Theta_B \subset \partial i \backslash j;\\  \Theta_S = \partial i \backslash (j \cup \Theta_A \cup \Theta_B)}} \hspace{-0.3cm} \sum_{\{\tau_k^A\}_{k \in \Theta_A}} \sum_{\{\tau_l^B\}_{l \in \Theta_B}} \left( \prod_{t' = 0}^{t-2} q^i_{S \to S} (t'; \Theta_A, \Theta_B; \{\tau_k^A, \tau_k^B\}_{k \in \partial i}) \right) q^i_{S \to A} (t-1; \Theta_A, \Theta_B; \{\tau_k^A, \tau_k^B\}_{k \in \partial i}) 
    \\
    & \hspace{-0.45cm}\times P^i_S(0) \prod_{k \in \Theta_A} m^{k \to i}_{t-1,t-1}(\tau_k^A,*) \prod_{l \in \Theta_B} m^{l \to i}_{t-1,t-1}(*,\tau_l^B) \prod_{n \in \Theta_S} m^{n \to i}_{t-1,t-1}(*,*), \\
    \notag
    m^{i \to j}_{t,t}&(*,t)
    \\
    \notag
    & \hspace{-0.45cm}= P^i_S(0) \hspace{-0.3cm} \sum_{\substack{\Theta_A \subset \partial i \backslash j;\\ \Theta_B \subset \partial i \backslash j;\\  \Theta_S = \partial i \backslash (j \cup \Theta_A \cup \Theta_B)}} \hspace{-0.3cm} \sum_{\{\tau_k^A\}_{k \in \Theta_A}} \sum_{\{\tau_l^B\}_{l \in \Theta_B}} \left( \prod_{t' = 0}^{t-2} q^i_{S \to S} (t'; \Theta_A, \Theta_B; \{\tau_k^A, \tau_k^B\}_{k \in \partial i})  \right) q^i_{S \to B} (t-1; \Theta_A, \Theta_B; \{\tau_k^A, \tau_k^B\}_{k \in \partial i})
    \\
    & \hspace{-0.45cm}\times \prod_{k \in \Theta_A} m^{k \to i}_{t-1,t-1}(\tau_k^A,*) \prod_{l \in \Theta_B} m^{l \to i}_{t-1,t-1}(*,\tau_l^B) \prod_{n \in \Theta_S} m^{n \to i}_{t-1,t-1}(*,*),
    \\
    \notag
    m^{i \to j}_{t,t}&(*,*) = P^i_S(0) \hspace{-0.3cm} \sum_{\substack{\Theta_A \subset \partial i \backslash j;\\ \Theta_B \subset \partial i \backslash j;\\  \Theta_S = \partial i \backslash (j \cup \Theta_A \cup \Theta_B)}} \hspace{-0.3cm} \sum_{\{\tau_k^A\}_{k \in \Theta_A}} \sum_{\{\tau_l^B\}_{l \in \Theta_B}} \left( \prod_{t' = 0}^{t-1} q^i_{S \to S} (t'; \Theta_A, \Theta_B; \{\tau_k^A, \tau_k^B\}_{k \in \partial i})  \right)
    \\
    & \times \prod_{k \in \Theta_A} m^{k \to i}_{t-1,t-1}(\tau_k^A,*) \prod_{l \in \Theta_B} m^{l \to i}_{t-1,t-1}(*,\tau_l^B) \prod_{n \in \Theta_S} m^{n \to i}_{t-1,t-1}(*,*).
\end{align}
In Section II of the Main Text, we show numerically that these equations are exact on tree graphs.
}

\section{Approximate DMP Equations for Mutually Exclusive Competitive Processes}
\label{app:competitive}

\new{
In the search of low-complexity equations that could be used on general graphs, in this Section we derive approximate DMP equations for mutually exclusive competitive processes. As outlined in the Main Text, our approximation scheme is based on the idea that in the absence of renormalization, the dynamics of each processes follows the dynamics of a simple SI process. Hence, we derive equations that perform \hanlin{the} renormalization procedure at the level of dynamic marginals.  
}

\new{
Let us start by reminding the equations for the SI model. We will closely follow the notations used in \cite{Lokhov2014, Lokhov2015}. For a single process $A$, let $\theta_A^{i \rightarrow j}(t)$ denote the probability that no infection message $A$ has been passed from node $i$ to node $j$ until time $t$, and  $\phi_A^{i \rightarrow j}(t)$ denote the probability that no infection message $A$ has been passed until time $t\!-\!1$, while node $i$ is infected with $A$ at time $t$. Exact DMP equations for a single-process SI model read \cite{Lokhov2015}:
\begin{align}
        \label{eq:SI_equations0}
        &P_{S}^{i}(t) = P_{S}^{i}(0) \prod_{k \in \partial i} \theta_A^{k \rightarrow i}(t),\\
        &P_{A}^{i}(t) = 1 - P_{S}^{i \rightarrow j}(t),\\
        &P_{S}^{i \rightarrow j}(t) = \prod_{k \in \partial i \backslash j} \theta_A^{k \rightarrow i}(t),\\
        &P_{A}^{i \rightarrow j}(t) = 1 - P_{S}^{i \rightarrow j}(t),\\
    	&\theta_A^{i \rightarrow j}(t) = \theta_A^{i \rightarrow j} (t-1)  - \alpha^A_{ji} \phi_A^{i \rightarrow j} (t-1),\\
    	&\phi_A^{i \rightarrow j}(t) = (1 - \alpha^A_{ji})\phi_A^{i \rightarrow j} (t-1) + \left[P_A^{i \rightarrow j}(t) - P_A^{i \rightarrow j} (t-1)\right].
    	\label{eq:SI_equations}
\end{align}
}

\new{
For the two competitive and mutually exclusive processes $A$ and $B$, we approximate the total spreading as a product of two independent spreading processes that follow equations \eqref{eq:SI_equations0}-\eqref{eq:SI_equations}. Our goal will be to derive approximate dynamic marginals, that we denote by $\widehat{P}_S^i(t)$, $\widehat{P}_A^i(t)$, and $\widehat{P}_B^i(t)$. For simplicity, let us assume that $\alpha^A_{ij} < 1$, and hence $\theta_A^{i \to j}$ type messages are non-zero (treatment of the case of deterministic spreading is straightforward and results in several additional equations). Under the chosen approximation, the \emph{non-normalized} probability of staying in the susceptible \hanlin{status} can be written as:}
\begin{equation}
    	\widetilde{P}_{S}^{i \rightarrow j}(t) \!=\! \widehat{P}_S^{i \rightarrow j} (t\!-\!1)\prod_{k \in \partial i \backslash j}\frac{\theta_A^{k \rightarrow i}(t)\theta_B^{k \rightarrow i}(t)}{\theta_A^{k \rightarrow i} (t\!-\!1)\theta_B^{k \rightarrow i} (t\!-\!1)}.
    	\label{eq:Mutually_exclusive_0}
\end{equation}
\new{
Under this approximation, the transition to \hanlin{status} $A$ happens when the target node is susceptible, and the $B$ infection is not transmitted:
}
\begin{eqnarray}
\label{eq:PmessagesOnCavity}
	\widetilde{P}_{A}^{i \rightarrow j}(t) &\!=\!& \widehat{P}_S^{i \rightarrow j} (t\!-\!1)\left[\prod_{k \in \partial i \backslash j}\left(1 - \frac{\alpha_B \phi_B^{k \rightarrow i} (t\!-\!1)}{\theta_B^{k \rightarrow i} (t\!-\!1)}\right)\left(1-\prod_{k \in \partial i \backslash j}\left(1-\frac{\alpha_A \phi_A^{k \rightarrow i} (t\!-\!1)}{\theta_A^{k \rightarrow i} (t\!-\!1)}\right)\right)\right] + \widehat{P}_A^{i \rightarrow j} (t\!-\!1),
\nonumber \\
	\widetilde{P}_{B}^{i \rightarrow j}(t) &\!=\!& \widehat{P}_S^{i \rightarrow j} (t\!-\!1)\left[\prod_{k \in \partial i \backslash j}\left(1 - \frac{\alpha_A \phi_A^{k \rightarrow i} (t\!-\!1)}{\theta_A^{k \rightarrow i} (t\!-\!1)}\right)\left(1-\prod_{k \in \partial i \backslash j}\left(1-\frac{\alpha_B \phi_B^{k \rightarrow i} (t\!-\!1)}{\theta_B^{k \rightarrow i} (t\!-\!1)}\right)\right)\right] + \widehat{P}_B^{i \rightarrow j} (t\!-\!1).
\end{eqnarray}
We then compute the renormalized messages for the three \hanlin{statuses} $\widehat{P}_{S/A/B}^{i \rightarrow j}(t)$:
\begin{equation}
	\widehat{P}_{S/A/B}^{i \rightarrow j}(t) \!=\! \frac{\widetilde{P}_{S/A/B}^{i \rightarrow j}(t)}{\widetilde{P}_{S}^{i \rightarrow j}(t) + \widetilde{P}_{A}^{i \rightarrow j}(t) + \widetilde{P}_{B}^{i \rightarrow j}(t)}.
\end{equation}
\new{
The dynamic marginals are calculated iteratively in a similar fashion:
}
\begin{eqnarray}
\label{eq:posteriorComp}
	\widetilde{P}_{S}^{i}(t) &\!=\!& \widehat{P}_S^i (t\!-\!1)\prod_{k \in \partial i}\frac{\theta_A^{k \rightarrow i}(t)\theta_B^{k \rightarrow i}(t)}{\theta_A^{k \rightarrow i} (t\!-\!1)\theta_B^{k \rightarrow i} (t\!-\!1)},
\nonumber \\
	\widetilde{P}_{A}^{i}(t) &\!=\!& \widehat{P}_S^{i} (t\!-\!1)\left[\prod_{k \in \partial i}\left(1 - \frac{\alpha_B \phi_B^{k \rightarrow i} (t\!-\!1)}{\theta_B^{k \rightarrow i} (t\!-\!1)}\right)\left(1-\prod_{k \in \partial i}\left(1-\frac{\alpha_A \phi_A^{k \rightarrow i} (t\!-\!1)}{\theta_A^{k \rightarrow i} (t\!-\!1)}\right)\right)\right] + \widehat{P}_A^{i} (t\!-\!1),
\nonumber \\
	\widetilde{P}_{B}^{i}(t) &\!=\!& \widehat{P}_S^{i} (t\!-\!1)\left[\prod_{k \in \partial i}\left(1 - \frac{\alpha_A \phi_A^{k \rightarrow i} (t\!-\!1)}{\theta_A^{k \rightarrow i} (t\!-\!1)}\right)\left(1-\prod_{k \in \partial i}\left(1-\frac{\alpha_B \phi_B^{k \rightarrow i} (t\!-\!1)}{\theta_B^{k \rightarrow i} (t\!-\!1)}\right)\right)\right] + \widehat{P}_B^{i} (t\!-\!1),
\nonumber \\
	P_{S/A/B}^{i}(t) &\!=\!& \frac{\widetilde{P}_{S/A/B}^{i}(t)}{\widetilde{P}_{S}^{i}(t) + \widetilde{P}_{A}^{i}(t) + \widetilde{P}_{B}^{i}(t)}.
	\label{eq:Mutually_exclusive}
\end{eqnarray}
\new{We numerically study the performance of these approximate DMP equations in the Main Text.}

\section{Exact DBP and DMP Equations for Collaborative Processes}
\label{app:collaborative_exactDMP}

\subsection{DBP Equations for Collaborative Processes}

\new{
Let us start by reminding the transition rules for the collaborative spreading, as discussed in the Main Text:
\begin{align}
    \label{eq:Dynamical_Rules0}
    S(i) + A(j) &\xrightarrow{\alpha^{A}_{ji}} A(i) + A(j),\\
    S(i) + B(j) &\xrightarrow{\alpha^{B}_{ji}} B(i) + B(j),\\
    A(i) + B(j) &\xrightarrow{\alpha^{BA}_{ji}} AB(i) + B(j),\\
    A(i) + B(j) &\xrightarrow{\alpha^{AB}_{ij}} A(i) + AB(j).
    \label{eq:Dynamical_Rules}
\end{align}
}

\new{
We start by specifying the transition kernel $W^i(\tau_i^A,\tau_i^B; \{\tau_k^A,\tau_k^B\}_{k \in \partial i})$ based on the dynamic rules \eqref{eq:Dynamical_Rules0}-\eqref{eq:Dynamical_Rules}:
\begin{align}
    W^i(\tau_i^A,\tau_i^B; \{\tau_k^A,\tau_k^B\}_{k \in \partial i}) &= \underbrace{P^i_S(0) \mathbbm{1}[\tau_i^A < \tau_i^B] \{S(i) \xrightarrow{\tau_i^A} A(i) \xrightarrow{\tau_i^B} AB(i) \}}_{(1)}\label{eq:transitionKernelD5}
    \\ &+ \underbrace{P^i_S(0) \mathbbm{1}[\tau_i^B < \tau_i^A] \{S(i) \xrightarrow{\tau_i^B} B(i) \xrightarrow{\tau_i^A} AB(i) \}}_{(2)}
    \\ &+ \underbrace{P^i_S(0) \mathbbm{1}[\tau_i^A = \tau_i^B] \{S(i) \xrightarrow{\tau_i^A = \tau_i^B} AB(i) \}}_{(3)}
    \\ &+ \underbrace{P^i_{A^*}(0) \mathbbm{1}[\tau_i^A = 0] \mathbbm{1}[\tau_i^B > 0] \{A(i) \xrightarrow{\tau_i^B} AB(i) \}}_{(4)}
    \\ &+ \underbrace{P^i_{B^*}(0) \mathbbm{1}[\tau_i^B = 0] \mathbbm{1}[\tau_i^A > 0] \{B(i) \xrightarrow{\tau_i^A} AB(i) \}}_{(5)}
    \\ &+ \underbrace{P^i_{AB}(0) \mathbbm{1}[\tau_i^A = \tau_i^B = 0]}_{(6)}.\label{eq:transitionKernelD10}
\end{align}
In the expression above, it is assumed that when $\tau_i^A = *$ or $\tau_i^B = *$, the corresponding transition does not happen during the given observation window; or, alternatively, happens \emph{eventually}, beyond the observation horizon. $P^i_S(0)$, $P^i_{A^*}(0)$, $P^i_{B^*}(0)$, and $P^i_{AB}(0)$ denote probabilities that at initial time, node $i$ is initialized in the \hanlin{statuses} $S$, $A$ exclusively, $B$ exclusively, or $AB$, respectively.
}

\new{
We can now provide explicit forms of the expressions that correspond to the transitions inside curly brackets\DS{; specifically, in the following we refer to the numbered items in~\eqref{eq:transitionKernelD5}-\eqref{eq:transitionKernelD10}}. In the case $\tau_i^A = *$ and $\tau_i^B = *$ we have:
\begin{align}
    (1) = &P^i_S(0) \mathbbm{1}[\tau_i^A < \tau_i^B] \mathbbm{1}[\tau_i^A > 0] \underbrace{\prod_{t'=0}^{\tau_i^A - 2} \prod_{k \in \partial i} (1-\alpha_{ki}^A\mathbbm{1}[\tau_k^A \leq t']) (1 - \prod_{k \in \partial i} (1-\alpha_{ki}^A\mathbbm{1}[\tau_k^A \leq \tau_i^A - 1]))}_{\text{$A$-activation with prob. $\alpha_{ki}^A$ at time $\tau_i^A$}} \nonumber
    \\
    &\times \underbrace{\prod_{t''=0}^{\tau_i^A-1} \prod_{k \in \partial i} (1-\alpha_{ki}^B\mathbbm{1}[\tau_k^B \leq t''])}_{\text{no $B$-activation with prob. $\alpha_{ki}^B$ by $\tau_i^A$}} \underbrace{\prod_{t'''=\tau_i^A}^{\tau_i^B-2} \prod_{k \in \partial i} (1-\alpha_{ki}^{BA}\mathbbm{1}[\tau_k^{B} \leq t''']) (1 - \prod_{k \in \partial i} (1-\alpha_{ki}^{BA}\mathbbm{1}[\tau_k^{B} \leq \tau_i^B-1]))}_{\text{$B$-activation with prob. $\alpha_{ki}^{BA}$ at time $\tau_i^B$}};
    \label{eq:term(1)}
    \\
    (2) = &P^i_S(0) \mathbbm{1}[\tau_i^B < \tau_i^A] \mathbbm{1}[\tau_i^B > 0] \underbrace{\prod_{t'=0}^{\tau_i^B - 2} \prod_{k \in \partial i} (1-\alpha_{ki}^B\mathbbm{1}[\tau_k^B \leq t']) (1 - \prod_{k \in \partial i} (1-\alpha_{ki}^B\mathbbm{1}[\tau_k^B \leq \tau_i^B - 1]))}_{\text{$B$-activation with prob. $\alpha_{ki}^B$ at time $\tau_i^B$}} \nonumber
    \\
    &\times \underbrace{\prod_{t''=0}^{\tau_i^B-1} \prod_{k \in \partial i} (1-\alpha_{ki}^A\mathbbm{1}[\tau_k^A \leq t''])}_{\text{no $A$-activation with prob. $\alpha_{ki}^A$ by $\tau_i^B$}} \underbrace{\prod_{t'''=\tau_i^B}^{\tau_i^A-2} \prod_{k \in \partial i} (1-\alpha_{ki}^{AB}\mathbbm{1}[\tau_k^{A} \leq t''']) (1 - \prod_{k \in \partial i} (1-\alpha_{ki}^{AB}\mathbbm{1}[\tau_k^{A} \leq \tau_i^A-1]))}_{\text{$A$-activation with prob. $\alpha_{ki}^{AB}$ at time $\tau_i^A$}};
    % \\
\end{align}
\begin{align}
    (3) = &P^i_S(0) \mathbbm{1}[\tau_i^A = \tau_i^B] \mathbbm{1}[\tau_i^A > 0] \underbrace{\prod_{t'=0}^{\tau_i^A - 2} \prod_{k \in \partial i} (1-\alpha_{ki}^A\mathbbm{1}[\tau_k^A \leq t']) (1 - \prod_{k \in \partial i} (1-\alpha_{ki}^A\mathbbm{1}[\tau_k^A \leq \tau_i^A - 1]))}_{\text{$A$-activation with prob. $\alpha_{ki}^A$ at time $\tau_i^A$}} \nonumber
    \\
    &\times \underbrace{\prod_{t'=0}^{\tau_i^A - 2} \prod_{k \in \partial i} (1-\alpha_{ki}^B\mathbbm{1}[\tau_k^B \leq t']) (1 - \prod_{k \in \partial i} (1-\alpha_{ki}^B\mathbbm{1}[\tau_k^B \leq \tau_i^A - 1]))}_{\text{$B$-activation with prob. $\alpha_{ki}^B$ at time $\tau_i^B$}};
    \\
    (4) = &P^i_{A^{*}}(0) \mathbbm{1}[\tau_i^B > 0] \mathbbm{1}[\tau_i^A = 0] \underbrace{\prod_{t'=0}^{\tau_i^B - 2} \prod_{k \in \partial i} (1-\alpha_{ki}^{BA}\mathbbm{1}[\tau_k^B \leq t']) (1 - \prod_{k \in \partial i} (1-\alpha_{ki}^{BA}\mathbbm{1}[\tau_k^B \leq \tau_i^B - 1]))}_{\text{$B$-activation with prob. $\alpha_{ki}^{BA}$ at time $\tau_i^B$}};
    \\
    (5) = &P^i_{B^{*}}(0) \mathbbm{1}[\tau_i^A > 0] \mathbbm{1}[\tau_i^B = 0] \underbrace{\prod_{t'=0}^{\tau_i^A - 2} \prod_{k \in \partial i} (1-\alpha_{ki}^{AB}\mathbbm{1}[\tau_k^A \leq t']) (1 - \prod_{k \in \partial i} (1-\alpha_{ki}^{AB}\mathbbm{1}[\tau_k^A \leq \tau_i^A - 1]))}_{\text{$A$-activation with prob. $\alpha_{ki}^{AB}$ at time $\tau_i^A$}};
    \\
    (6) = &P^i_{AB}(0) \underbrace{\mathbbm{1}[\tau_i^A = 0] \mathbbm{1}[\tau_i^B = 0]}_{\text{no activation in \hanlin{status} $AB$}}.
\end{align}
}

\new{
If either $\tau_i^A = *$ or $\tau_i^B = *$, the \emph{activation} above is replaced by an \emph{absence of activation}. For instance, in the case \DS{of $\tau_i^B = *$}, the last term in \DS{Eq. \eqref{eq:term(1)}} would read:
\begin{equation}
    \underbrace{\prod_{t'''=\tau_i^A}^{t} \prod_{k \in \partial i} (1-\alpha_{ki}^{BA}\mathbbm{1}[\tau_k^{B} \leq t'''])}_{\text{absence of $B$-activation with prob. $\alpha_{ki}^{BA}$ at time $\tau_i^B$}}
\end{equation}
for the $(t+1,t+1)$-trajectory of node $i$.
}

\new{
The DBP equations for collaborative processes are given by \eqref{eq:DBP_marginals}-\eqref{eq:DBP_messages} using the $W^i(\tau_i^A,\tau_i^B; \{\tau_k^A,\tau_k^B\}_{k \in \partial i})$ defined above. In what follows, we explain how \hanlin{to} derive the equivalent low-complexity DMP equations for the collaborative processes from the DBP equations \eqref{eq:DBP_marginals}-\eqref{eq:DBP_messages}, state their full iterative form, and finally expand the exact DMP equations around the non-interacting point, which yields \emph{approximate} DMP equations that have a more compact form and a lower algorithmic complexity.
}

\subsection{Definitions of Dynamic Messages}

\new{
Let us introduce the mathematical definitions for the dynamic messages as the functions of \hanlin{fundamental probabilities} on time trajectories, $m_{T_A,T_B}^{i \to j}(\tau_i^A,\tau_i^B) \equiv m_{T_A,T_B}^{i \to j}(\tau_i^A,\tau_i^B \Vert *,*)$, that will be used in the computations scheme below. First, it is useful to consider reduced marginals:
\begin{align}
    \mu^{i}_A(t) = \sum_{\tau_i^B} m_{t,t}^i(t,\tau_i^B) = P^i_A(t) - P^i_A(t-1),\\
    \mu^{i}_B(t) = \sum_{\tau_i^A} m_{t,t}^i(\tau_i^A,t) = P^i_B(t) - P^i_B(t-1).
    \label{eq:reduced_marginals}
\end{align}
They have the physical meaning of probabilities of individual activation with one of the processes, when we do not care about the activation by the other process. In a similar way, we also define reduced messages:
\begin{align}
    \mu^{i \to j}_A(t) = \sum_{\tau_i^B} m_{t,t}^{i \to j}(t,\tau_i^B) = P^{i \to j}_A(t) - P^{i \to j}_A(t-1),\\
    \mu^{i \to j}_B(t) = \sum_{\tau_i^A} m_{t,t}^{i \to j}(\tau_i^A,t) = P^{i \to j}_B(t) - P^{i \to j}_B(t-1).
    \label{eq:reduced_messages}
\end{align}
}

\new{
Similarly to dynamic marginals \eqref{eq:Marginal_Probability_S}-\eqref{eq:Marginal_Probability_AB}, one can define the aggregated dynamic messages. In principle, we can defined them for a general fixed dynamics $(\tau_j^A, \tau_j^B)$ of the ``cavity'' node $j$, but in what follows we will only use the messages for the fixed dynamics $(\tau_j^A, \tau_j^B) = (*,*)$. 
\begin{align}
    & P^{i \to j}_S(t) = \hanlin{\sum_{\tau_i^A > t}\sum_{\tau_i^B > t}} m_{T_A,T_B}^{i \to j}(\tau_i^A,\tau_i^B) = m_{t,t}^{i \to j}(*,*),
    \label{eq:Conditional_Probability_S}
    \\
    & P^{i \to j}_A(t) = \hanlin{\sum_{\tau_i^B}\sum_{\tau_i^A \leq t}} m_{T_A,T_B}^{i \to j}(\tau_i^A,\tau_i^B) = \sum_{\tau_i^A \leq t} \mu^{i \to j}_A(\tau_i^A),
    \label{eq:Conditional_Probability_A}
    \\
    & P^{i \to j}_B(t) = \hanlin{\sum_{\tau_i^B}\sum_{\tau_i^B \leq t}} m_{T_A,T_B}^{i \to j}(\tau_i^A,\tau_i^B) = \sum_{\tau_i^B \leq t} \mu^{i \to j}_B(\tau_i^B),
    \label{eq:Conditional_Probability_B}
    \\
    & P^{i \to j}_{AB}(t) = \hanlin{\sum_{\tau_i^A \leq t}\sum_{\tau_i^B \leq t}} m_{t,t}^{i \to j}(\tau_i^A,\tau_i^B).
    \label{eq:Conditional_Probability_AB}
\end{align}
These quantities have the same physical interpretation, except that they are defined on the ``cavity'' graph where node $j$ is following a fixed $(\tau_j^A, \tau_j^B)$-dynamics. 
}

\new{
Let us also define the following dynamic messages that are weighted combinations of messages:
\begin{align}
        \theta_{A,B}^{k \rightarrow i}(t_A,t_B) = \sum_{\tau_k^A} \sum_{\tau_k^B} \prod_{t'=0}^{t_A-1} (1-\alpha_{ki}^A\mathbbm{1}[\tau_k^A \leq t']) \prod_{t''=0}^{t_B-1} (1-\alpha_{ki}^B\mathbbm{1}[\tau_k^B \leq t'']) m^{k \rightarrow i}_{T_A,T_B}(\tau_k^A,\tau_k^B),
        \label{eq:theta_A_B_definition}
\end{align}
\begin{align}
        \theta_{A,B,AB}^{k \rightarrow i}(t_A,t_B,t_{AB}; \tau_i^A) = \sum_{\tau_k^A} \sum_{\tau_k^B} \prod_{t'=0}^{t_A-1} (1-\alpha_{ki}^A\mathbbm{1}[\tau_k^A \leq t']) &\prod_{t''=0}^{t_B-1}\nonumber (1-\alpha_{ki}^B\mathbbm{1}[\tau_k^B \leq t''])\\ 
        \times \prod_{t'''=\tau_i^A}^{t_{AB}-1} &(1-\alpha_{ki}^{BA}\mathbbm{1}[\tau_k^{B} \leq t''']) m^{k \rightarrow i}_{T_A,T_B}(\tau_k^A,\tau_k^B),
        \label{eq:theta_A_B_AB_definition}
\end{align}
\begin{align}
        \theta_{B,A,BA}^{k \rightarrow i}(t_B,t_A,t_{BA}; \tau_i^B) = \sum_{\tau_k^A} \sum_{\tau_k^B} \prod_{t'=0}^{t_B-1} (1-\alpha_{ki}^B\mathbbm{1}[\tau_k^B \leq t']) &\prod_{t''=0}^{t_A-1}\nonumber (1-\alpha_{ki}^A\mathbbm{1}[\tau_k^A \leq t''])\\ 
        \times \prod_{t'''=\tau_i^B}^{t_{BA}-1} &(1-\alpha_{ki}^{AB}\mathbbm{1}[\tau_k^{A} \leq t''']) m^{k \rightarrow i}_{T_A,T_B}(\tau_k^A,\tau_k^B),
        \label{eq:theta_B_A_AB_definition}
\end{align}
\begin{equation}
    \theta^{k \to i}_{AB}(t_B) = \theta_{A,B,AB}^{k \rightarrow i}(0,0,t_B; 0) =\sum_{\tau_k^B} \prod_{t'=0}^{t_{B}-1} (1-\alpha_{ki}^{BA}\mathbbm{1}[\tau_k^{B} \leq t']) \mu_B^{k \rightarrow i}(\tau_k^B),
\end{equation}
\begin{equation}
    \theta^{k \to i}_{BA}(t_A) = \theta_{B,A,BA}^{k \rightarrow i}(0,0,t_A; 0) = \sum_{\tau_k^A} \prod_{t'=0}^{t_{A}-1} (1-\alpha_{ki}^{AB}\mathbbm{1}[\tau_k^{A} \leq t']) \mu_A^{k \rightarrow i}(\tau_k^A).
\end{equation}
These messages have the meaning of the probabilities that node $k$ did not send $A$ and $B$ activation signals to node $i$ on the ``cavity'' graph where node $i$ is following a fixed $(*,*)$-dynamics.
}

\new{
Next, we define the following quantities:
\begin{equation}
    \phi_{A}^{k \rightarrow i}(t_A,t_B) = \sum_{\tau_k^A} \sum_{\tau_k^B} \mathbbm{1}[\tau_k^A \leq t_A] \prod_{t'=0}^{t_A-1} (1-\alpha_{ki}^A\mathbbm{1}[\tau_k^A \leq t']) \prod_{t''=0}^{t_B-1} (1-\alpha_{ki}^B\mathbbm{1}[\tau_k^B \leq t'']) m^{k \rightarrow i}_{T_A,T_B}(\tau_k^A,\tau_k^B),
    \label{eq:phi_A_definition}
\end{equation}
\begin{equation}
    \phi_{B}^{k \rightarrow i}(t_A,t_B) = \sum_{\tau_k^A} \sum_{\tau_k^B} \mathbbm{1}[\tau_k^B \leq t_B] \prod_{t'=0}^{t_A-1} (1-\alpha_{ki}^A\mathbbm{1}[\tau_k^A \leq t']) \prod_{t''=0}^{t_B-1} (1-\alpha_{ki}^B\mathbbm{1}[\tau_k^B \leq t'']) m^{k \rightarrow i}_{T_A,T_B}(\tau_k^A,\tau_k^B),
    \label{eq:phi_B_definition}
\end{equation}
\begin{equation}
    \phi_{A,B}^{k \rightarrow i}(t_A,t_B) = \sum_{\tau_k^A} \sum_{\tau_k^B} \mathbbm{1}[\tau_k^A \leq t_A] \mathbbm{1}[\tau_k^B \leq t_B] \prod_{t'=0}^{t_A-1} (1-\alpha_{ki}^A\mathbbm{1}[\tau_k^A \leq t']) \prod_{t''=0}^{t_B-1} (1-\alpha_{ki}^B\mathbbm{1}[\tau_k^B \leq t'']) m^{k \rightarrow i}_{T_A,T_B}(\tau_k^A,\tau_k^B).
    \label{eq:phi_A_B_definition}
\end{equation}
\begin{align}
        \phi_{A,B,AB}^{k \rightarrow i}(t_A,t_B,t_{AB}; \tau_i^A) = \sum_{\tau_k^A} \sum_{\tau_k^B} \mathbbm{1}[\tau_k^{B} \leq t_{AB}] \prod_{t'=0}^{t_A-1} (1-\alpha_{ki}^A\mathbbm{1}[\tau_k^A \leq t']) &\prod_{t''=0}^{t_B-1} (1-\alpha_{ki}^B\mathbbm{1}[\tau_k^B \leq t'']) \nonumber \\ 
        \times \prod_{t'''=\tau_i^A}^{t_{AB}-1} (1-\alpha_{ki}^{BA}\mathbbm{1}[\tau_k^{B} \leq t'''])& m^{k \rightarrow i}_{T_A,T_B}(\tau_k^A,\tau_k^B),
        \label{eq:phi_A_B_AB_definition}
\end{align}
\begin{align}
        \phi_{B,A,BA}^{k \rightarrow i}(t_B,t_A,t_{BA}; \tau_i^B) = \sum_{\tau_k^A} \sum_{\tau_k^B} \mathbbm{1}[\tau_k^{A} \leq t_{BA}] \prod_{t'=0}^{t_B-1}  (1-\alpha_{ki}^B\mathbbm{1}[\tau_k^B \leq t']) &\prod_{t''=0}^{t_A-1} (1-\alpha_{ki}^A\mathbbm{1}[\tau_k^A \leq t'']) \nonumber \\ 
        \times \prod_{t'''=\tau_i^B}^{t_{BA}-1} (1-\alpha_{ki}^{AB}\mathbbm{1}[\tau_k^{A} \leq t''']) & m^{k \rightarrow i}_{T_A,T_B}(\tau_k^A,\tau_k^B),
        \label{eq:phi_B_A_AB_definition}
\end{align}
\begin{equation}
    \phi^{k \to i}_{AB}(t_B) = \sum_{\tau_k^B} \mathbbm{1}[\tau_k^{B} \leq t_B] \prod_{t'=0}^{t_{B}-1} (1-\alpha_{ki}^{BA}\mathbbm{1}[\tau_k^{B} \leq t']) \mu_B^{k \rightarrow i}(\tau_k^B),
\end{equation}
\begin{equation}
    \phi^{k \to i}_{BA}(t_A) = \sum_{\tau_k^A} \mathbbm{1}[\tau_k^{A} \leq t_A] \prod_{t'=0}^{t_{A}-1} (1-\alpha_{ki}^{AB}\mathbbm{1}[\tau_k^{A} \leq t']) \mu_A^{k \rightarrow i}(\tau_k^A).
\end{equation}
These messages have the meaning of the probabilities that node $k$ did not send $A$ and $B$ activation signals to node $i$ on the ``cavity'' graph where node $i$ is following a fixed $(*,*)$-dynamics, but node $k$ is in an active \hanlin{status} itself.
}

\new{
Finally, we define auxiliary expressions:
\begin{equation}
    \theta_B^{k \rightarrow i}(t,t) = \sum_{\tau_k^B} \prod_{t''=0}^{t-1} (1-\alpha_{ki}^B\mathbbm{1}[\tau_k^B \leq t'']) m^{k \rightarrow i}_{T_A,T_B}(t,\tau_k^B )
\end{equation}
\begin{equation}
    \theta_A^{k \rightarrow i}(t,t) = \sum_{\tau_k^A} \prod_{t'=0}^{t-1} (1-\alpha_{ki}^A\mathbbm{1}[\tau_k^A \leq t'']) m^{k \rightarrow i}_{T_A,T_B}(\tau_k^A,t )
\end{equation}
}

\subsection{Update Equations for Dynamic Messages}

\new{
We start with the definition
\begin{equation}
    P_S^i(t+1) = m_{t+1,t+1}^i(*,*).
\end{equation}
From the DBP equations on time trajectories \eqref{eq:DBP_marginals}, we get
\begin{equation}
    m_{t+1,t+1}^i(*,*) = \sum_{\{\tau_k^A,\tau_k^B\}_{k \in \partial i}} W^i(*,*; \{\tau_k^A,\tau_k^B\}_{k \in \partial i}) \prod_{k \in \partial i} m^{k \to i}_{t,t}(\tau_k^A,\tau_k^B \Vert *,*),
\end{equation}
where
\begin{equation}
    W^i(*,*; \{\tau_k^A,\tau_k^B\}_{k \in \partial i}) = \prod_{k \in \partial i} \prod_{t'=0}^{t} (1-\alpha_{ki}^A\mathbbm{1}[\tau_k^A \leq t']) \prod_{t''=0}^{t} (1-\alpha_{ki}^B\mathbbm{1}[\tau_k^B \leq t''])
\end{equation}
Using the definition \eqref{eq:theta_A_B_definition}
\begin{equation}
    P_S^i(t+1) = P_S^i(0)\prod_{k \in \partial i \backslash j}\theta_{A,B}^{k \rightarrow i}(t+1, t+1 \Vert *,*)
\end{equation}
For simplicity, let us denote $\theta_{A,B}^{k \rightarrow i}(t+1, t+1) \equiv \theta_{A,B}^{k \rightarrow i}(t+1, t+1 \Vert *,*)$. With this simplified notation, we finally write
\begin{equation}
    P_S^i(t+1) = P_S^i(0)\prod_{k \in \partial i \backslash j}\theta_{A,B}^{k \rightarrow i}(t+1, t+1).
\end{equation}
We will use the same simplified notation for all quantities of the type $\#^{k \to i}(...) \equiv \#^{k \to i}(... \Vert *,*)$. Using the previously defined quantities $\phi_A$, $\phi_B$ and $\phi_{A,B}$, through \eqref{eq:phi_A_definition}, \eqref{eq:phi_B_definition} and \eqref{eq:phi_A_B_definition}, we derive the following update equations:
\begin{equation}
    \theta_{A,B}^{k \rightarrow i}(t+1, t+1) = \theta_{A,B}^{k \rightarrow i}(t,t) - \alpha_{ki}^A \phi_{A}^{k \rightarrow i}(t, t) - \alpha_{ki}^{B} \phi_B^{k \rightarrow i}(t, t) + \alpha_{ki}^A \alpha_{ki}^B \phi_{A,B}^{k \rightarrow i}(t,t).
\end{equation}
}

\new{
Using the equality $\mathbbm{1}[\tau_k^A \leq t_A] = \mathbbm{1}[\tau_k^A \leq t_A-1] + \mathbbm{1}[\tau_k^A = t_A]$, we further get
\begin{align}
        \phi_A^{k \rightarrow i}(t,t)=(1-\alpha_{ki}^A)\phi_A^{k \rightarrow i}(t-1,t) + \theta_B^{k \rightarrow i}(t,t),\\
        \phi_B^{k \rightarrow i}(t,t)=(1-\alpha_{ki}^B)\phi_B^{k \rightarrow i}(t-1,t-1) + \theta_A^{k \rightarrow i}(t,t), 
\end{align}
where $\theta_A$ and $\theta_B$ have been defined above. Finally, using
\begin{align}
    \phi_A^{k \rightarrow i}(t-1,t) = \phi_A^{k \rightarrow i}(t-1,t-1) + (1-\alpha_{ki}^B) \phi_{A,B}^{k \rightarrow i}(t-1,t-1),\\
    \phi_B^{k \rightarrow i}(t,t-1) = \phi_B^{k \rightarrow i}(t-1,t-1) + (1-\alpha_{ki}^A) \phi_{A,B}^{k \rightarrow i}(t-1,t-1),
\end{align}
we get in the end:
\begin{equation}
        \phi_A^{k \rightarrow i}(t,t)=(1-\alpha_{ki}^A)\phi_A^{k \rightarrow i}(t-1,t-1) - \alpha_{ki}^B(1-\alpha_{ki}^A)\phi_{A,B}^{k \rightarrow i}(t-1,t-1) + \theta_B^{k \rightarrow i}(t,t),
\end{equation}
\begin{equation}
        \phi_B^{k \rightarrow i}(t,t)=(1-\alpha_{ki}^B)\phi_B^{k \rightarrow i}(t-1,t-1) - \alpha_{ki}^A(1-\alpha_{ki}^B)\phi_{A,B}^{k \rightarrow i}(t-1,t-1) + \theta_A^{k \rightarrow i}(t,t).
\end{equation}
% \begin{align}
%     &\phi_A^{k \rightarrow i}(t,t) = (1-\alpha_{ki}^A)\phi_A^{k \rightarrow i}(t-1,t) + \theta_B^{k \rightarrow i}(t,t),
%     \\
%     &\phi_A^{k \rightarrow i}(t-1,t) = -\alpha_{ki}^B\phi_{A,B}^{k \rightarrow i}(t-1,t-1) + \phi_A^{k \rightarrow i}(t-1,t-1), 
% \end{align}
Additionally,
\begin{align}
    \phi_{A,B}^{k \rightarrow i}(t,t)& = (1-\alpha_{ki}^A)(1-\alpha_{ki}^B)\phi_{A,B}^{k \rightarrow i}(t-1,t-1) - m_{t,t}^{k \rightarrow i}(t,t)\\
    & + \sum_{\tau_{k}^A \leq t}\prod_{t'=0}^{t-1}(1-\alpha_{ki}^A\mathbbm{1}[t' \geq \tau_k^A])m_{t,t}^{k \rightarrow i}(\tau_k^A,t)\\
    & +  \sum_{\tau_{k}^B \leq t}\prod_{t'=0}^{t-1}(1-\alpha_{ki}^B\mathbbm{1}[t' \geq \tau_k^B])m_{t,t}^{k \rightarrow i}(t, \tau_k^B),
\end{align}
where $m_{t,t}^{k \rightarrow i}(\tau_k^A,\tau_k^B) \equiv m_{t,t}^{k \rightarrow i}(\tau_k^A,\tau_k^B \Vert *,*)$. Finally, we need to compute iteratively $\theta_A^{k \rightarrow i}(t,t)$ and $\theta_B^{k \rightarrow i}(t,t)$:
\begin{align}
    \theta^{k \to i}_A(t,t) = \theta^{k \to i}_A(t-1,t) - \alpha_{ki}^A \sum_{\tau_k^A \leq t-1}\prod_{t'=0}^{t-2}(1-\alpha_{ki}^A\mathbbm{1}[t' \geq \tau_k^A])m_{t,t}^{k \rightarrow i}(\tau_k^A,t),\\
    \theta^{k \to i}_B(t,t) = \theta^{k \to i}_B(t,t-1) -\alpha_{ki}^B \sum_{\tau_k^B \leq t-1}\prod_{t'=0}^{t-2}(1-\alpha_{ki}^B\mathbbm{1}[t' \geq \tau_k^B])m_{t,t}^{k \rightarrow i}(t,\tau_k^B),
\end{align}
where by definition $\theta^{k \to i}_A(0,t) = \mu^{k \to i}_B(t) = P^{k \to i}_B(t) - P^{k \to i}_B(t-1)$, and similarly $\theta^{k \to i}_B(t,0) = \mu^{k \to i}_A(t) = P^{k \to i}_A(t) - P^{k \to i}_A(t-1)$. Initial and border conditions of that type can be directly obtained from the definitions of the dynamic messages and their connections to messages on time trajectories.
}

\new{
From the scheme above, we still need to compute the update equations for the marginals and the messages on time trajectories of the type $m_{t,t}^i(\tau_i^A,\tau_i^B)$ and $m_{t,t}^{i \rightarrow j}(\tau_i^A,\tau_i^B)$. The derivation of the equations for these messages is given below. Once the full or reduced marginals on time trajectories are computed, the dynamic marginals of interest are easy to compute according to the following equations:
\begin{equation}
    P_A^i(t) = \sum_{t' \leq t}\mu_A^i(t'),
\end{equation}
\begin{equation}
    P_B^i(t) = \sum_{t' \leq t}\mu_B^i(t'),
\end{equation}
\begin{equation}
    P_{AB}^i(t) = \sum_{t' \leq t, t'' \leq t} m_{t',t''}^i(t',t'') = P_A^i(t) + P_B^i(t) + P_S^i(t) - 1.
\end{equation}
}

\subsection{Computation of $m_{t,t}^{i}(\tau_{i}^{A},\tau_{i}^{B})$ and $m_{t,t}^{i \rightarrow j}(\tau_{i}^{A},\tau_{i}^{B})$}

\new{
It is convenient to break down the computation depending on a particular combination of $(\tau_i^A,\tau_i^B)$, including cases of times $0$ and $*$. We will provide a detailed derivation of one of the most general cases, all the others being can be treated similarly. Let us consider the case of finite-time $(\tau_i^A,\tau_i^B)$. Several cases of the type depicted in Fig.~\ref{fig:three_cases_for_finite_time} need to be considered. For the sake of illustration, we will consider the subcase $* \neq \tau_i^B > \tau_i^A > 0$.
}

\begin{figure}[!ht]
    \centering
    \includegraphics[width=\textwidth]{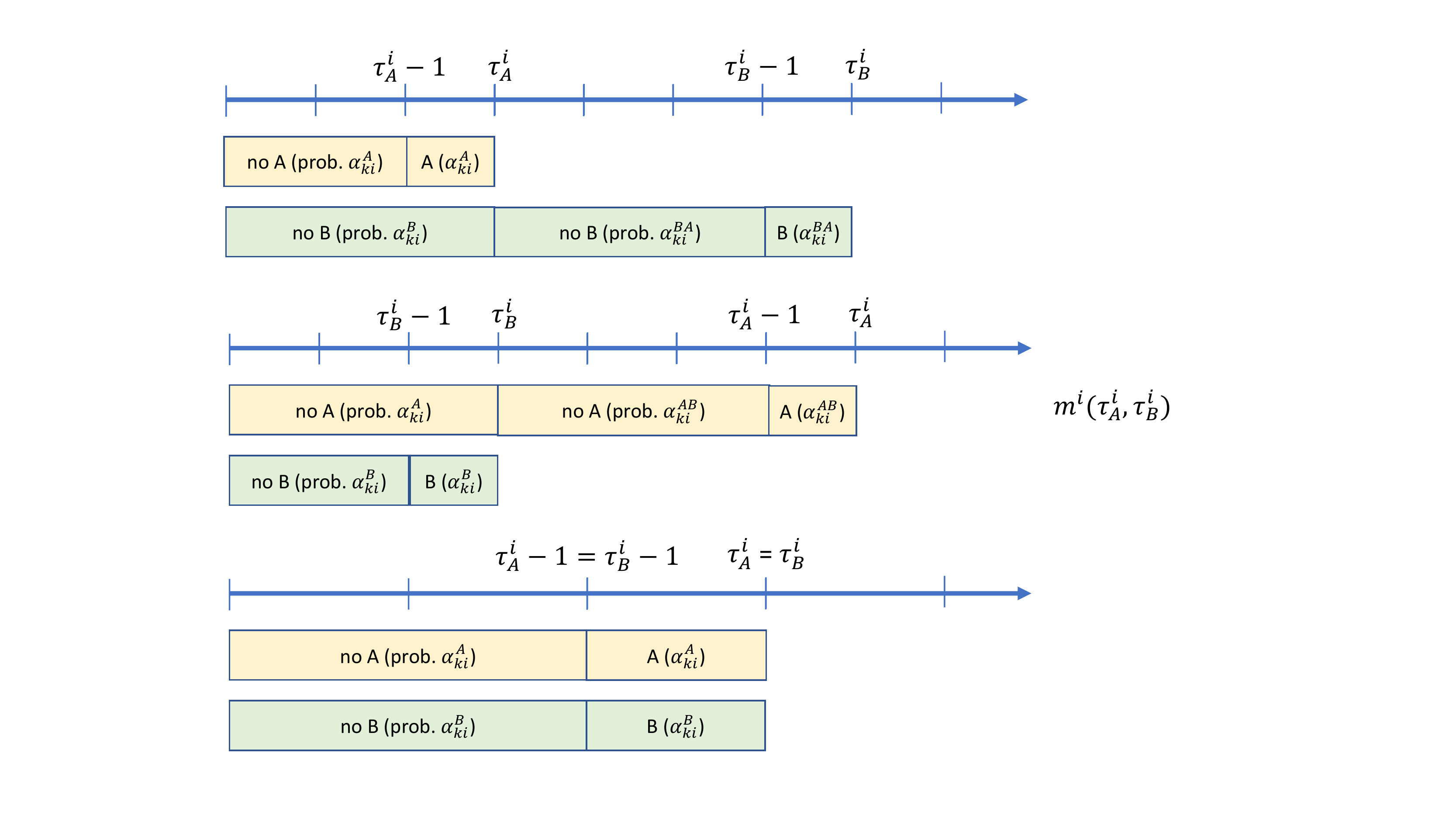}
    \caption{\new{Three cases to consider for the marginals and messages for the time trajectory $(\tau_i^A,\tau_i^B)$.}}
    \label{fig:three_cases_for_finite_time}
\end{figure}

\new{
Using Fact \ref{fact:simplified_DBP}, we the simplified DBP equations read
\begin{equation}
    m_{t,t}^{i}(\tau_i^A,\tau_i^B) = \sum_{\{\tau_k^A,\tau_k^B\}_{k \in \partial i}} W^i(\tau_i^A,\tau_i^B; \{\tau_k^A,\tau_k^B\}_{k \in \partial i}) \prod_{k \in \partial i} m^{k \to i}_{t-1,t-1}(\tau_k^A,\tau_k^B),
    \label{eq:simplified_DBP_marginal}
\end{equation}
and
\begin{equation}
    m_{t,t}^{i \to j}(\tau_i^A,\tau_i^B) = \sum_{\{\tau_k^A,\tau_k^B\}_{k \in \partial i \backslash j}} W^i(\tau_i^A,\tau_i^B; \{\tau_k^A,\tau_k^B\}_{k \in \partial i \backslash j}, \tau_j^A = *, \tau_j^B = *) \prod_{k \in \partial i \backslash j} m^{k \to i}_{t-1,t-1}(\tau_k^A,\tau_k^B).
    \label{eq:simplified_DBP_message}
\end{equation}
Using the definitions for the dynamics in the case considered, i.e.\hanlin{,} term (1) in $W^i(\tau_i^A,\tau_i^B; \{\tau_k^A,\tau_k^B\}_{k \in \partial i})$,  as well as for the dynamic messages \eqref{eq:theta_A_B_AB_definition}, we obtain
\begin{align}
    m_{t,t}^{i}(\tau_i^A,\tau_i^B) = P^i_{S}(0)
    \Big[
    \prod_{k \in \partial i} \theta_{A,B,AB}^{k \to i}(\tau_i^A - 1, \tau_i^A, \tau_i^B - 1; \tau_i^A)
    - \prod_{k \in \partial i} \theta_{A,B,AB}^{k \to i}(\tau_i^A - 1, \tau_i^A, \tau_i^B; \tau_i^A) \nonumber
    \\ 
    - \prod_{k \in \partial i} \theta_{A,B,AB}^{k \to i}(\tau_i^A, \tau_i^A, \tau_i^B - 1; \tau_i^A)
    + \prod_{k \in \partial i} \theta_{A,B,AB}^{k \to i}(\tau_i^A, \tau_i^A, \tau_i^B; \tau_i^A)
    \Big],
\end{align}
and, in the same way,
\begin{align}
    m_{t,t}^{i \to j}(\tau_i^A,\tau_i^B) = P^i_{S}(0)
    \Big[
    \prod_{k \in \partial i \backslash j} \theta_{A,B,AB}^{k \to i}(\tau_i^A - 1, \tau_i^A, \tau_i^B - 1; \tau_i^A)
    - \prod_{k \in \partial i \backslash j} \theta_{A,B,AB}^{k \to i}(\tau_i^A - 1, \tau_i^A, \tau_i^B; \tau_i^A) \nonumber
    \\
    - \prod_{k \in \partial i \backslash j} \theta_{A,B,AB}^{k \to i}(\tau_i^A, \tau_i^A, \tau_i^B - 1; \tau_i^A)
    + \prod_{k \in \partial i \backslash j} \theta_{A,B,AB}^{k \to i}(\tau_i^A, \tau_i^A, \tau_i^B; \tau_i^A)
    \Big].
\end{align}
Through the definition, we immediately get 
\begin{equation}
    \theta_{A,B,AB}^{k \to i}(t_A, t_B, t_{AB} + 1; \tau_i^A) = \theta_{A,B,AB}^{k \to i}(t_A, t_B, t_{AB}; \tau_i^A) - \mathbbm{1}[\tau_i^A \leq t_{AB}]  \alpha_{ki}^{BA} \phi_{A,B,AB}^{k \to i}(t_A, t_B, t_{AB}; \tau_i^A),
\end{equation}
where the former quantity is defined through \eqref{eq:phi_A_B_AB_definition}. Working with this definition, we get the following update equation for $\phi_{A,B,AB}^{k \to i}(t_A, t_B, t_{AB}; \tau_i^A)$:
\begin{align}
    \phi_{A,B,AB}^{k \to i}(t_A, t_B, t_{AB}; \tau_i^A \Vert \tau_i^A, \tau_i^B) &=
    (1 - \alpha_{ki}^{BA}) \phi_{A,B,AB}^{k \to i}(t_A, t_B, t_{AB} -1; \tau_i^A \Vert \tau_i^A, \tau_i^B) \nonumber \\ 
    & + \sum_{\tau_k^A} \prod_{t'=0}^{t_A-1} (1 - \alpha_{ki}^{A} \mathbbm{1}[\tau_k^A \leq t']) \prod_{t''=0}^{t_B-1} (1 - \alpha_{ki}^{B}\mathbbm{1}[\tau_k^B \leq t'']) m_{t_{AB},t_{AB}}^{k \to i}(\tau_k^A, t_{AB}).
\end{align}
The term $\prod_{t''=0}^{t_B-1} (1 - \alpha_{ki}^{B}\mathbbm{1}[\tau_k^B \leq t''])$ is not active for the times considered here and is equal to one because $\tau_k^B = t_{AB}$ and $t_{AB} \geq t_B$, and hence we get the final update equation
\begin{align}
    \phi_{A,B,AB}^{k \to i}(t_A, t_B, t_{AB}; \tau_i^A)
    & = (1 - \alpha_{ki}^{BA}) \phi_{A,B,AB}^{k \to i}(t_A, t_B, t_{AB} -1; \tau_i^A)\nonumber \\
    & + \sum_{\tau_k^A \leq t_{AB}} \prod_{t'=0}^{t_A-1} (1 - \alpha_{ki}^{A} \mathbbm{1}[\tau_k^A \leq t']) m_{t_{AB},t_{AB}}^{k \to i}(\tau_k^A, t_{AB}) + m_{t_{AB},t_{AB}}^{k \to i}(*, t_{AB}),
\end{align}
where all $m_{t_{AB},t_{AB}}^{k \to i}(t',t'')$ are known from previous step computations.
}

\subsection{Exact DMP Equations for Collaborative Processes}

\new{
In this Section, we summarize the derivation procedure outlined above, and state the full form of the DMP equations for collaborative processes.
}
\\
\new{
\underline{Initial conditions from the problem definition:}
\begin{equation}
    P^i_S(0), P^i_{A^*}(0), P^i_{B^*}(0), P^i_{AB}(0).
\end{equation}
}
\new{
\underline{Initialization of dynamic messages:}
\begin{align}
    m_{0,0}^{i}(0,0) = P^i_{AB}(0), \quad m_{0,0}^{i}(0,*) = P^i_{A^*}(0), \quad m_{0,0}^{i}(*,0) = P^i_{B^*}(0)\\
    m_{0,0}^{i \rightarrow j}(0,0) = P^i_{AB}(0), \quad m_{0,0}^{i \rightarrow j}(0,*) = P^i_{A^*}(0), \quad m_{0,0}^{i \rightarrow j}(*,0) = P^i_{B^*}(0)\\
    \mu_A^i(0) = P^i_A(0), \quad \mu_B^i(0) = P^i_B(0)\\
    \theta_{A,B}^{k \to i}(0,0) = \theta_{A,B,AB}^{k \to i}(0,0) = \theta_{B,A,BA}^{k \to i}(0,0) = \theta_{AB}^{k \to i}(0,0) = \theta_{BA}^{k \to i}(0,0) = 1\\
    \phi_{A}^{k \rightarrow i}(0,0) = P^k_A(0), \quad \phi_{B}^{k \rightarrow i}(0,0) = P^k_B(0), \quad \phi_{A,B}^{k \rightarrow i}(0,0) = P^k_{AB}(0)\\
    \phi_{AB}^{k \to i}(0) = P^k_{B}(0), \quad \phi_{BA}^{k \to i}(0) = P^k_{A}(0)
\end{align}
}
\new{
\underline{Border conditions for dynamic messages:}
\begin{align}
    \theta^{k \to i}_A(0,t) = \mu^{k \to i}_B(t), \quad \theta^{k \to i}_B(t,0) = \mu^{k \to i}_A(t)\\
    \theta_{A,B,AB}^{k \rightarrow i}(t_A,t_B,t_{AB}; \tau_i^A) = \theta_{A,B}^{k \rightarrow i}(t_A,t_B) \, \text{for } t_{AB} \leq \tau_i^A\\
    \theta_{B,A,BA}^{k \rightarrow i}(t_B,t_A,t_{BA}; \tau_i^B) = \theta_{A,B}^{k \rightarrow i}(t_A,t_B) \, \text{for } t_{BA} \leq \tau_i^B\\
    \phi_{A,B,AB}^{k \to i}(t_A, t_B, t_{AB}; \tau_i^A) = \phi_{B}^{k \rightarrow i}(t_A, t_B) \, \text{for } t_{AB} = t_B \text{ and } t_{AB} \leq \tau_i^A\\
    \phi_{B,A,BA}^{k \to i}(t_B, t_A, t_{BA}; \tau_i^B) = \phi_{A}^{k \rightarrow i}(t_A, t_B) \, \text{for } t_{BA} = t_A \text{ and } t_{BA} \leq \tau_i^B\\
    \notag
    \text{(true but no need to enforce)} \quad
    \theta^{k \to i}_{AB}(t_B) = \theta_{A,B,AB}^{k \rightarrow i}(0,0,t_B; 0), \quad \theta^{k \to i}_{BA}(t_A) = \theta_{B,A,BA}^{k \rightarrow i}(0,0,t_A; 0)
\end{align}
}

\new{
\underline{Expression for the dynamic marginals:}
\begin{align}
%\begin{equation}
    P_A^i(t) = \sum_{t' \leq t}\mu_A^i(t') \\
%\end{equation}
%\begin{equation}
    P_B^i(t) = \sum_{t' \leq t}\mu_B^i(t') \\
%\end{equation}
%\begin{equation}
    P_S^i(t) = m_{t,t}^i(*,*) = P_S^i(0)\prod_{k \in \partial i \backslash j }{\theta_{A,B}^{k \rightarrow i}(t, t)} \\
%\end{equation}
%\begin{equation}
    P_{AB}^i(t) = \sum_{t' \leq t, t'' \leq t} m_{t',t''}^i(t',t'') = P_A^i(t) + P_B^i(t) + P_S^i(t) - 1.
%\end{equation}
\end{align}
}

\new{
\underline{Update equations:}
\begin{align}
%\begin{equation}
    \mu^{i}_A(t) % = m_{t,0}^{i}(t,0) + m_{t,0}^{i}(t,*)
    = \sum_{t' \leq t} m_{t,t}^i(t,t') + m_{t,t}^i(t,*) \\
%\end{equation}
%\begin{equation}
    \mu^{i}_B(t) % = m_{0,t}^{i}(0,t) + m_{0,t}^{i}(*,t)
    = \sum_{t' \leq t} m_{t,t}^i(t',t) + m_{t,t}^i(*,t) \\
%\end{equation}
%\begin{equation}
    \mu^{i \to j}_A(t) % = m_{t,0}^{i \to j}(t,0) + m_{t,0}^{i \to j}(t,*)
    = \sum_{t' \leq t} m_{t,t}^{i \to j}(t,t') + m_{t,t}^{i \to j}(t,*) \\
%\end{equation}
%\begin{equation}
    \mu^{i \to j}_B(t) % = m_{0,t}^{i \to j}(0,t) + m_{0,t}^{i \to j}(*,t)
    = \sum_{t' \leq t} m_{t,t}^{i \to j}(t',t) + m_{t,t}^{i \to j}(*,t)
%\end{equation}
\end{align}
}

\new{
\begin{align}
    \notag
    &m_{t,t}^{i}(\tau_i^A,\tau_i^B) = \\
    \notag
    & \\
    &\begin{cases}
    P^i_{A^*}(0) \left[ \prod_{k \in \partial i} \theta_{AB}^{k \to i}(\tau_i^B-1) - \prod_{k \in \partial i} \theta_{AB}^{k \to i}(\tau_i^B) \right], & \text{if}\ \tau_i^A = 0, \tau_i^B \neq 0 \\
    & \\
    %%%%%
     P^i_{B^*}(0) \left[ \prod_{k \in \partial i} \theta_{BA}^{k \to i}(\tau_i^A-1) - \prod_{k \in \partial i} \theta_{BA}^{k \to i}(\tau_i^A) \right], & \text{if}\ \tau_i^B = 0, \tau_i^A \neq 0 \\
    & \\
    %%%%%
    P^i_{S}(0)
    \Big[
    \prod_{k \in \partial i} \theta_{A,B,AB}^{k \to i}(\tau_i^A - 1, \tau_i^A, \tau_i^B - 1; \tau_i^A)
    - \prod_{k \in \partial i} \theta_{A,B,AB}^{k \to i}(\tau_i^A - 1, \tau_i^A, \tau_i^B; \tau_i^A)
    \\
    - \prod_{k \in \partial i} \theta_{A,B,AB}^{k \to i}(\tau_i^A, \tau_i^A, \tau_i^B - 1; \tau_i^A)
    + \prod_{k \in \partial i} \theta_{A,B,AB}^{k \to i}(\tau_i^A, \tau_i^A, \tau_i^B; \tau_i^A)
    \Big], & \text{if}\ \tau_i^B > \tau_i^A > 0 \\
    & \\
    %%%%%
    P^i_{S}(0)
    \Big[
    \prod_{k \in \partial i} \theta_{B,A,BA}^{k \to i}(\tau_i^B - 1, \tau_i^B, \tau_i^A - 1; \tau_i^B)
    - \prod_{k \in \partial i} \theta_{B,A,BA}^{k \to i}(\tau_i^B - 1, \tau_i^B, \tau_i^A; \tau_i^B)
    \\
    - \prod_{k \in \partial i} \theta_{B,A,BA}^{k \to i}(\tau_i^B, \tau_i^B, \tau_i^A - 1; \tau_i^B)
    + \prod_{k \in \partial i} \theta_{B,A,BA}^{k \to i}(\tau_i^B, \tau_i^B, \tau_i^A; \tau_i^B)
    \Big], & \text{if}\ \tau_i^A > \tau_i^B > 0 \\
    & \\
    %%%%%
    P^i_{S}(0)
    \Big[
    \prod_{k \in \partial i} \theta_{A,B}^{k \to i}(\tau_i^A - 1, \tau_i^A - 1)
    - \prod_{k \in \partial i} \theta_{A,B}^{k \to i}(\tau_i^A - 1, \tau_i^A)
    \\
    - \prod_{k \in \partial i} \theta_{A,B}^{k \to i}(\tau_i^A, \tau_i^A - 1)
    + \prod_{k \in \partial i} \theta_{A,B}^{k \to i}(\tau_i^A, \tau_i^A)
    \Big], & \text{if}\ \tau_i^A = \tau_i^B \neq 0 \\
    & \\
    %%%%%
    P^i_{A^*}(0) \prod_{k \in \partial i} \theta_{AB}^{k \to i}(t), & \text{if}\ \tau_i^A = 0, \tau_i^B = * \\
    & \\
    %%%%%
    P^i_{B^*}(0) \prod_{k \in \partial i} \theta_{BA}^{k \to i}(t), & \text{if}\ \tau_i^B = 0, \tau_i^A = * \\
    & \\
    %%%%%
    P^i_{S}(0)
    \Big[
    \prod_{k \in \partial i} \theta_{A,B,AB}^{k \to i}(\tau_i^A - 1, \tau_i^A, t; \tau_i^A)
    - \prod_{k \in \partial i} \theta_{A,B,AB}^{k \to i}(\tau_i^A, \tau_i^A, t; \tau_i^A)
    \Big], & \text{if}\ \tau_i^A > 0, \tau_i^B = * \\
    & \\
    %%%%%
    P^i_{S}(0)
    \Big[
    \prod_{k \in \partial i} \theta_{B,A,BA}^{k \to i}(\tau_i^B - 1, \tau_i^B, t; \tau_i^B)
    - \prod_{k \in \partial i} \theta_{B,A,BA}^{k \to i}(\tau_i^B, \tau_i^B, t; \tau_i^B)
    \Big], & \text{if}\ \tau_i^B > 0, \tau_i^A = * \\
    & \\
    %%%%%
    P^i_{S}(0) \prod_{k \in \partial i} \theta_{A,B}^{k \to i}(t, t), & \text{if}\ \tau_i^A = *, \tau_i^B = *
    \label{eq:marginal_on_trajectories_update}
    \end{cases}
\end{align}
}

\new{
\noindent
The update expressions for the messages $m_{t,t}^{i \to j}(\tau_i^A,\tau_i^B)$ are similar to the expressions for the marginals $m_{t,t}^{i}(\tau_i^A,\tau_i^B)$ above, except $\partial i$ is replaced with $\partial i \backslash j$:
}

\new{
\begin{align}
    \notag
    &m_{t,t}^{i \to j}(\tau_i^A,\tau_i^B) = \\
    \notag
    & \\
    &\begin{cases}
    P^i_{A^*}(0) \left[ \prod_{k \in \partial i \backslash j} \theta_{AB}^{k \to i}(\tau_i^B-1) - \prod_{k \in \partial i \backslash j} \theta_{AB}^{k \to i}(\tau_i^B) \right], & \text{if}\ \tau_i^A = 0, \tau_i^B \neq 0 \\
    & \\
    %%%%%
     P^i_{B^*}(0) \left[ \prod_{k \in \partial i \backslash j} \theta_{BA}^{k \to i}(\tau_i^A-1) - \prod_{k \in \partial i \backslash j} \theta_{BA}^{k \to i}(\tau_i^A) \right], & \text{if}\ \tau_i^B = 0, \tau_i^A \neq 0 \\
    & \\
    %%%%%
    P^i_{S}(0)
    \Big[
    \prod_{k \in \partial i \backslash j} \theta_{A,B,AB}^{k \to i}(\tau_i^A - 1, \tau_i^A, \tau_i^B - 1; \tau_i^A)
    - \prod_{k \in \partial i \backslash j} \theta_{A,B,AB}^{k \to i}(\tau_i^A - 1, \tau_i^A, \tau_i^B; \tau_i^A)
    \\
    - \prod_{k \in \partial i \backslash j} \theta_{A,B,AB}^{k \to i}(\tau_i^A, \tau_i^A, \tau_i^B - 1; \tau_i^A)
    + \prod_{k \in \partial i \backslash j} \theta_{A,B,AB}^{k \to i}(\tau_i^A, \tau_i^A, \tau_i^B; \tau_i^A)
    \Big], & \text{if}\ \tau_i^B > \tau_i^A > 0 \\
    & \\
    %%%%%
    P^i_{S}(0)
    \Big[
    \prod_{k \in \partial i \backslash j} \theta_{B,A,BA}^{k \to i}(\tau_i^B - 1, \tau_i^B, \tau_i^A - 1; \tau_i^B)
    - \prod_{k \in \partial i \backslash j} \theta_{B,A,BA}^{k \to i}(\tau_i^B - 1, \tau_i^B, \tau_i^A; \tau_i^B)
    \\
    - \prod_{k \in \partial i \backslash j} \theta_{B,A,BA}^{k \to i}(\tau_i^B, \tau_i^B, \tau_i^A - 1; \tau_i^B)
    + \prod_{k \in \partial i \backslash j} \theta_{B,A,BA}^{k \to i}(\tau_i^B, \tau_i^B, \tau_i^A; \tau_i^B)
    \Big], & \text{if}\ \tau_i^A > \tau_i^B > 0 \\
    & \\
    %%%%%
    P^i_{S}(0)
    \Big[
    \prod_{k \in \partial i \backslash j} \theta_{A,B}^{k \to i}(\tau_i^A - 1, \tau_i^A - 1)
    - \prod_{k \in \partial i \backslash j} \theta_{A,B}^{k \to i}(\tau_i^A - 1, \tau_i^A)
    \\
    - \prod_{k \in \partial i \backslash j} \theta_{A,B}^{k \to i}(\tau_i^A, \tau_i^A - 1)
    + \prod_{k \in \partial i \backslash j} \theta_{A,B}^{k \to i}(\tau_i^A, \tau_i^A)
    \Big], & \text{if}\ \tau_i^A = \tau_i^B \neq 0 \\
    & \\
    %%%%%
    P^i_{A^*}(0) \prod_{k \in \partial i \backslash j} \theta_{AB}^{k \to i}(t), & \text{if}\ \tau_i^A = 0, \tau_i^B = * \\
    & \\
    %%%%%
    P^i_{B^*}(0) \prod_{k \in \partial i \backslash j} \theta_{BA}^{k \to i}(t), & \text{if}\ \tau_i^B = 0, \tau_i^A = * \\
    & \\
    %%%%%
    P^i_{S}(0)
    \Big[
    \prod_{k \in \partial i \backslash j} \theta_{A,B,AB}^{k \to i}(\tau_i^A - 1, \tau_i^A, t; \tau_i^A)
    - \prod_{k \in \partial i \backslash j} \theta_{A,B,AB}^{k \to i}(\tau_i^A, \tau_i^A, t; \tau_i^A)
    \Big], & \text{if}\ \tau_i^A > 0, \tau_i^B = * \\
    & \\
    %%%%%
    P^i_{S}(0)
    \Big[
    \prod_{k \in \partial i \backslash j} \theta_{B,A,BA}^{k \to i}(\tau_i^B - 1, \tau_i^B, t; \tau_i^B)
    - \prod_{k \in \partial i \backslash j} \theta_{B,A,BA}^{k \to i}(\tau_i^B, \tau_i^B, t; \tau_i^B)
    \Big], & \text{if}\ \tau_i^B > 0, \tau_i^A = * \\
    & \\
    %%%%%
    P^i_{S}(0) \prod_{k \in \partial i \backslash j} \theta_{A,B}^{k \to i}(t, t), & \text{if}\ \tau_i^A = *, \tau_i^B = *
    \label{eq:message_on_trajectories_update}
    \end{cases}
\end{align}
}

\new{
The dynamic messages entering the expressions above are updated as follows.
\begin{align}
    &\theta_{A,B}^{k \rightarrow i}(t+1, t+1) = \theta_{A,B}^{k \rightarrow i}(t,t) - \alpha_{ki}^A \phi_{A}^{k \rightarrow i}(t, t) - \alpha_{ki}^{B} \phi_B^{k \rightarrow i}(t, t) + \alpha_{ki}^A \alpha_{ki}^B \phi_{A,B}^{k \rightarrow i}(t,t)
    \label{eq:theta_A_B_equal_times_update}
    \\
    &\theta^{k \to i}_{A,B}(t+1,t) = \theta^{k \to i}_{A,B}(t,t) - \alpha_{ki}^A\phi_A^{k \rightarrow i}(t,t),
    \\
    &\theta^{k \to i}_{A,B}(t,t+1) = \theta^{k \to i}_{A,B}(t,t) - \alpha_{ki}^B\phi_B^{k \rightarrow i}(t,t),
    \label{eq:theta_AB_nonequal_times_update}
\end{align}
\begin{align}
    \phi_{A,B}^{k \rightarrow i}(t,t) = &(1-\alpha_{ki}^A)(1-\alpha_{ki}^B)\phi_{A,B}^{k \rightarrow i}(t-1,t-1) - m_{t,t}^{k \rightarrow i}(t,t) \nonumber \\ 
    & + \sum_{\tau_{k}^A \leq t}\prod_{t'=0}^{t-1}(1-\alpha_{ki}^A\mathbbm{1}[t' \geq \tau_k^A])m_{t,t}^{k \rightarrow i}(\tau_k^A,t) +  \sum_{\tau_{k}^B \leq t}\prod_{t'=0}^{t-1}(1-\alpha_{ki}^B\mathbbm{1}[t' \geq \tau_k^B])m_{t,t}^{k \rightarrow i}(t,\tau_k^B).
\end{align}
\begin{align}
    \phi_A^{k \rightarrow i}(t-1,t) = \phi_A^{k \rightarrow i}(t-1,t-1) + (1-\alpha_{ki}^B) \phi_{A,B}^{k \rightarrow i}(t-1,t-1),\\
    \phi_B^{k \rightarrow i}(t,t-1) = \phi_B^{k \rightarrow i}(t-1,t-1) + (1-\alpha_{ki}^A) \phi_{A,B}^{k \rightarrow i}(t-1,t-1),
\end{align}
\begin{equation}
        \phi_A^{k \rightarrow i}(t,t)=(1-\alpha_{ki}^A)\phi_A^{k \rightarrow i}(t-1,t-1) - \alpha_{ki}^B(1-\alpha_{ki}^A)\phi_{A,B}^{k \rightarrow i}(t-1,t-1) + \theta_B^{k \rightarrow i}(t,t),
\end{equation}
\begin{equation}
        \phi_B^{k \rightarrow i}(t,t)=(1-\alpha_{ki}^B)\phi_B^{k \rightarrow i}(t-1,t-1) - \alpha_{ki}^A(1-\alpha_{ki}^B)\phi_{A,B}^{k \rightarrow i}(t-1,t-1) + \theta_A^{k \rightarrow i}(t,t).
\end{equation}
\begin{align}
    \theta^{k \to i}_A(t,t) = \theta^{k \to i}_A(t-1,t) - \alpha_{ki}^A \sum_{\tau_k^A \leq t-1}\prod_{t'=0}^{t-2}(1-\alpha_{ki}^A\mathbbm{1}[t' \geq \tau_k^A])m_{t,t}^{k \rightarrow i}(\tau_k^A,t),\\
    \theta^{k \to i}_B(t,t) = \theta^{k \to i}_B(t,t-1) -\alpha_{ki}^B \sum_{\tau_k^B \leq t-1}\prod_{t'=0}^{t-2}(1-\alpha_{ki}^B\mathbbm{1}[t' \geq \tau_k^B])m_{t,t}^{k \rightarrow i}(t,\tau_k^B),
\end{align}
\begin{equation}
    \theta_{AB}^{k \to i}(t + 1) = \theta_{AB}^{k \to i}(t) - \alpha_{ki}^{BA} \phi_{AB}^{k \to i}(t),
\end{equation}
\begin{equation}
    \theta_{BA}^{k \to i}(t + 1) = \theta_{BA}^{k \to i}(t) - \alpha_{ki}^{AB} \phi_{BA}^{k \to i}(t),
\end{equation}
\begin{equation}
    \phi_{AB}^{k \to i}(t) = (1 - \alpha_{ki}^{BA}) \phi_{AB}^{k \to i}(t-1) + \mu^{k \to i}_B(t).
\end{equation}
\begin{equation}
    \phi_{BA}^{k \to i}(t) = (1 - \alpha_{ki}^{AB}) \phi_{BA}^{k \to i}(t-1) + \mu^{k \to i}_A(t).
\end{equation}
\begin{equation}
    \theta_{A,B,AB}^{k \to i}(t_A, t_B, t_{AB} + 1; \tau_i^A) = \theta_{A,B,AB}^{k \to i}(t_A, t_B, t_{AB}; \tau_i^A) - \mathbbm{1}[\tau_i^A \leq t_{AB}]  \alpha_{ki}^{BA} \phi_{A,B,AB}^{k \to i}(t_A, t_B, t_{AB}; \tau_i^A),
    \label{eq:theta_A_B_AB_update}
\end{equation}
\begin{equation}
    \theta_{B,A,BA}^{k \to i}(t_B, t_A, t_{BA} + 1; \tau_i^B) = \theta_{B,A,BA}^{k \to i}(t_B, t_A, t_{BA}; \tau_i^B) - \mathbbm{1}[\tau_i^B \leq t_{BA}]  \alpha_{ki}^{AB} \phi_{B,A,BA}^{k \to i}(t_B, t_A, t_{BA}; \tau_i^B),
    \label{eq:theta_B_A_AB_update}
\end{equation}
\begin{align}
    \phi_{A,B,AB}^{k \to i}(t_A, t_B, t_{AB}; \tau_i^A)
    & = (1 - \alpha_{ki}^{BA}) \phi_{A,B,AB}^{k \to i}(t_A, t_B, t_{AB} -1; \tau_i^A \nonumber \\ 
    & + \sum_{\tau_k^A \leq t_{AB}} \prod_{t'=0}^{t_A-1} (1 - \alpha_{ki}^{A} \mathbbm{1}[\tau_k^A \leq t']) m_{t_{AB},t_{AB}}^{k \to i}(\tau_k^A, t_{AB}) + m_{t_{AB},t_{AB}}^{k \to i}(*, t_{AB}),
\end{align}
\begin{align}
    \phi_{B,A,BA}^{k \to i}(t_B, t_A, t_{BA}; \tau_i^B)
    & = (1 - \alpha_{ki}^{AB}) \phi_{B,A,BA}^{k \to i}(t_B, t_A, t_{BA} - 1; \tau_i^B) \nonumber \\ 
    & + \sum_{\tau_k^B \leq t_{BA}} \prod_{t'=0}^{t_B-1} (1 - \alpha_{ki}^{B} \mathbbm{1}[\tau_k^B \leq t']) m_{t_{BA},t_{BA}}^{k \to i}(t_{BA},\tau_k^B) + m_{t_{BA},t_{BA}}^{k \to i}(t_{BA},*).
\end{align}
}

\new{
\underline{Order of the updates:}
\begin{enumerate}
    \item Only once, at time zero: initialize messages
    \item If initialization is not given for a message, it means that its initial value is fixed through a border condition (reduction to a different message), and will need to be initialized in this way for each update
    \item Update $\phi$'s
    \item Update $\theta$'s
    \item Update $m$ messages and marginals
    \item Update $\mu$ messages and marginals
    \item Compute dynamic marginals $P_S^i(t), P_A^i(t), P_B^i(t), P_{AB}^i(t)$.
\end{enumerate}
}

\section{Approximate DMP Equations for Collaborative Processes}
\label{app:collaborative}

\new{
From the dynamic rules for the interacting spreading processes \eqref{eq:Dynamical_Rules0}-\eqref{eq:Dynamical_Rules}, we notice that the special case of transmission probabilities $\alpha_{ij}^{AB} = \alpha_{ij}^{A}$ and $\alpha_{ij}^{BA} = \alpha_{ij}^{B}$ corresponds to non-interacting spreading processes: activation by one process does not change the activation dynamics for the other. In this case, the DMP equations should simplify into the product of two independent SI-like processes:
\begin{equation}
    m_{t,t}^i(\tau_i^A, \tau_i^B) = \mu_A^i(\tau_i^A) \mu_B^i(\tau_i^B).
\end{equation}
We use this observation to produce a simplified version of DMP equations, expanding exact equations for interacting spreading processes around the non-interacting point. Several expansions are possible, including the most straightforward one where we expand to a certain order in $\alpha_{ij}^{A} - \alpha_{ij}^{AB}$ and $\alpha_{ij}^{B} - \alpha_{ij}^{BA}$, which will result in bulky expressions. Instead, here we produce a simplified set of DMP equations by keeping certain first-order corrections in the update equations only, so that the resulting equations are similar to the SI-type equations. Using the approximation
\begin{equation}
     m_{t,t}^i(\tau_i^A, \tau_i^B) \approx \mu_A^i(\tau_i^A) \mu_B^i(\tau_i^B)
\end{equation}
and the general expression
\begin{equation}
    \mu^{i}_A(t)
    = \widehat{P}^{i \to j}_A(t) - \widehat{P}^{i \to j}_A(t-1)
    = \sum_{t' \leq t} m_{t,t}^i(t,t') + m_{t,t}^i(t,*)
    = \sum_{t' \leq t-1} m_{t,t-1}^{i}(t,t') + m_{t,t-1}^{i}(t,*),
\end{equation}
we compute the corresponding contribution of each term. For simplicity, in what follows we assume that $\alpha_{ij} < 1$ and hence $\theta^{k \to i}$ type messages are non-zero (treatment of the case of deterministic spreading is straightforward and results in several additional equations).
}

\new{
\underline{Simplified approximate computation of $m_{t,t-1}^{i}(t,*)$:}
We start with the following approximations in the dynamic messages:
\begin{align}
    \theta_{A,B}^{k \rightarrow i}(t_A,t_B) = \sum_{\tau_k^A} \sum_{\tau_k^B} \prod_{t'=0}^{t_A-1} (1-\alpha_{ki}^A\mathbbm{1}[\tau_k^A \leq t']) \prod_{t''=0}^{t_B-1} (1-\alpha_{ki}^B\mathbbm{1}[\tau_k^B \leq t'']) m^{k \rightarrow i}_{T_A,T_B}(\tau_k^A,\tau_k^B) \approx \widehat{\theta}_A^{k \rightarrow i}(t_A) \widehat{\theta}_B^{k \rightarrow i}(t),
\end{align}
\begin{align}
    &\phi_{A}^{k \rightarrow i}(t_A,t_B) \nonumber
    \\
    & = \sum_{\tau_k^A} \sum_{\tau_k^B} \mathbbm{1}[\tau_k^A \leq t_A] \prod_{t'=0}^{t_A-1} (1-\alpha_{ki}^A\mathbbm{1}[\tau_k^A \leq t']) \prod_{t''=0}^{t_B-1} (1-\alpha_{ki}^B\mathbbm{1}[\tau_k^B \leq t'']) m^{k \rightarrow i}_{T_A,T_B}(\tau_k^A,\tau_k^B)
    \approx \widehat{\phi}_A^{k \rightarrow i}(t_A) \widehat{\theta}_B^{k \rightarrow i}(t).
\end{align}
The update equation
\begin{equation}
    \theta^{k \to i}_{A,B}(t+1,t) = \theta^{k \to i}_{A,B}(t,t) - \alpha_{ki}^A\phi_A^{k \rightarrow i}(t,t)
\end{equation}
becomes under this approximation
\begin{equation}
    \widehat{\theta}_A^{k \rightarrow i}(t+1) \widehat{\theta}_B^{k \rightarrow i}(t) = \widehat{\theta}_A^{k \rightarrow i}(t) \widehat{\theta}_B^{k \rightarrow i}(t) - \alpha_{ki}^A\widehat{\phi}_A^{k \rightarrow i}(t) \widehat{\theta}_B^{k \rightarrow i}(t),
\end{equation}
or simply
\begin{equation}
    \widehat{\theta}_A^{k \rightarrow i}(t+1) = \widehat{\theta}_A^{k \rightarrow i}(t)  - \alpha_{ki}^A\widehat{\phi}_A^{k \rightarrow i}(t).
\end{equation}
Using these simplified expressions, we get after some algebra
\begin{align}
    m_{t,t-1}^i(t,*) = P^i_{S}(0)
    \Big[
    \prod_{k \in \partial i} \theta_{A,B}^{k \to i}(t - 1, t-1)
    - \prod_{k \in \partial i} \theta_{A,B}^{k \to i}(t, t-1)
    \Big]
    \\
    \approx P^i_{S}(0)
    \Big[
    \prod_{k \in \partial i} \widehat{\theta}_A^{k \rightarrow i}(t - 1) \widehat{\theta}_B^{k \rightarrow i}(t-1)
    - \prod_{k \in \partial i} \widehat{\theta}_A^{k \rightarrow i}(t) \widehat{\theta}_B^{k \rightarrow i}(t-1)
    \Big]
    \\
    \approx P^i_{S}(0)
    \Big[
    \prod_{k \in \partial i} \widehat{\theta}_A^{k \rightarrow i}(t - 1) \widehat{\theta}_B^{k \rightarrow i}(t-1)
    - \prod_{k \in \partial i} \widehat{\theta}_A^{k \rightarrow i}(t) \widehat{\theta}_B^{k \rightarrow i}(t-1)
    \Big]
    \\
    \approx P^i_{S}(0)
    \prod_{k \in \partial i} \widehat{\theta}_A^{k \rightarrow i}(t - 1) \widehat{\theta}_B^{k \rightarrow i}(t-1)
    \left[
    1 - \prod_{k \in \partial i} \frac{\widehat{\theta}_A^{k \rightarrow i}(t)}{\widehat{\theta}_A^{k \rightarrow i}(t - 1)}  
    \right]
    \\
    \approx P^i_{S}(t-1)
    \left[
    1 - \prod_{k \in \partial i} \left( 1 - \frac{\alpha_{ij}^{A} \widehat{\phi}^{i \to j}_A(t-1)}{\widehat{\theta}^{k \to i}_A(t-1)} \right) 
    \right].
\end{align}
}

\new{
\underline{Simplified approximate computation of $\sum_{t' \leq t-1} m_{t,t-1}^{i}(t,t')$:}
Similarly, we start with the approximations in the dynamic messages:
\begin{align}
        \theta_{B,A,BA}^{k \rightarrow i}(t_B,t',t_{BA}; t') =& \sum_{\tau_k^A} \sum_{\tau_k^B} \prod_{t=0}^{t_B-1} (1-\alpha_{ki}^B\mathbbm{1}[\tau_k^B \leq t]) \prod_{t''=0}^{t'-1} (1-\alpha_{ki}^A\mathbbm{1}[\tau_k^A \leq t'']) \nonumber \\ 
        & \times \prod_{t'''=t'}^{t_{BA}-1} (1-\alpha_{ki}^{AB}\mathbbm{1}[\tau_k^{A} \leq t''']) m^{k \rightarrow i}_{T_A,T_B}(\tau_k^A,\tau_k^B)
        \\
        & \approx \widehat{\theta}_A^{k \rightarrow i}(t_{BA}) \widehat{\theta}_B^{k \rightarrow i}(t_B),
        \\
%\end{align}
%\begin{align}
        \phi_{B,A,BA}^{k \rightarrow i}(t_B,t',t_{BA}; t') =& \sum_{\tau_k^A} \sum_{\tau_k^B} \mathbbm{1}[\tau_k^{A} \leq t_{BA}] \prod_{t=0}^{t_B-1}  (1-\alpha_{ki}^B\mathbbm{1}[\tau_k^B \leq t]) \prod_{t''=0}^{t'-1} (1-\alpha_{ki}^A\mathbbm{1}[\tau_k^A \leq t'']) \nonumber \\ 
        & \times \prod_{t'''=t'}^{t_{BA}-1} (1-\alpha_{ki}^{AB}\mathbbm{1}[\tau_k^{A} \leq t''']) m^{k \rightarrow i}_{T_A,T_B}(\tau_k^A,\tau_k^B)
        \\
        &\approx \widehat{\phi}_A^{k \rightarrow i}(t_{BA}) \widehat{\theta}_B^{k \rightarrow i}(t_B)&,
        \\
%\end{align}
%\begin{align}
    \theta^{k \to i}_{BA}(t_A) =& \sum_{\tau_k^A} \prod_{t'=0}^{t_{A}-1} (1-\alpha_{ki}^{AB}\mathbbm{1}[\tau_k^{A} \leq t']) \mu_A^{k \rightarrow i}(\tau_k^A) \approx \widehat{\theta}_A^{k \rightarrow i}(t_{A}).
\end{align}
The approximate version of the update equations
\begin{equation}
    \theta_{B,A,BA}^{k \rightarrow i}(t_B,t',t_{BA} + 1; t') = \theta_{B,A,BA}^{k \rightarrow i}(t_B,t',t_{BA}; t') - \mathbbm{1}[t' \leq t_{BA}]  \alpha_{ki}^{AB} \phi_{B,A,BA}^{k \rightarrow i}(t_B,t',t_{BA}; t')
\end{equation}
then reads
\begin{equation}
    \widehat{\theta}_A^{k \rightarrow i}(t_{BA}+1) \widehat{\theta}_B^{k \rightarrow i}(t_B) = \widehat{\theta}_A^{k \rightarrow i}(t_{BA}) \widehat{\theta}_B^{k \rightarrow i}(t_B) - \mathbbm{1}[t' \leq t_{BA}]  \alpha_{ki}^{AB} \widehat{\phi}_A^{k \rightarrow i}(t_{BA}) \widehat{\theta}_B^{k \rightarrow i}(t_B),
\end{equation}
or simply
\begin{equation}
    \widehat{\theta}_A^{k \rightarrow i}(t_{BA}+1) = \widehat{\theta}_A^{k \rightarrow i}(t_{BA}) - \mathbbm{1}[t' \leq t_{BA}]  \alpha_{ki}^{AB} \widehat{\phi}_A^{k \rightarrow i}(t_{BA}).
\end{equation}
Using these simplified expressions, we get after some algebra
\begin{align}
    \sum_{t' \leq t-1} m_{t,t-1}^{i}(t,t') =
    \sum_{0 < t' \leq t-1}
    P^i_{S}(0)
    \Big[
    \prod_{k \in \partial i} \theta_{B,A,BA}^{k \to i}(t' - 1, t', t - 1; t')
    - \prod_{k \in \partial i} \theta_{B,A,BA}^{k \to i}(t' - 1, t', t; t') \nonumber
    \\ 
    - \prod_{k \in \partial i} \theta_{B,A,BA}^{k \to i}(t', t', t - 1; t')
    + \prod_{k \in \partial i} \theta_{B,A,BA}^{k \to i}(t', t', t; t')
    \Big] \nonumber
    \\ 
    +
    P^i_{B^*}(0) \left[ \prod_{k \in \partial i} \theta_{BA}^{k \to i}(t-1) - \prod_{k \in \partial i} \theta_{BA}^{k \to i}(t) \right]
    \\
    = \sum_{0 < t' \leq t-1}
    P^i_{S}(0) \Big[
    \prod_{k \in \partial i} \widehat{\theta}_A^{k \rightarrow i}(t-1) \widehat{\theta}_B^{k \rightarrow i}(t'-1) - \prod_{k \in \partial i} \widehat{\theta}_A^{k \rightarrow i}(t) \widehat{\theta}_B^{k \rightarrow i}(t'-1) - \prod_{k \in \partial i} \widehat{\theta}_A^{k \rightarrow i}(t-1) \widehat{\theta}_B^{k \rightarrow i}(t')\nonumber
    \\ 
    + \prod_{k \in \partial i} \widehat{\theta}_A^{k \rightarrow i}(t) \widehat{\theta}_B^{k \rightarrow i}(t')
    + P^i_{B^*}(0) \left[ \prod_{k \in \partial i} \widehat{\theta}_A^{k \rightarrow i}(t-1) - \prod_{k \in \partial i} \widehat{\theta}_A^{k \rightarrow i}(t) \right]
    \Big]
    \\
    = \sum_{0 < t' \leq t-1}
    P^i_{S}(0)
    \left(\prod_{k \in \partial i} \widehat{\theta}_B^{k \rightarrow i}(t'-1) - \prod_{k \in \partial i} \widehat{\theta}_B^{k \rightarrow i}(t') \right) \left( \prod_{k \in \partial i} \widehat{\theta}_A^{k \rightarrow i}(t-1) - \prod_{k \in \partial i} \widehat{\theta}_A^{k \rightarrow i}(t) \right) \nonumber
    \\ 
    + P^i_{B^*}(0) \left[ \prod_{k \in \partial i} \widehat{\theta}_A^{k \rightarrow i}(t-1) - \prod_{k \in \partial i} \widehat{\theta}_A^{k \rightarrow i}(t) \right]
    \\
    = \left[ \prod_{k \in \partial i} \widehat{\theta}_A^{k \rightarrow i}(t-1) - \prod_{k \in \partial i} \widehat{\theta}_A^{k \rightarrow i}(t) \right] \left[ \sum_{0 < t' \leq t-1} P^i_{S}(0) \left( \prod_{k \in \partial i} \widehat{\theta}_B^{k \rightarrow i}(t'-1) - \prod_{k \in \partial i} \widehat{\theta}_B^{k \rightarrow i}(t') \right) + P^i_{B^*}(0) \right]
    \\
    = \left[ \prod_{k \in \partial i} \widehat{\theta}_A^{k \rightarrow i}(t-1) - \prod_{k \in \partial i} \widehat{\theta}_A^{k \rightarrow i}(t) \right] \left[ P^i_{S}(0) \left( \prod_{k \in \partial i} \widehat{\theta}_B^{k \rightarrow i}(0) - \prod_{k \in \partial i} \widehat{\theta}_B^{k \rightarrow i}(t-1) \right) + P^i_{B^*}(0) \right]
    \\
    \approx P^i_{B^*}(t-1) \left[ \prod_{k \in \partial i} \widehat{\theta}_A^{k \rightarrow i}(t-1) - \prod_{k \in \partial i} \widehat{\theta}_A^{k \rightarrow i}(t) \right].
\end{align}
}

\new{
\underline{Final form of the approximate DMP equations:}
The final update equations that we will need,
\begin{align}
    \phi_A^{k \rightarrow i}(t,t)=(1-\alpha_{ki}^A)\phi_A^{k \rightarrow i}(t-1,t) + \theta_B^{k \rightarrow i}(t,t),
\end{align}
with
\begin{align}
    \theta_B^{k \rightarrow i}(t,t) \approx \mu_A^{k \rightarrow i}(t)\widehat{\theta}_B^{k \rightarrow i}(t),
\end{align}
gives
\begin{equation}
    \widehat{\phi}^{i \to j}_A(t) = \widehat{\phi}^{i \to j}_A(t-1) - \alpha_{ij}^{A} \widehat{\phi}^{i \to j}_A(t-1) + \widehat{P}^{i \to j}_A(t) - \widehat{P}^{i \to j}_A(t-1).
\end{equation}
}

\new{
\noindent Combining all of the computation above, we finally obtain the following approximate form of the DMP equations that should provide a good approximation around the non-interacting point:
}
\begin{align}%\begin{equation}
    \widehat{P}_S^i(t) = P_S^i(0)\prod_{k \in \partial i \backslash j}\widehat{\theta}_{A}^{k \rightarrow i}(t)\widehat{\theta}_{B}^{k \rightarrow i}(t),
    \label{eq:Collaborative_0}
%\end{equation}
\\
%\begin{equation}
    \widehat{\theta}^{i \to j}_A(t) = \widehat{\theta}^{i \to j}_A(t-1) - \alpha_{ij}^{A} \widehat{\phi}^{i \to j}_A(t-1),
%\end{equation}
\\
%\begin{equation}
    \widehat{\theta}^{i \to j}_B(t) = \widehat{\theta}^{i \to j}_B(t-1) - \alpha_{ij}^{B} \widehat{\phi}^{i \to j}_B(t-1),
%\end{equation}
\\
%\begin{equation}
    \widehat{\phi}^{i \to j}_A(t) = \widehat{\phi}^{i \to j}_A(t-1) - \alpha_{ij}^{A} \widehat{\phi}^{i \to j}_A(t-1) + \widehat{P}^{i \to j}_A(t) - \widehat{P}^{i \to j}_A(t-1),
%\end{equation}
\\
%\begin{equation}
    \widehat{\phi}^{i \to j}_B(t) = \widehat{\phi}^{i \to j}_B(t-1) - \alpha_{ij}^{B} \widehat{\phi}^{i \to j}_B(t-1) + \widehat{P}^{i \to j}_B(t) - \widehat{P}^{i \to j}_B(t-1),
%\end{equation}
\end{align}
\begin{align}
    \widehat{P}^{i \to j}_A(t) = \widehat{P}^{i \to j}_A(t-1)
    & + \widehat{P}_S^i(t-1) \left[1 - \prod_{k \in \partial i \backslash j} \left( 1 - \frac{\alpha_{ij}^{A} \widehat{\phi}^{i \to j}_A(t-1)}{\widehat{\theta}^{k \to i}_A(t-1)} \right) \right] \nonumber
    \\
    & + \widehat{P}^{i \to j}_{B*}(t-1) \left[1 - \prod_{k \in \partial i \backslash j} \left( 1 - \frac{\alpha_{ij}^{AB} \widehat{\phi}^{i \to j}_A(t-1)}{\widehat{\theta}^{k \to i}_A(t-1)} \right) \right],
\end{align}
\begin{align}
    \widehat{P}^{i \to j}_B(t) = \widehat{P}^{i \to j}_B(t-1)
    & + \widehat{P}_S^i(t-1) \left[1 - \prod_{k \in \partial i \backslash j} \left( 1 - \frac{\alpha_{ij}^{B} \widehat{\phi}^{i \to j}_B(t-1)}{\widehat{\theta}^{k \to i}_B(t-1)} \right) \right]  \nonumber
    \\
    & + \widehat{P}^{i \to j}_{A*}(t-1) \left[1 - \prod_{k \in \partial i \backslash j} \left( 1 - \frac{\alpha_{ij}^{BA} \widehat{\phi}^{i \to j}_B(t-1)}{\widehat{\theta}^{k \to i}_B(t-1)} \right) \right]. 
    \label{eq:Collaborative}
\end{align}
\new{
In the Main Text, we numerically establish the approximation power of these equations. In what follows, we use these approximate DMP equations for performing optimization due to their simpler form and lower computational complexity.
}

\section{Optimization of Mutually Exclusive Competitive Processes}
\label{app:OptCompetitive}
For competitive processes we study two optimization problems, the multi-agent seeding problem and disease containment, the only difference between the two is the objective function. For disease containment the objective function to be maximized is 
$
	\mathcal{O} \!=\! \sum_i\left(1-P_i^B(T)\right)
$
and for the multi-agent seeding problem 
$
	\mathcal{O} \!=\! \sum_i(1-P_i^S(T))~.
$
The budget constraint at time zero is enforced by the Lagrange multiplier $\lambda^{Bu}$ is:
\begin{equation}
	\mathcal{B} \!=\! \lambda^{Bu}\left(B_{\nu}-\sum_i{\nu^i(0)}\right)
\end{equation}
where permitted $\underline{\nu} <\nu^i < \overline{\nu} $ value restrictions are enforced by the term 
\begin{equation}
	\mathcal{P} \!=\! \epsilon \sum_i\left(\log(\overline{\nu} - \nu^i(0)) + \log(\nu^i(0)-\underline{\nu})\right)~.\label{Restriction}
\end{equation}
In this case, the restrictions used are $\overline{\nu}\!=\!1$ and $\underline{\nu}\!=\!0$.

Initial conditions are forced through a set of Lagrange multipliers $\lambda$ for the various parameter values
\begin{equation}
\label{eq:Icondition}
	\begin{split}
	\mathcal{I }\!=\! & \sum_i \lambda_i^{\widehat{A}}(0)\left(\widehat{P}_{A}^i(0)-\nu^i(0)\left(1-\delta_{\sigma_i(0), B}\right)\right)+\sum_i \lambda_i^{{A}}(0)\left({P_A}^i(0)-\nu^i(0)\left(1-\delta_{\sigma_i(0), B}\right)\right)\\
	     & +\sum_i \lambda_i^{\widehat{B}}(0)\left(\widehat{P}_{B}^i(0)-\delta_{\sigma_i(0),B}\right) + \sum_i \lambda_i^{{B}}(0)\left({P_B}^i(0)-\delta_{\sigma_i(0),B}\right)\\
	    & + \sum_i\lambda_{i}^{\widehat{S}}(0)\left(\widehat{P}_S^i(0)-1+v^i(0)\left(1-\delta_{\sigma_i(0), B}\right)+\delta_{\sigma_i(0), B}\right) \\
     	& + \sum_i\lambda_{i}^{S}(0)\left({P}_S^i(0)-1+v^i(0)\left(1-\delta_{\sigma_i(0), B}\right)+\delta_{\sigma_i(0), B}\right) + \sum_{ij}\lambda_{ij}^{\theta_B}(0)\left(\theta_B^{i \rightarrow j}(0)-1\right) \\
     	& + \sum_{ij}\lambda_{ij}^{\theta_B}(0)\left(\theta_B^{i \rightarrow j}(0)-1\right) + \sum_{ij}\lambda_{ij}^{\phi_A}(0)\left(\phi_A^{i \rightarrow j}(0)-\nu^i(0)\left(1-\delta_{\sigma_i(0), B}\right)\right)\\
     	& + \sum_{ij}\lambda_{ij}^{\phi_B}(0)\left(\phi_B^{i \rightarrow j}(0)-\delta_{\sigma_i(0), B}\right)+ \sum_{ij}\lambda_{ij}^{\widehat{A}}(0)\left(\widehat{P}_{A}^{i \rightarrow j}(0)-\nu^i(0)\left(1-\delta_{\sigma_i(0), B}\right)\right)\\
     	& + \sum_{ij}\lambda_{ij}^{\widehat{B}}(0)\left(\widehat{P}_{B}^{i \rightarrow j}(0)-\delta_{\sigma_i(0), B}\right) +\sum_{ij}\lambda_{ij}^{{A}}(0)\left({P_A}^{i \rightarrow j}(0)-\nu^i(0)\left(1-\delta_{\sigma_i(0), B}\right)\right)\\
     	& + \sum_{ij}\lambda_{ij}^{{B}}(0)\left({P_B}^{i \rightarrow j}(0)-\delta_{\sigma_i(0), B}\right) + \sum_{ij}\lambda_{ij}^{\widehat{S}}(0)\left(\widehat{P}_S^{i \rightarrow j}(0)-1+\delta_{\sigma_i(0), B}+\nu^i(0)\left(1-\delta_{\sigma_i(0), B}\right)\right)\\
     	& + \sum_{ij}\lambda_{ij}^{S}(0)\left({P}_S^{i \rightarrow j}(0)-1+\delta_{\sigma_i(0), B}+\nu^i(0)\left(1-\delta_{\sigma_i(0), B}\right)\right)
	\end{split}
\end{equation}
	
The DMP equations for the dynamics are forced through a set of Lagrange multipliers 
{\small
\begin{equation}
\label{eq:Dcompetitive}
	\begin{split}
	 \mathcal{D} \!=\! &\sum_{ij}\sum_{t\!=\!0}^{T-1}\lambda_{ij}^{\theta_A} (t\!+\!1)[\theta_A^{i \rightarrow j} (t\!+\!1) \!-\! \theta_A^{i \rightarrow j}(t) + \alpha_A \phi_A^{i \rightarrow j}(t)]\\
		 & \!+\! \sum_{ij}\sum_{t\!=\!0}^{T-1}\lambda_{ij}^{\theta_B} (t\!+\!1)[\theta_B^{i \rightarrow j} (t\!+\!1) \!-\! \theta_B^{i \rightarrow j}(t) \!+\! \alpha_B \phi_B^{i \rightarrow j}(t)]\\
		 & \!+\! \sum_{ij}\sum_{t\!=\!0}^{T-1}\lambda_{ij}^{\phi_A} (t\!+\!1)[\phi_A^{i \rightarrow j} (t\!+\!1) \!+\! (\alpha_A\!-\!1)\phi_A^{i \rightarrow j}(t)\!-\![P_A^{i \rightarrow j} (t\!+\!1)\!-\!P_A^{i \rightarrow j}(t)]]\\
		 & \!+\! \sum_{ij}\sum_{t\!=\!0}^{T-1}\lambda_{ij}^{\phi_B} (t\!+\!1)[\phi_B^{i \rightarrow j} (t\!+\!1) \!+\! (\alpha_B\!-\!1)\phi_B^{i \rightarrow j}(t)\!-\![P_B^{i \rightarrow j} (t\!+\!1)\!-\!P_B^{i \rightarrow j}(t)]]\\
		& \!+\! \sum_{ij}\sum_{t\!=\!0}^{T-1}\lambda_{ij}^{\widehat{S}} (t\!+\!1)\left[\widehat{P}_{S}^{i \rightarrow j} (t\!+\!1) \!-\! P_S^{i \rightarrow j}(t)\prod_{k \in \partial i \backslash j}\frac{\theta_A^{k \rightarrow i} (t\!+\!1)\theta_B^{k \rightarrow i} (t\!+\!1)}{\theta_A^{k \rightarrow i}(t)\theta_B^{k \rightarrow i}(t)}\right]\\
		& \!+\! \sum_{ij}\sum_{t\!=\!0}^{T-1}\lambda_{ij}^{\widehat{A}} (t\!+\!1)\left[\widehat{P}_{A}^{i \rightarrow j} (t\!+\!1)\!-\!{P_A}^{i \rightarrow j}(t)\!-\!{P_S}^{i \rightarrow j}(t)\prod_{k \in \partial i \backslash j}\left(1\!-\!\frac{\alpha_B \phi_B^{k \rightarrow i}(t)}{\theta_B^{k \rightarrow i}(t)}\right)\left(1\!-\!\prod_{k \in \partial i \backslash j}\left(1\!-\!\frac{\alpha_A \phi_A^{k \rightarrow i}(t)}{\theta_A^{k \rightarrow i}(t)}\right)\right)\right]\\
		& \!+\! \sum_{ij}\sum_{t\!=\!0}^{T-1}\lambda_{ij}^{\widehat{B}} (t\!+\!1)\left[\widehat{P}_{B}^{i \rightarrow j} (t\!+\!1)\!-\!P_B^{i \rightarrow j}(t)\!-\!{P_S}^{i \rightarrow j}(t)\prod_{k \in \partial i \backslash j}\left(1\!-\!\frac{\alpha_A \phi_A^{k \rightarrow i}(t)}{\theta_A^{k \rightarrow i}(t)}\right)\left(1\!-\!\prod_{k \in \partial i \backslash j}\left(1\!-\!\frac{\alpha_B \phi_B^{k \rightarrow i}(t)}{\theta_{B}^{k \rightarrow i}(t)}\right)\right)\right]\\
		& \!+\! \sum_{ij}\sum_{t\!=\!0}^{T-1}\lambda_{ij}^{S} (t\!+\!1)\left[P_S^{i \rightarrow j} (t\!+\!1) \!-\! \frac{\widehat{P}_{S}^{i \rightarrow j} (t\!+\!1)}{\widehat{P}_{S}^{i \rightarrow j} (t\!+\!1) \!+\!\widehat{P}_{A}^{i \rightarrow j} (t\!+\!1) \!+\! \widehat{P}_{B}^{i \rightarrow j} (t\!+\!1)}\right]\\
		& \!+\! \sum_{ij}\sum_{t\!=\!0}^{T-1}\lambda_{ij}^{A} (t\!+\!1)\left[P_{A}^{i \rightarrow j} (t\!+\!1) \!-\! \frac{\widehat{P}_{A}^{i \rightarrow j} (t\!\!+\!\!1)}{\widehat{P}_{S}^{i \rightarrow j} (t\!+\!1) \!+\!\widehat{P}_{A}^{i \rightarrow j} (t\!+\!1) \!+\! \widehat{P}_{B}^{i \rightarrow j} (t\!+\!1)}\right]\\
		& \!+\! \sum_{ij}\sum_{t\!=\!0}^{T-1}\lambda_{ij}^{B} (t\!+\!1)\left[P_B^{i \rightarrow j} (t\!+\!1) \!-\! \frac{\widehat{P}_{B}^{i \rightarrow j} (t\!+\!1)}{\widehat{P}_{S}^{i \rightarrow j} (t\!+\!1) \!+\!\widehat{P}_{A}^{i \rightarrow j} (t\!+\!1) \!+\! \widehat{P}_{B}^{i \rightarrow j} (t\!+\!1)}\right]\\
		& \!+\! \sum_{i}\sum_{t\!=\!0}^{T-1}\lambda_{i}^{\widehat{S}} (t\!+\!1)\left[\widehat{P}_{S}^{i} (t\!+\!1) \!-\! P_S^i(t)\prod_{k \in \partial i}\frac{\theta_A^{k \rightarrow i} (t\!+\!1)\theta_B^{k \rightarrow i} (t\!+\!1)}{\theta_A^{k \rightarrow i}(t)\theta_B^{k \rightarrow i}(t)}\right]\\
		& \!+\! \sum_{i}\sum_{t\!=\!0}^{T-1}\lambda_{i}^{\widehat{A}} (t\!+\!1)\left[\widehat{P}_{A}^{i} (t\!+\!1)\!-\!{P_A}^{i}(t)\!-\!{P_S}^{i}(t)\prod_{k \in \partial i}\left(1\!-\!\frac{\alpha_B \phi_B^{k \rightarrow i}(t)}{\theta_B^{k \rightarrow i}(t)}\right)\left(1\!-\!\prod_{k \in \partial i}\left(1\!-\!\frac{\alpha_A \phi_A^{k \rightarrow i}(t)}{\theta_A^{k \rightarrow i}(t)}\right)\right)\right]\\
		& \!+\! \sum_{i}\sum_{t\!=\!0}^{T-1}\lambda_{i}^{\widehat{A}} (t\!+\!1)\left[\widehat{P}_{A}^{i} (t\!+\!1)\!-\!{P_A}^{i}(t)\!-\!{P_S}^{i}(t)\prod_{k \in \partial i}\left(1\!-\!\frac{\alpha_B \phi_B^{k \rightarrow i}(t)}{\theta_B^{k \rightarrow i}(t)}\right)\left(1\!-\!\prod_{k \in \partial i}\left(1\!-\!\frac{\alpha_A \phi_A^{k \rightarrow i}(t)}{\theta_{A}^{k \rightarrow i}(t)}\right)\right)\right]\\
		& \!+\! \sum_{i}\sum_{t\!=\!0}^{T-1}\lambda_{i}^{\widehat{B}} (t\!+\!1)\left[\widehat{P}_{B}^{i} (t\!+\!1)\!-\!{P_B}^{i}(t)\!-\!{P_S}^{i}(t)\prod_{k \in \partial i}\left(1\!-\!\frac{\alpha_A \phi_A^{k \rightarrow i}(t)}{\theta_A^{k \rightarrow i}(t)}\right)\left(1\!-\!\prod_{k \in \partial i}\left(1\!-\!\frac{\alpha_B \phi_B^{k \rightarrow i}(t)}{\theta_{B}^{k \rightarrow i}(t)}\right)\right)\right]\\
		& \!+\! \sum_{i}\sum_{t\!=\!0}^{T-1}\lambda_{i}^{S} (t\!+\!1)\left[P_S^{i} (t\!+\!1) \!-\! \frac{\widehat{P}_{S}^{i} (t\!+\!1)}{\widehat{P}_{S}^{i} (t\!+\!1) \!+\!\widehat{P}_{A}^{i} (t\!+\!1) \!+\! \widehat{P}_{B}^{i} (t\!+\!1)}\right]\\
		& \!+\! \sum_{i}\sum_{t\!=\!0}^{T-1}\lambda_{i}^{A} (t\!+\!1)\left[P_A^{i} (t\!+\!1) \!-\! \frac{\widehat{P}_{A}^{i} (t\!+\!1)}{\widehat{P}_{S}^{i} (t\!+\!1) \!+\!\widehat{P}_{A}^{i} (t\!+\!1) \!+\! \widehat{P}_{B}^{i} (t\!+\!1)}\right]\\
		& \!+\! \sum_{i}\sum_{t\!=\!0}^{T-1}\lambda_{i}^{B} (t\!+\!1)\left[P_B^{i} (t\!+\!1) \!-\! \frac{\widehat{P}_{B}^{i} (t\!+\!1)}{\widehat{P}_{S}^{i} (t\!+\!1) \!+\!\widehat{P}_{A}^{i} (t\!+\!1) \!+\! \widehat{P}_{B}^{i} (t\!+\!1)}\right]
	\end{split}
\end{equation}
}
Derivatives with respect to the dynamics parameters give rise to the optimization (dynamical) equations for the Lagrange multipliers. The case described here is that of the containment problem (minimizing spread of an adversarial process).
{\small 
\begin{equation}
\label{eq:derivativesComp1}
	\begin{split}
		\frac{\partial \mathcal{L}}{\partial\widehat{P}_S^{i \rightarrow j}(t)} & \!=\! \lambda_{ij}^{\widehat{S}}(t) + 1\left[t \neq 0\right]\left[-\lambda_{ij}^S(t)\frac{\widehat{P}_A^{i \rightarrow j}(t)+\widehat{P}_B^{i \rightarrow j}(t)}{(\widehat{P}_S^{i \rightarrow j}(t)+\widehat{P}_A^{i \rightarrow j}(t)+\widehat{P}_B^{i \rightarrow j}(t))^2}+\lambda_{ij}^A(t)\frac{\widehat{P}_A^{i \rightarrow j}(t)}{(\widehat{P}_S^{i \rightarrow j}(t)+\widehat{P}_A^{i \rightarrow j}(t)+\widehat{P}_B^{i \rightarrow j}(t))^2} \right. \\
		& \left. +\lambda_{ij}^B(t)\frac{\widehat{P}_B^{i \rightarrow j}(t)}{(\widehat{P}_S^{i \rightarrow j}(t)+\widehat{P}_A^{i \rightarrow j}(t)+\widehat{P}_B^{i \rightarrow j}(t))^2}\right] 
 \\
		\frac{\partial \mathcal{L}}{\partial\widehat{P}_S^{i}(t)} & \!=\! \lambda_{i}^{\widehat{S}}(t)+1\left[t \neq 0\right]\left[-\lambda_{i}^S(t)\frac{\widehat{P}_A^{i}(t)+\widehat{P}_B^{i}(t)}{(\widehat{P}_S^{i}(t)+\widehat{P}_A^{i}(t)+\widehat{P}_B^{i}(t))^2}+\lambda_{i}^A(t)\frac{\widehat{P}_A^{i}(t)}{(\widehat{P}_S^{i}(t)+\widehat{P}_A^{i}(t)+\widehat{P}_B^{i}(t))^2} \right. \\
		& \left. +\lambda_{i}^B(t)\frac{\widehat{P}_B^{i}(t)}{(\widehat{P}_S^{i}(t)+\widehat{P}_A^{i}(t)+\widehat{P}_B^{i}(t))^2}\right]
\\
		\frac{\partial \mathcal{L}}{\partial P_S^{i \rightarrow j}(t)} 
		& \!=\! \left[-\lambda_{ij}^{\widehat{S}} (t\!+\!1)\prod_{k \in \partial i \backslash j}\frac{\theta_A^{k \rightarrow i} (t\!+\!1)\theta_B^{k \rightarrow i} (t\!+\!1)}{\theta_A^{k \rightarrow i}(t)\theta_B^{k \rightarrow i}(t)} \right. 
		 -\lambda_{ij}^{\widehat{A}} (t\!+\!1)\prod_{k \in \partial i \backslash j}\left(1-\frac{\alpha_B \phi_B^{k \rightarrow i}(t)}{\theta_B^{k \rightarrow i}(t)}\right)\left(1-\prod_{k \in \partial i \backslash j}\left(1-\frac{\alpha_A \phi_A^{k \rightarrow i}(t)}{\theta_A^{k \rightarrow i}(t)}\right)\right)\\
		& \left. - \lambda_{ij}^{\widehat{B}} (t\!+\!1)\prod_{k \in \partial i \backslash j}\left(1-\frac{\alpha_A \phi_A^{k \rightarrow i}(t)}{\theta_A^{k \rightarrow i}(t)}\right)\left(1-\prod_{k \in \partial i \backslash j}\left(1-\frac{\alpha_B \phi_B^{k \rightarrow i}(t)}{\theta_B^{k \rightarrow i}(t)}\right)\right)\right]1\left[t \neq T\right] + \lambda_{ij}^S(t)
\\
		\frac{\partial \mathcal{L}}{\partial P_S^{i}(t)} 
		& \!=\! \left[-\lambda_{i}^{\widehat{S}} (t\!+\!1)\prod_{k \in \partial i}\frac{\theta_A^{k \rightarrow i} (t\!+\!1)\theta_B^{k \rightarrow i} (t\!+\!1)}{\theta_A^{k \rightarrow i}(t)\theta_B^{k \rightarrow i}(t)}-\lambda_{i}^{\widehat{A}} (t\!+\!1)\prod_{k \in \partial i}\left(1-\frac{\alpha_B \phi_B^{k \rightarrow i}(t)}{\theta_B^{k \rightarrow i}(t)}\right)\left(1-\prod_{k \in \partial i}\left(1-\frac{\alpha_A \phi_A^{k \rightarrow i}(t)}{\theta_A^{k \rightarrow i}(t)}\right)\right) \right.\\
		& \left. - \lambda_{i}^{\widehat{B}} (t\!+\!1)\prod_{k \in \partial i}\left(1-\frac{\alpha_A \phi_A^{k \rightarrow i}(t)}{\theta_A^{k \rightarrow i}(t)}\right)\left(1-\prod_{k \in \partial i}\left(1-\frac{\alpha_B \phi_B^{k \rightarrow i}(t)}{\theta_B^{k \rightarrow i}(t)}\right)\right)\right]1\left[t \neq T\right] + \lambda_{i}^S(t)
 \\
	\frac{\partial \mathcal{L}}{\partial P_A^{i \rightarrow j}(t)} & \!=\! \left[\lambda_{ij}^{\phi_A} (t\!+\!1) - \lambda_{ij}^{\widehat{A}} (t\!+\!1)\right]1\left[t \neq T\right] + \lambda_{ij}^A(t)-1\left[t \neq 0\right]\left[\lambda_{ij}^{\phi_A}(t)\right]
 \\
	\frac{\partial \mathcal{L}}{\partial P_A^{i}(t)} &\!=\! \left[-\lambda_{i}^{\widehat{A}} (t\!+\!1)\right]1\left[t \neq T\right] + \lambda_i^A(t)
\\
 		\frac{\partial \mathcal{L}}{\partial\widehat{P}_A^{i \rightarrow j}(t)} & \!=\! \lambda_{ij}^{\widehat{A}}(t)+ 1\left[t \neq 0\right]\left[-\lambda_{ij}^A(t)\frac{\widehat{P}_S^{i \rightarrow j}(t)+\widehat{P}_B^{i \rightarrow j}(t)}{(\widehat{P}_S^{i \rightarrow j}(t)+\widehat{P}_A^{i \rightarrow j}(t)+\widehat{P}_B^{i \rightarrow j}(t))^2}+\lambda_{ij}^S(t)\frac{\widehat{P}_S^{i \rightarrow j}(t)}{(\widehat{P}_S^{i \rightarrow j}(t)+\widehat{P}_A^{i \rightarrow j}(t)+\widehat{P}_B^{i \rightarrow j}(t))^2} \right.\\ 
		& \left. +\lambda_{ij}^B(t)\frac{\widehat{P}_B^{i \rightarrow j}(t)}{(\widehat{P}_S^{i \rightarrow j}(t)+\widehat{P}_A^{i \rightarrow j}(t)+\widehat{P}_B^{i \rightarrow j}(t))^2}\right]
\\	
		\frac{\partial \mathcal{L}}{\partial\widehat{P}_A^{i}(t)} & \!=\! \lambda_{i}^{\widehat{A}}(t)+1[t \neq 0]\left[-\lambda_{i}^A(t)\frac{\widehat{P}_S^{i}(t)+\widehat{P}_B^{i}(t)}{(\widehat{P}_S^{i}(t)+\widehat{P}_A^{i}(t)+\widehat{P}_B^{i}(t))^2}+\lambda_{i}^S(t)\frac{\widehat{P}_S^{i}(t)}{(\widehat{P}_S^{i}(t)+\widehat{P}_A^{i}(t)+\widehat{P}_B^{i}(t))^2} \right.\\
		& \left. +\lambda_{i}^B(t)\frac{\widehat{P}_B^{i}(t)}{(\widehat{P}_S^{i}(t)+\widehat{P}_A^{i}(t)+\widehat{P}_B^{i}(t))^2}\right]
\\			 
		\frac{\partial \mathcal{L}}{\partial \theta_A^{i \rightarrow j}(t)} 
		& \!=\! \lambda_{ij}^{\theta_A}(t) - 1[t \neq T]\left(\lambda_{ij}^{\theta_A} (t\!+\!1)\right) 
		+\sum_{a \in \partial j \backslash i}\lambda_{ja}^{\widehat{S}} (t\!+\!1)P_S^{j \rightarrow a}(t)\frac{1}{\theta_A^{i \rightarrow j}(t)}\prod_{l \in \partial j \backslash a}\frac{\theta_A^{l \rightarrow j} (t\!+\!1)\theta_B^{l \rightarrow j} (t\!+\!1)}{\theta_A^{l \rightarrow j}(t)\theta_B^{l \rightarrow j}(t)}1[t \neq T]\\
		& - \sum_{a \in \partial j \backslash i}\lambda_{ja}^{\widehat{S}}(t)P_S^{j \rightarrow a} (t\!-\!1)\frac{1}{\theta_A^{i \rightarrow j}(t)}\prod_{l \in \partial j \backslash a}\frac{\theta_A^{l \rightarrow j}(t)\theta_B^{l \rightarrow j}(t)}{\theta_A^{l \rightarrow j} (t\!-\!1)\theta_B^{l \rightarrow j} (t\!-\!1)}1[t \neq 0]\\
		& + \lambda_j^{\widehat{S}} (t\!+\!1)P_S^{j}(t)\frac{1}{\theta_A^{i \rightarrow j}(t)}\prod_{l \in \partial j}\frac{\theta_A^{l \rightarrow j} (t\!+\!1)\theta_B^{l \rightarrow j} (t\!+\!1)}{\theta_A^{l \rightarrow j}(t)\theta_B^{l \rightarrow j}(t)}1[t \neq T] 
		 - \lambda_j^{\widehat{S}}(t)P_S^j (t\!-\!1)\frac{1}{\theta_A^{i \rightarrow j}(t)}\prod_{l \in \partial j}\frac{\theta_A^{l \rightarrow j}(t)\theta_B^{l \rightarrow j}(t)}{\theta_A^{l \rightarrow j} (t\!-\!1)\theta_B^{l \rightarrow j} (t\!-\!1)}1[t \neq 0]\\
		& - \sum_{a \in \partial j \backslash i}\lambda_{ja}^{\widehat{A}} (t\!+\!1)P_S^{j \rightarrow a}(t)\left[1-\frac{\alpha_A \phi_A^{i \rightarrow j}}{(\theta_A^{i \rightarrow j}(t))^2}\prod_{l \in \partial j \backslash a,i}\left(1-\frac{\alpha_A \phi_A^{l \rightarrow j}(t)}{\theta_A^{l \rightarrow j}(t)}\right)\right]\prod_{l \in \partial j \backslash a}\left(1-\frac{\alpha_B \phi_B^{l \rightarrow j}(t)}{\theta_B^{l \rightarrow j}(t)}\right)1[t \neq T]\\
		& - \sum_{a \in \partial j \backslash i}\lambda_{ja}^{\widehat{B}} (t\!+\!1)P_S^{j \rightarrow a}(t)\left[\frac{\alpha_A \phi_A^{i \rightarrow j}}{(\theta_A^{i \rightarrow j}(t))^2}\prod_{l \in \partial j \backslash a,i}\left(1-\frac{\alpha_A \phi_A^{l \rightarrow j}(t)}{\theta_A^{l \rightarrow j}(t)}\right)\right]\left(1-\prod_{l \in \partial j \backslash a}\left(1-\frac{\alpha_B \phi_B^{l \rightarrow j}(t)}{\theta_B^{l \rightarrow j}(t)}\right)\right)1[t \neq T]\\
		& - \lambda_{j}^{\widehat{A}} (t\!+\!1)P_S^{j}(t)\left[1-\frac{\alpha_A \phi_A^{i \rightarrow j}}{(\theta_A^{i \rightarrow j}(t))^2}\prod_{l \in \partial j \backslash i}\left(1-\frac{\alpha_A \phi_A^{l \rightarrow j}(t)}{\theta_A^{l \rightarrow j}(t)}\right)\right]\prod_{l \in \partial j}\left(1-\frac{\alpha_B \phi_B^{l \rightarrow j}(t)}{\theta_B^{l \rightarrow j}(t)}\right)1[t \neq T]\\
		& - \lambda_{j}^{\widehat{B}} (t\!+\!1)P_S^{j}(t)\left[\frac{\alpha_A \phi_A^{i \rightarrow j}}{(\theta_A^{i \rightarrow j}(t))^2}\prod_{l \in \partial j \backslash i}\left(1-\frac{\alpha_A \phi_A^{l \rightarrow j}(t)}{\theta_A^{l \rightarrow j}(t)}\right)\right]\left(1-\prod_{l \in \partial j}\left(1-\frac{\alpha_B \phi_B^{l \rightarrow j}(t)}{\theta_B^{l \rightarrow j}(t)}\right)\right)1[t \neq T]\\
\end{split}
\end{equation}
\begin{equation}
\label{eq:derivativesComp2}
\begin{split}
		\frac{\partial \mathcal{L}}{\partial \phi_A^{i \rightarrow j}(t)} & \!=\! \lambda_{ij}^{\phi_A}(t)+\left[p\lambda_{ij}^{\theta_A} (t\!+\!1) + (\alpha_A -1)\lambda_{ij}^{\phi_A} (t\!+\!1)\right]1[t \neq T]\\
		& -\sum_{a \in \partial j \backslash i}\lambda_{ja}^{\widehat{A}} (t\!+\!1)P_S^{j \rightarrow a}(t)\left[\prod_{l \in \partial j \backslash a}\left(1-\frac{\alpha_B \phi_B^{l \rightarrow j}(t)}{\theta_B^{l \rightarrow j}(t)}\right)\right]\left(1 + \frac{\alpha_A}{\theta_A^{i \rightarrow j}(t)}\prod_{l \in \partial j \backslash a,i}\left(1-\frac{\alpha_A \phi_A^{l \rightarrow j}(t)}{\theta_A^{l \rightarrow j}(t)}\right)\right) 1[t \neq T] \\
		& + \sum_{a \in \partial j \backslash i}\lambda_{ja}^{\widehat{B}} (t\!+\!1)P_S^{j \rightarrow a}(t)\frac{\alpha_A}{\theta_A^{i \rightarrow j}(t)}\left[\prod_{l \in \partial j \backslash a,i}\left(1-\frac{\alpha_A \phi_A^{l \rightarrow j}(t)}{\theta_A^{l \rightarrow j}(t)}\right)\right]\left(1-\prod_{l \in \partial j \backslash a}\left(1-\frac{\alpha_B \phi_B^{l \rightarrow j}(t)}{\theta_B^{l \rightarrow j}(t)}\right)\right)1[t \neq T]\\
		& - \lambda_{j}^{\widehat{A}} (t\!+\!1)P_S^{j}(t)\left[\prod_{l \in \partial j}\left(1-\frac{\alpha_B \phi_B^{l \rightarrow j}(t)}{\theta_B^{l \rightarrow j}(t)}\right)\right]\left(1 + \frac{\alpha_A}{\theta_A^{i \rightarrow j}(t)}\prod_{l \in \partial j \backslash i}\left(1-\frac{\alpha_A \phi_A^{l \rightarrow j}(t)}{\theta_A^{l \rightarrow j}(t)}\right)\right)1[t \neq T]\\
		&+ \lambda_{j}^{\widehat{B}} (t\!+\!1)P_S^{j}(t)\frac{\alpha_A}{\theta_A^{i \rightarrow j}(t)}\left[\prod_{l \in \partial j \backslash i}\left(1-\frac{\alpha_A \phi_A^{l \rightarrow j}(t)}{\theta_A^{l \rightarrow j}(t)}\right)\right]\left(1-\prod_{l \in \partial j}\left(1-\frac{\alpha_B \phi_B^{l \rightarrow j}(t)}{\theta_B^{l \rightarrow j}(t)}\right)\right)1[t \neq T]\\
	\end{split}
\end{equation}

\begin{equation}
	\begin{split}
	\frac{\partial \mathcal{L}}{\partial \nu^i(0)} \!=\! & -\lambda^{Bu}(0)-\lambda_i^{\widehat{A}}(0)-\lambda_i^A(0)-\sum_j\lambda_{ij}^{\widehat{A}}(0)-\sum_j\lambda_{ij}^A(0)-\sum_j\lambda_{ij}^{\phi_A}(0) \\
	& + \lambda_{i}^{\widehat{S}}(0)+\lambda_{i}^S(0)+\sum_j\lambda_{ij}^{\widehat{S}}(0)+\sum_j\lambda_{ij}^S(0)+\epsilon\left(\frac{1}{\nu^i(0)}-\frac{1}{1-\nu^i(0)}\right)\!=\!0\label{qua}
	\end{split}
\end{equation}
}

The equations for ${\partial \mathcal{L}}/{\partial P_B^{i \rightarrow j}(t)}$, ${\partial \mathcal{L}}/{\partial P_B^{i}(t)}$, ${\partial \mathcal{L}}/{\partial\widehat{P}_B^{i \rightarrow j}(t)}$, ${\partial \mathcal{L}}/{\partial\widehat{P}_B^{i}(t)}$, ${\partial \mathcal{L}}/{\partial \theta_B^{i \rightarrow j}(t)} $, ${\partial \mathcal{L}}/{\partial \phi_B^{i \rightarrow j}(t)}$ are similar to their $A$ process counterparts with an exchange of variables $A  \leftrightarrow B$.

For simplicity one can write equation~\eqref{qua} as:
\begin{equation}
	\frac{\partial \mathcal{L}}{\partial \nu^i(0)} \!=\! -\lambda^{Bu}(0) + \psi_i + \epsilon\left(\frac{1}{\nu^i(0)}-\frac{1}{1-\nu^i(0)}\right)\label{simqua}\!=\!0
\end{equation}
The same method as in~\cite{Lokhov2017} is used here to solve the quadratic equation~\eqref{simqua}, writing $\nu^i(0)$ as a function of $\psi_i$:
\begin{equation}
	\nu^i(0) \!=\! \frac{\lambda^{Bu}(0) + \psi_i - 2\epsilon \pm \sqrt{(-\lambda^{Bu}(0) + \psi_i)^2 + 4\epsilon^2}}{-2\lambda^{Bu}(0) + 2\psi_i}\label{numerical1}
\end{equation}
In this scenario, the positive square root solution satisfies restriction~\eqref{Restriction}. Given the budget restriction one can obtain $\nu^i(0)$ numerically through
\begin{equation}
	\sum_i{\nu^i(0)} \!=\! B_{\nu}\label{numerical2}
\end{equation}

One can use the following update procedure to obtain optimal solution iteratively:

\begin{enumerate}
\item[\textbf{Step 1:}] Carry out the forward iteration using given initial conditions, Eqs.~\eqref{eq:Mutually_exclusive_0}-\eqref{eq:Mutually_exclusive} 
\item[\textbf{Step 2:}] Compute all  Lagrangian multipliers at $t \!=\! T$. 
\item[\textbf{Step 3:}] Compute $\lambda_{ij}^S (t\!-\!1), \lambda_i^S (t\!-\!1), \lambda_i^A (t\!-\!1), \lambda_i^B (t\!-\!1), \lambda_{ij}^{\phi_A} (t\!-\!1), \lambda_{ij}^{\phi_B} (t\!-\!1)$ using obtained Lagrangian multipliers at time $t$ from Eqs.(\ref{eq:derivativesComp1})-(\ref{eq:derivativesComp2}).
\item[\textbf{Step 4:}] Compute
	$\lambda_{ij}^A (t\!-\!1), \lambda_{ij}^B (t\!-\!1), \lambda_i^{\widehat{A}} (t\!-\!1),\lambda_i^{\widehat{B}} (t\!-\!1),\lambda_i^{\widehat{S}} (t\!-\!1)$ using obtained Lagrangian multipliers and Eqs.(\ref{eq:derivativesComp1})-(\ref{eq:derivativesComp2}).
\item[\textbf{Step 5:}] Compute
	$\lambda_{ij}^{\widehat{A}} (t\!-\!1), \lambda_{ij}^{\widehat{B}} (t\!-\!1), \lambda_{ij}^{\widehat{S}} (t\!-\!1)$ using obtained Lagrangian multipliers and Eqs.(\ref{eq:derivativesComp1})-(\ref{eq:derivativesComp2}).
\item[\textbf{Step 6:}] Compute $\lambda_{ij}^{\theta_A} (t\!-\!1), \lambda_{ij}^{\theta_B} (t\!-\!1)$ using obtained Lagrangian multipliers.
\item[\textbf{Step 7:}] Repeat Steps 1-5 back in time until all Lagrangian multipliers have been obtained for the range $t\!=\!0,\ldots,T$.
\item[\textbf{Step 8:}] Solve Eqs.~\eqref{numerical1} and \eqref{numerical2} numerically and update $\nu^i$.
\item[\textbf{Step 9:}] Repeat Step 1 to Step 8 until convergence.
\item[\textbf{Step 10:}] Calculate the posterior marginals on the basis of the messages using Eqs.~(\ref{eq:posteriorComp}).
\end{enumerate}

A similar approach is used in collaborative processes using the corresponding equations for the dynamics and the Lagrange multipliers' dynamics.

\section{Optimization of Competitive Processes - Example}
\label{app:WTNexample}
We also demonstrate the efficacy of DMP-optimal algorithm on the denser network of the 1994 world metal trade network~\cite{nooy2011} .The network consists of 80 countries; we will ignore their respective trade volumes and use the dense topology for the current example. The infection probabilities used for the two processes are $\alpha_A \!=\! \alpha_B \!=\! 0.5$; the budget for $B$ is 1, allocated at node 1 (Argentina) at $t\!=\!0$, and a budget of 1 per time step is assigned using the DMP-optimal strategy for process $A$ within the time window of $T\!=\!3$. The results are shown in Fig.~\ref{fig:t123}, where the subfigures represent the containment of process $B$ at different times. The heat bar represents the dominating process per node through the value $P_A^i(t)\!-\!P_B^i(t)$. Red/blue represent dominating processes $A/B$, respectively.
	\begin{figure}[!ht]
		\centering  
		\includegraphics[width=0.8\textwidth]{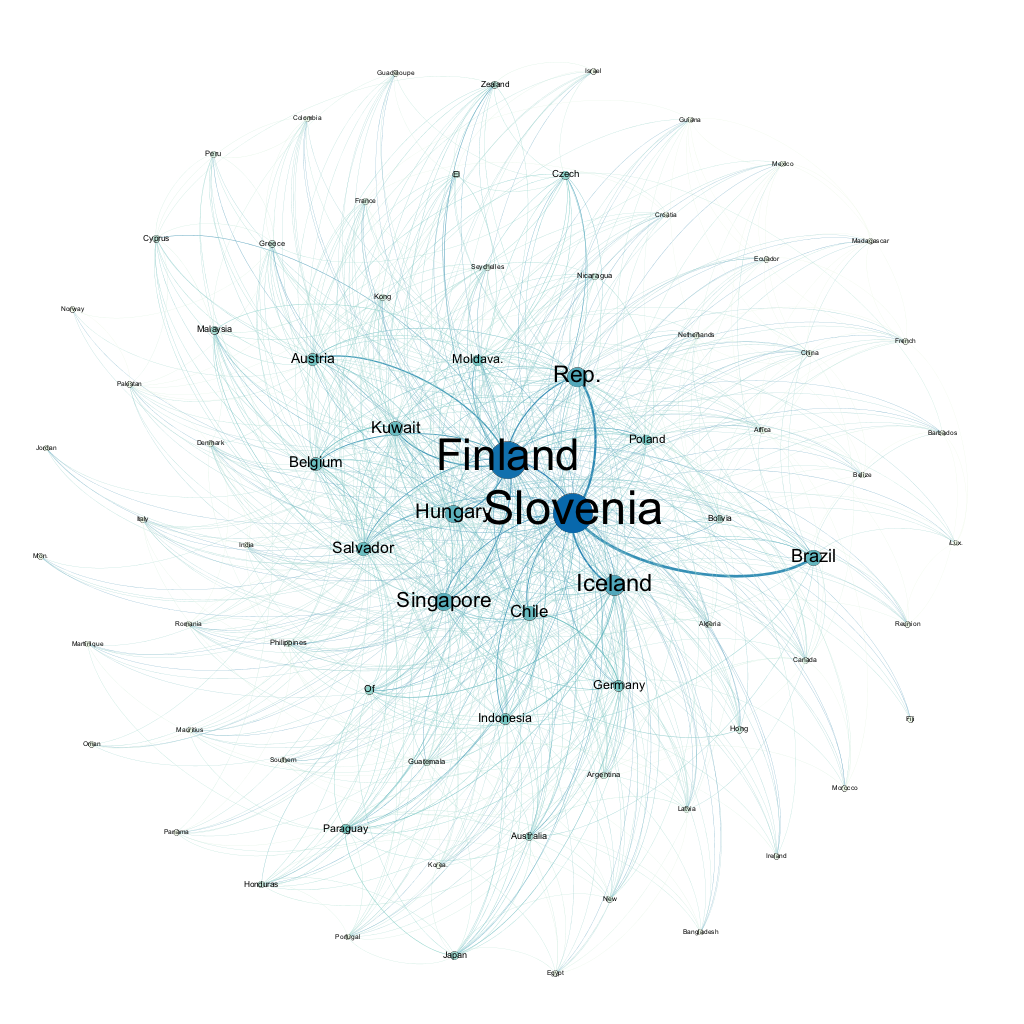}  
		\caption{The 1994 network of world trade in metal. The size of nodes represents their degree~\cite{gephi_WTN}. Figure courtesy of Hanyu Zhang.\label{fig:worldtrade}}
	\end{figure} 
	
	\begin{figure} [!ht]
		\centering 
		\subfigure[~$t\!=\!1$]{ 
			\label{fig:subfig:t1} 
			\includegraphics[width=2.7in]{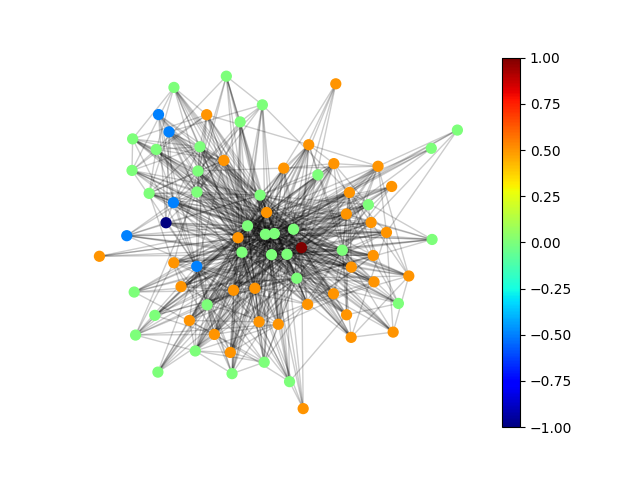}} 
		\hspace{-0.35in} 
		\subfigure[~$t\!=\!2$]{ 
			\label{fig:subfig:t2} 
			\includegraphics[width=2.7in]{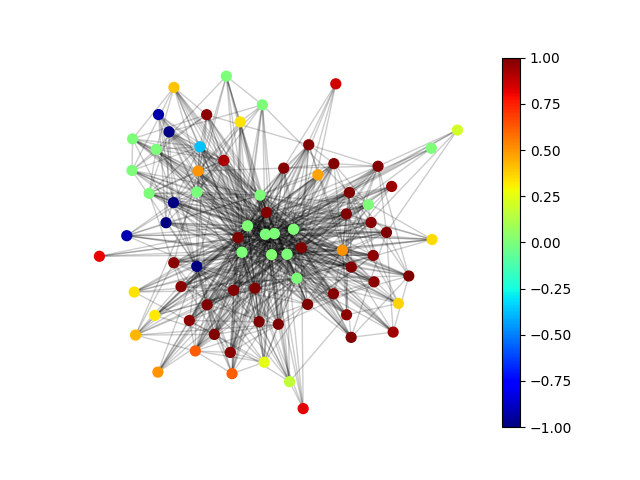}} 
		\hspace{-0.35in} 
		\subfigure[~$t\!=\!3$]{ 
			\label{fig:subfig:t3}
			\includegraphics[width=2.7in]{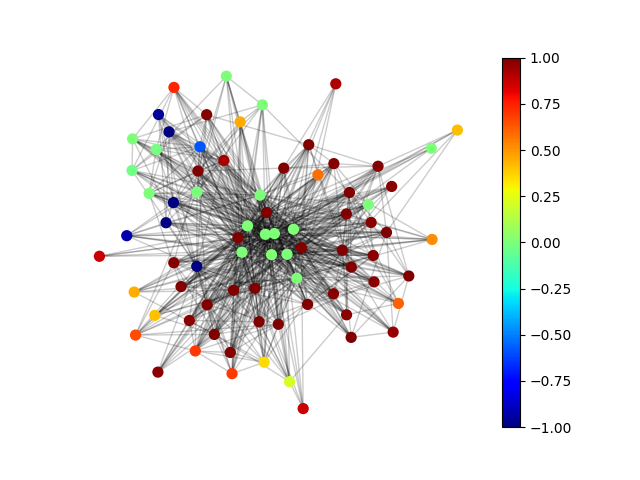}} 
		\caption{The 1994  world metal trade network. The infection parameters used for the two processes are $\alpha_A \!=\! \alpha_B \!=\! 0.5$, the budget for $B$ is 1, allocated at node 1 (Argentina) at $t\!=\!0$. A budget of 1 per time step is deployed using DMP-optimal for process $A$ for a time window of $T\!=\!3$. The subfigures represent the containment of process $B$ at the different times. The heat bar represents the dominating process per node through the value $P_A^i(t)\!-\!P_B^i(t)$. Red/blue represent dominating processes $A/B$, respectively.}
		\label{fig:t123} 
			\end{figure}

\section{Optimization of Collaborative Processes}
\label{app:OptCollaborative}
Optimization in collaborative processes is highly relevant in many health-related problems such as the co-epidemic of HIV and tuberculosis, and in the marketing of correlated products or opinions. In the co-epidemic case, identifying the most catastrophic health-risk scenario is an important aspect of evaluating the risk of a co-epidemic event; while in the marketing area, it aims at the best joint advertising campaign. 

We consider two main scenarios in our model: (a) The multi-agent seeding problem, i.e., computing the optimal allocation of one agent which minimizes the number of susceptible vertices at a certain time $T$ for a given spread of the second agent. (b) Optimal allocation of vaccines against one of the processes for maximizing the impact on the spread of both processes. Assume we do not have any vaccines for HIV but vaccines for tuberculosis are available (such as Bacillus Calmette-Guérin~\cite{wiki:xxx1}); the optimization problem would be articulated as: given a certain budget of vaccines for tuberculosis, what is the best vaccination policy to minimize the spreading of both?

The Lagrangian used is similar to the one of the competitive case but with different dynamics and initial conditions. The dynamics equations are enforced through a set of Lagrange multipliers $\lambda$ as before
{\small
\begin{equation}
	\begin{split}
		\mathcal{D} & \!=\! \sum_{ij}\sum_{t\!=\!0}^{T-1}\lambda_{ij}^{\theta_A} (t\!+\!1)\left[\theta_A^{i \rightarrow j} (t\!+\!1)-\theta_A^{i \rightarrow j}(t)+\alpha_A\phi_A^{i \rightarrow j}(t)\right] \\
		  &\!+\!\sum_{ij}\sum_{t\!=\!0}^{T-1}\lambda_{ij}^{\theta_B} (t\!+\!1)\left[\theta_B^{i \rightarrow j} (t\!+\!1)-\theta_B^{i \rightarrow j}(t)+\alpha_B\phi_B^{i \rightarrow j}(t)\right]\\
		  &\!+\!\sum_{ij}\sum_{t\!=\!0}^{T-1}\lambda_{ij}^S (t\!+\!1)\left[P_S^{i \rightarrow j} (t\!+\!1)-P_S^i(0)\prod_{k \in \partial i \backslash j}\theta_A^{k \rightarrow i}  (t\!+\!1)\theta_B^{k \rightarrow i} (t\!+\!1)\right]\\
		  &\!+\!\sum_{ij}\sum_{t\!=\!0}^{T-1}\lambda_{ij}^{\phi_A} (t\!+\!1)\left[\phi_A^{i \rightarrow j} (t\!+\!1)-\phi_A^{i \rightarrow j}(t) +\alpha_A\phi_A^{i \rightarrow j}(t)-P_A^{i \rightarrow j} (t\!+\!1)\!+\!P_A^{i \rightarrow j}(t)\right]\\
		  &\!+\!\sum_{ij}\sum_{t\!=\!0}^{T-1}\lambda_{ij}^{\phi_B} (t\!+\!1)\left[\phi_B^{i \rightarrow j} (t\!+\!1)-\phi_B^{i \rightarrow j}(t)\!+\!\alpha_B\phi_B^{i \rightarrow j}(t)-P_B^{i \rightarrow j} (t\!+\!1)\!+\!P_B^{i \rightarrow j}(t)\right]\\
		  &\!+\!\sum_{ij}\sum_{t\!=\!0}^{T-1}\lambda_{ij}^A (t\!+\!1)\left[P_A^{i \rightarrow j} (t\!+\!1)-P_A^{i \rightarrow j}(t)-P_S^{i \rightarrow j}(t)\left[1\!-\!\prod_{k \in \partial i \backslash j}\left(1\!-\!\frac{\alpha_A\phi_A^{k \rightarrow i}(t)}{\theta_A^{k \rightarrow i}(t)}\right)\right] \right.\\
		  &\left. \!-\! P_{B_{\mbox{*}}}^{i \rightarrow j}(t)\left[1\!-\!\prod_{k \in \partial i \backslash j}\left(1\!-\!\frac{\alpha_{AB}\phi_A^{k \rightarrow i}(t)}{\theta_A^{k \rightarrow i}(t)}\right)\right]\right]\\
		  &\!+\!\sum_{ij}\sum_{t\!=\!0}^{T-1}\lambda_{ij}^B (t\!+\!1)\left[P_B^{i \rightarrow j} (t\!+\!1)\!-\!P_B^{i \rightarrow j}(t)\!-\!P_S^{i \rightarrow j}(t)\left[1\!-\!\prod_{k \in \partial i \backslash j}\left(1\!-\!\frac{\alpha_B\phi_B^{k \rightarrow i}(t)}{\theta_B^{k \rightarrow i}(t)}\right)\right] \right. \\
		  & \left. \!-\!P_{A_{\mbox{*}}}^{i \rightarrow j}(t)\left[1\!-\!\prod_{k \in \partial i \backslash j}\left(1-\frac{\alpha_{BA}\phi_B^{k \rightarrow i}(t)}{\theta_B^{k \rightarrow i}(t)}\right)\right]\right]\\
		  &\!+\!\sum_{ij}\sum_{t\!=\!0}^{T-1}\lambda_{ij}^{AB} (t\!+\!1)\left[P_{AB}^{i \rightarrow j} (t\!+\!1)\!-\!P_A^{i \rightarrow j} (t\!+\!1)\!-\!P_B^{i \rightarrow j} (t\!+\!1)\!-\!P_S^{i \rightarrow j} (t\!+\!1)\!+\!1\right]\\
		  &\!+\!\sum_{ij}\sum_{t\!=\!0}^{T-1}\lambda_{ij}^{A_{\mbox{*}}} (t\!+\!1)\left[P_{A_{\mbox{*}}^{i \rightarrow j}} (t\!+\!1)\!-\!P_A^{i \rightarrow j} (t\!+\!1)\!+\!P_{AB}^{i \rightarrow j} (t\!+\!1)\right]\\
		  &\!+\!\sum_{ij}\sum_{t\!=\!0}^{T-1}\lambda_{ij}^{B_{\mbox{*}}} (t\!+\!1)\left[P_{B_{\mbox{*}}}^{i \rightarrow j} (t\!+\!1)\!-\!P_B^{i \rightarrow j} (t\!+\!1)\!+\!P_{AB}^{i \rightarrow j} (t\!+\!1)\right]\\
		  &\!+\!\sum_{i}\sum_{t\!=\!0}^{T-1}\lambda_i^S (t\!+\!1)\left[P_S^i (t\!+\!1)-P_S^i(0)\prod_{k \in \partial i}\theta_A^{k \rightarrow i} (t\!+\!1)\theta_B^{k \rightarrow i} (t\!+\!1)\right]\\
		  &\!+\!\sum_{i}\sum_{t\!=\!0}^{T-1}\lambda_{i}^{AB} (t\!+\!1)\left[P_{AB}^{i} (t\!+\!1)\!-\!P_A^{i} (t\!+\!1)\!-\!P_B^{i} (t\!+\!1)\!-\!P_S^{i} (t\!+\!1)\!+\!1\right]\\
		  &\!+\!\sum_{i}\sum_{t\!=\!0}^{T-1}\lambda_{i}^A (t\!+\!1)\left[P_A^{i} (t\!+\!1)\!-\!P_A^{i}(t)\!-\!P_S^{i}(t)\left(1\!-\!\prod_{k \in \partial i}\left(1\!-\!\frac{\alpha_A\phi_A^{k \rightarrow i}(t)}{\theta_A^{k \rightarrow i}(t)}\right)\right)\!-\!P_{B_{\mbox{*}}}^{i}(t)\left(1\!-\!\prod_{k \in \partial i}\left(1\!-\!\frac{\alpha_{AB}\phi_A^{k \rightarrow i}(t)}{\theta_A^{k \rightarrow i}(t)}\right)\right)\right]\\
		  &\!+\!\sum_{i}\sum_{t\!=\!0}^{T-1}\lambda_{i}^B (t\!+\!1)\left[P_B^{i} (t\!+\!1)\!-\!P_B^{i}(t)\!-\!P_S^{i}(t)\left(1\!-\!\prod_{k \in \partial i}\left(1\!-\!\frac{\alpha_B\phi_B^{k \rightarrow i}(t)}{\theta_B^{k \rightarrow i}(t)}\right)\right) \!-\!P_{A_{\mbox{*}}}^{i}(t)\left(1\!-\!\prod_{k \in \partial i}\left(1\!-\!\frac{\alpha_{BA}\phi_B^{k \rightarrow i}(t)}{\theta_B^{k \rightarrow i}(t)}\right)\right)\right]\\
		  &\!+\!\sum_{i}\sum_{t\!=\!0}^{T-1}\lambda_{i}^{A_{\mbox{*}}} (t\!+\!1)\left[P_{A_{\mbox{*}}}^{i} (t\!+\!1)\!-\!P_A^{i} (t\!+\!1)\!+\!P_{AB}^{i} (t\!+\!1)\right]
		  \!+\!\sum_{i}\sum_{t\!=\!0}^{T-1}\lambda_{i}^{B_{\mbox{*}}} (t\!+\!1)\left[P_{B_{\mbox{*}}}^{i} (t\!+\!1)\!-\!P_B^{i} (t\!+\!1)\!+\!P_{AB}^{i} (t\!+\!1)\right]\\
	\end{split}
\end{equation}}

The initial conditions in this case are of the form:
\begin{equation}
	\begin{split}
		\mathcal{I} & \!=\! \sum_{ij}\lambda_{ij}^{\theta_A}(0)\left(\theta_A^{i \rightarrow j}(0)-1\right) + \sum_{ij}\lambda_{ij}^{\theta_B}(0)\left(\theta_B^{i \rightarrow j}(0)-1\right)\\
		& + \sum_{ij}\lambda_{ij}^{\phi_A}(0)\left(\phi_A^{i \rightarrow j}(0)-\nu^i(0)\left(1-\delta_{\sigma_i(0), B}\right)\right) + \sum_{ij}\lambda_{ij}^{\phi_B}(0)\left(\phi_B^{i \rightarrow j}(0)-\delta_{\sigma_i(0), B}\right)\\
		& + \sum_{ij}\lambda_{ij}^S(0)\left(P_S^{i \rightarrow j}(0)-1+\nu^i(0)\left(1-\delta_{\sigma_i(0), B}\right)+\delta_{\sigma_i(0), B}\right)\\
		& + \sum_{ij}\lambda_{ij}^A(0)\left(P_A^{i \rightarrow j}(0)-\nu^i(0)\left(1-\delta_{\sigma_i(0), B}\right)\right) + \sum_{ij}\lambda_{ij}^B(0)\left(P_B^{i \rightarrow j}(0)-\delta_{\sigma_i(0),B}\right)+\sum_{ij}\lambda_{ij}^{AB}(0)P_{AB}^{i \rightarrow j}(0)\\
		& + \sum_{ij}\lambda_{ij}^{A_{\mbox{*}}}(0)\left(P_{A_{\mbox{*}}}^{i \rightarrow j}(0)-\nu^i(0)\left(1-\delta_{\sigma_i(0), B}\right)\right) + \sum_{ij}\lambda_{ij}^{B_{\mbox{*}}}(0)\left(P_{B_{\mbox{*}}}^{i \rightarrow j}(0)-\delta_{\sigma_i(0), B}\right)\\
		& + \sum_{i}\lambda_i^S(0)\left(P_S^i(0)-1+\nu^i(0)\left(1-\delta_{\sigma_i(0), B}\right)+\delta_{\sigma_i(0), B}\right)\\
		& + \sum_{i}\lambda_{i}^{A_{\mbox{*}}}(0)\left(P_{A_{\mbox{*}}}^{i}(0)-\nu^i(0)\left(1-\delta_{\sigma_i(0), B}\right)\right) + \sum_{i}\lambda_{i}^{B_{\mbox{*}}}(0)\left(P_{B_{\mbox{*}}}^{i}(0)-\delta_{\sigma_i(0), B}\right)\\	
		& + \sum_{i}\lambda_{i}^A(0)\left(P_A^{i}(0)-\nu^i(0)\left(1-\delta_{\sigma_i(0), B}\right)\right) + \sum_{i}\lambda_{i}^B(0)\left(P_B^{i}(0)-\delta_{\sigma_i(0),B}\right)+\sum_{i}\lambda_{i}^{AB}(0)P_{AB}^{i}(0)\\
		\end{split}
\end{equation}

\section{Optimization of Vaccine Allocation}
\label{app:Vaccine Allocation}

The multi-process seeding problem in the collaborative case has a similar structure to that of the containment task in competitive processes. The vaccine allocation problem is of a slightly different nature. There are many ways in which it can be formulated, the model illustrated below is the one we have chosen to use.

If a node receives a unit of vaccine before being exposed to infected neighbors, it will be immune to that process (denoted as $B$ in our model), which means the infection 
message from its neighbor will decrease to 0. However, sometime a node receives part of a unit of vaccine; specifically, if one allocates a certain amount of vaccine, say $b$ to a node, the probability of it being infected will decrease to $\alpha_B \!-\! b$ (bounded by 0 from below), where $\alpha_B$ is the initial infection parameter. Meanwhile, the parameter $\alpha_{BA}$ will also decrease to $\alpha_{BA} \!-\! b$. Therefore the budget for vaccines can be expressed by the budget for decreasing $\alpha_B$ and $\alpha_{BA}$ in Eqs. \eqref{eq:Collaborative_0}-\eqref{eq:Collaborative}, which now take the following form:
\begin{eqnarray}
	\theta_B^{i \rightarrow j}(t)&\!-\!&\theta_B^{i \rightarrow j} (t\!-\!1) \!=\! \!-\!(\alpha_B\!-\!b(j))\phi_B^{i \rightarrow j} (t\!-\!1)
\\
	\phi_B^{i \rightarrow j}(t)&\!-\!&\phi_B^{i \rightarrow j} (t\!-\!1)\!=\! \!-\!(\alpha_B\!-\!b(j))\phi_B^{i \rightarrow j} (t\!-\!1) \!+\! P_B^{i \rightarrow j}(t) \!-\! P_B^{i \rightarrow j} (t\!-\!1)
\nonumber \\
	P_B^{i}(t) &\!=\!& P_B^{i} (t\!-\!1) \!+\! P_S^{i} (t\!-\!1)\left(1\!-\!\prod_{k \in \partial i}\left(1\!-\!\frac{(\alpha_B\!-\!b(i))\phi_B^{k \rightarrow i} (t\!-\!1)}{\theta_B^{k \rightarrow i} (t\!-\!1)}\right)\right)\nonumber \\ 
	&\!+\!& P_{A_{\mbox{*}}}^{i} (t\!-\!1)\left(1\!-\!\prod_{k \in \partial i}\left(1\!-\!\frac{(\alpha_{BA}\!-\!b(i))\phi_B^{k \rightarrow i} (t\!-\!1)}{\theta_B^{k \rightarrow i} (t\!-\!1)}\right)\right) 
\nonumber \\
	P_B^{i \rightarrow j}(t) &\!=\!& P_B^{i \rightarrow j} (t\!-\!1) \!+\! P_S^{i \rightarrow j} (t\!-\!1)\left(1\!-\!\prod_{k \in \partial i \backslash j}\left(1\!-\!\frac{(\alpha_B\!-\!b(i))\phi_B^{k \rightarrow i} (t\!-\!1)}{\theta_B^{k \rightarrow i} (t\!-\!1)}\right)\right) \nonumber \\
	&\!+\!& P_{A_{\mbox{*}}}^{i \rightarrow j} (t\!-\!1)\left(1\!-\!\prod_{k \in \partial i \backslash j}\left(1\!-\!\frac{(\alpha_{BA}\!-\!b(i))\phi_B^{k \rightarrow i} (t\!-\!1)}{\theta_B^{k \rightarrow i} (t\!-\!1)}\right)\right) \nonumber
\end{eqnarray}

The budget restriction in this case is:
\begin{equation}
	\mathcal{B} \!=\! \lambda^{Bu}\sum_i \left(b(i)-Bu\right)
\end{equation} 
where $Bu$ is the total vaccination budget. Another restriction for $b(i)$ allocated on a node (not exceeding the infection probability) is expressed similarly to before as:
\begin{equation}
	\mathcal{P} \!=\! \epsilon\left[\log(b(i))\!+\!\log(\alpha_B - b(i))\right]
\end{equation}

Finally, the derivative of the part of the Lagrangian that enforces the dynamics $\mathcal{D}$ is differentiated with respect to $b(i)$, required to complete the set of equations provides
\begin{equation}
	\begin{split}
		\frac{\partial \mathcal{D}}{\partial b(i)} & \!=\!\sum_{t\!=\!0}^{T-1}\left[-\sum_{k \in \partial i}\lambda_{ki}^{\theta_B} (t\!+\!1)\phi_B^{k \rightarrow i}(t)-\sum_{k \in \partial i}\lambda_{ki}^{\phi_B} (t\!+\!1)\phi_B^{k \rightarrow i}(t) \right.\\
		& + \sum_j\lambda_{ij}^B (t\!+\!1)P_S^{i \rightarrow j}(t)\left(\sum_{l \in \partial i \backslash j}\frac{\phi_B^{l \rightarrow i}(t)}{\theta_B^{l \rightarrow i}(t)}\prod_{k \in \partial i \backslash j, l}\left(1 - \frac{(\alpha_B - b(i))\phi_B^{k \rightarrow i}(t)}{\theta_B^{k \rightarrow i}(t)}\right)\right)\\
		& + \sum_j\lambda_{ij}^B (t\!+\!1)P_{A_{\mbox{*}}}^{i \rightarrow j}(t)\left(\sum_{l \in \partial i \backslash j}\frac{\phi_B^{l \rightarrow i}(t)}{\theta_B^{l \rightarrow i}(t)}\prod_{k \in \partial i \backslash j, l}\left(1 - \frac{(\alpha_{BA} - b(i))\phi_B^{k \rightarrow i}(t)}{\theta_B^{k \rightarrow i}(t)}\right)\right)\\
		& + \lambda_{i}^B (t\!+\!1)P_S^{i}(t)\left(\sum_{l \in \partial i}\frac{\phi_B^{l \rightarrow i}(t)}{\theta_B^{l \rightarrow i}(t)}\prod_{k \in \partial i \backslash l}\left(1 - \frac{(\alpha_B - b(i))\phi_B^{k \rightarrow i}(t)}{\theta_B^{k \rightarrow i}(t)}\right)\right)\\
		&\left. + \lambda_{i}^B (t\!+\!1)P_{A_{\mbox{*}}}^{i}(t)\left(\sum_{l \in \partial i}\frac{\phi_B^{l \rightarrow i}(t)}{\theta_B^{l \rightarrow i}(t)}\prod_{k \in \partial i \backslash l}\left(1 - \frac{(\alpha_{BA} - b(i))\phi_B^{k \rightarrow i}(t)}{\theta_B^{k \rightarrow i}(t)}\right)\right)\right]\\
	\end{split}
\end{equation}

The same procedure (following after Eq.\eqref{simqua}) can be implemented for the optimization in this case.
\end{widetext}
\bibliography{./reference,./model}

\end{document}